\documentclass{jfm}

\usepackage{graphicx}
\usepackage{caption}
\usepackage{subcaption}
\usepackage{subcaption}
\usepackage{newtxtext}
\usepackage{newtxmath}
\usepackage{natbib}
\usepackage{hyperref}

\hypersetup{
    colorlinks = true,
    urlcolor   = blue,
    citecolor  = black,
}

\newcommand{\RomanNumeralCaps}[1]
\linenumbers

\usepackage{cleveref}
\crefformat{section}{\S#2#1#3}

\title{The onset and saturation of the Faraday instability in miscible fluids in a rotating environment.
}

\author{Narinder Singh\aff{1}
 and Anikesh Pal\aff{1}
  \corresp{\email{pala@iitk.ac.in}}}

\affiliation{\aff{1}Department of Mechanical Engineering, Indian Institute of Technology Kanpur, Kanpur 208016, U.P., India}

\begin{document}
\maketitle

\begin{abstract}
We investigate the influence of rotation on the onset and saturation of the Faraday instability in a vertically oscillating two-layer miscible fluid using a theoretical model and direct numerical simulations (DNS). Our analytical approach utilizes the Floquet analysis to solve a set of the Mathieu equations obtained from the linear stability analysis. The solution of the Mathieu equations comprises stable and harmonic, and sub-harmonic unstable regions in a three-dimensional stability diagram. We find that the Coriolis force stabilizes the flow and delays the onset of the sub-harmonic instability responsible for turbulent mixing at lower forcing amplitudes. However, at higher forcing amplitudes, the flow is energetic enough to mitigate the stabilizing effect of rotation, and the evolution of the turbulent mixing zone is similar in both rotating and non-rotating environments. These results are corroborated by DNS at different Coriolis frequencies and forcing amplitudes. We also observe that for  $\left(f/\omega\right)^2<0.25$, where $f$ is the Coriolis frequency, and $\omega$ is the forcing frequency, the instability, and the turbulent mixing zone size-$L$ saturates.  When $\left(f/\omega\right)^2\geq0.25$, the turbulent mixing zone size-$L$ never saturates and continues to grow.  
\end{abstract}

  
%{\bf MSC Codes }  {\it(Optional)} Please enter your MSC Codes here

 \section{Introduction}
 \label{sec:intro} 
The interface of two immiscible fluids enclosed in a vertically vibrating container upon reaching a certain frequency, and acceleration becomes unstable. This unstable phase is known as Faraday instability and results in the formation of nonlinear standing waves at the interface of the two fluids. These waves are termed as Faraday waves and were first described by Michael Faraday in 1831 \citep{faraday1831xvii}. The onset of Faraday instability in immiscible liquids was theoretically demonstrated by \citet{benjamin1954stability} using linear stability analysis of the interface of an ideal fluid governed by the set of equations relevant to the system of Mathieu equations. Linear stability of the immiscible, viscous and finite depth fluid problem for one and two frequency excitation was study by \citet{kumar1994parametric} and \citet{besson1996two} respectively, using the Floquet analysis. Apart from theoretical investigations, experimental \citep{gollub1983symmetry, douady1988pattern, simonelli1989surface, douady1990experimental, muller1993periodic, edwards1994patterns, binks1997nonlinear, kudrolli1998superlattice, arbell2002pattern, westra2003patterns, kityk2005spatiotemporal, rajchenbach2013observation, shao2020experimental, shao2021surface} and numerical \citep{perinet2009numerical, takagi2011numerical,kahouadji2015numerical,takagi2015numerical} studies were also performed by many researchers to understand the Faraday wave patterns at the interface of immiscible liquids. \\

Majority of the studies on Faraday waves deals with immiscible liquids. However, in the past few decades the researchers have shifted their focus to understand the dynamics of Faraday waves in miscible liquids.  \citet{zoueshtiagh2009experimental} performed experiments and two-dimensional (2D) numerical simulations to investigate the Faraday instability of diffuse interfaces between pairs of miscible
liquids of different densities in a rectangular cell. They observed that, above a certain forcing amplitude, the standing waves that appear on the diffuse interface are highly disorganized and interact with each other leading to the mixing of fluids followed by the disappearance of the waviness on the interface when the two fluids are mixed. They also found the mixing phase to be sub-harmonic, similar to the sub-harmonic nature of the Faraday waves in immiscible fluids. Further this finding was supported by \citet{diwakar2015faraday} who used Floquet theory in conjunction with a quasi-steady approximation to carry out linear analysis of Faraday instability in miscible fluids. An experimental and numerical framework similar to \citet{zoueshtiagh2009experimental} has been used by \citet{amiroudine2012mixing} to report the exponential growth of the mixing layer thickness owing to fingering at the interface, followed by a small growth rate that demonstrates the saturation of mixing. Recently, \cite{grea2018final} conducted three-dimensional (3D) numerical simulations and demonstrated that Faraday instability can generate turbulent mixing zones for strong forcing parameter and/or for sufficiently random initial condition at the interface of two miscible fluids. Further, they formulated a system of equations based on the second-order correlation spectra for turbulent quantities and used perturbation analysis to estimate the final size of the mixing zone:  
\begin{equation}
\label{Lsat}
L_{sat}=\frac{2\mathcal{A}g_0}{\omega^2}\left(2F+4\right),
\end{equation}
where $L_{sat}$ is the final size of mixing zone (defined later in the equation \ref{mz width}) in saturation state, $\mathcal{A}=\left(\rho_1-\rho_2\right)/\left(\rho_1+\rho_2\right)$ is the Atwood number expressing the density contrast between heavy fluid ($\rho_1$) and light fluid ($\rho_2$), $g_0$ is the mean acceleration, $F$ is the oscillation amplitude and $\omega$ is the forcing frequency. \cite{grea2018final} validated the predicted $L_{sat}$ against their simulations for a wide range of parameters within the homogeneous framework and full-inhomogeneous systems of two miscible fluids. They reported that the instability of the diffuse interface begins with a small harmonic phase, but the main instability phase is dominantly sub-harmonic, where the turbulent mixing zone develops. The irreversible mixing associated with the growth of the mixing zone owing to the harmonic to sub-harmonic transition was numerically quantified by \citet{briard2019harmonic} using potential energies. The total potential energy was splitted into background potential energy (BPE) and available potential energy (APE). They demonstrated that the BPE, which signifies the measure of irreversible mixing, increases after transition. The APE, which denotes the fraction of the total potential energy, that can be converted to BPE through irreversible mixing, peaks at saturation and is partially released in the flow as BPE. This increase in the BPE causes the numerically obtained final mixing zone size $L$ to exceed the theoretically predicted $L_{sat}$ (\ref{Lsat}). To further elucidate the dynamics of the turbulent mixing zone driven by the Faraday instability, \citet{briard2020turbulent} performed experiments with fresh and salty water, and supported their finding with DNS and theoretical predictions. They concluded that when the instability is triggered, a natural wavelength appears at the interface between the two fluids. As the amplitude of this wave increases, well-defined structures break-up to produce a turbulent mixing layer. At the saturation of the instability turbulence is inhibited, and a mixing layer of final size consistent with the analytically predicted $L_{sat}$ was obtain for a wide range of parameters. \citet{mondal2004effect}, performed a linear stability analysis to investigate the effects of Coriolis force on the Faraday waves in a thin sheet of viscous fluid placed on the vibrating plate. They used the Floquet theory to solve the set of equations with a form of Mathieu equations and explained the effect of rotation rate on the unstable and stable regions of the stability diagrams. They found that the sub-harmonic waves get suppressed with increasing rotation rates resulting in the delay of the onset of the Faraday waves. They also reported the existence of a tri-critical point at the onset of the instabilities, where sub-harmonic, harmonic, and super-harmonic waves are simultaneously excited . \\

Many researchers explored the physics associated with the onset of Faraday instabilities and the subsequent turbulent mixing layer in two miscible fluids. However, the effect of the Coriolis force on the onset and saturation of the turbulent mixing zone driven by the Faraday waves in miscible fluids, is yet to be investigated. Unprecedented to the previous studies, we present a linear stability analysis of the Faraday waves in miscible fluids under the influence of the Coriolis force. We aim to investigate the effects of the Coriolis force on the harmonic to sub-harmonic transition, the onset of the sub-harmonic instability phase, and the saturation phase of the instability. Our analytical model accounts for the vertical inhomogeneity in the background density profile and the Coriolis effect. We followed the classical approach of \citet{yih1960gravity} to derive the linear equations of the form of the Mathieu equations and use the Floquet theory to solve them. We draw the stability diagrams for the corresponding stable and unstable solutions of the Mathieu equations. DNS are also performed  at different forcing amplitudes and Coriolis frequencies to check the validity of our analytical model.\\ 

The problem formulation, which includes the governing equations and theoretical framework for linear stability analysis, is provided in \cref{sec:Problem formulation}. The details of the numerical methodology, and simulation parameters, are given \cref{sec:numerical methods}. The predictions of the linear stability analysis, and the numerical results, are explained in \cref{sec:results}, and we conclude \cref{sec:conclusions}. \\
 
\section{\textbf{Problem formulation}}\label{sec:Problem formulation}
We consider two miscible fluids of small contrasting density contrast where the lighter fluid of density $\rho_2$ is above the denser fluid of density $\rho_1$ in a cubical domain (see figure \ref{fig:domain}). This domain is subject to periodic vertical oscillations with acceleration $g(t)=g_0(1+F\cos{(\omega t))}$, where $g_0$ is the mean acceleration, $F$ is the oscillation amplitude, and $\omega$ is the forcing frequency. We are assuming that the density of the mixture is linearly varying with mass concentration, with $C(\rho_1)=1$ and $C(\rho_2)=0$. The three-dimensional conservation equations for mass, momentum, and concentration field under vertically periodic forcing $g(t)$, for unsteady incompressible flow with Boussinesq approximation, are: 
 \begin{subequations}
    \begin{equation}
    \label{continuity}
    \nabla \cdot \boldsymbol{U} = 0,
    \end{equation}
    \begin{equation}
    \label{momentum}
     \frac{\partial \boldsymbol{U}}{\partial t} + \boldsymbol{U}\cdot \nabla \boldsymbol{U} = -\nabla P - f\hat{k} \times \boldsymbol{U}  + \nu {\nabla}^2 \boldsymbol{U} - 2\mathcal{A}{C}g(t) \hat{k},
    \end{equation}
    \begin{equation}
    \label{concentration}
     \frac{\partial{C}}{\partial t} + \boldsymbol{U}\cdot \nabla {C} = \kappa {\nabla}^2 {C}.
    \end{equation}
 \end{subequations}
Here $\boldsymbol{U}$ represent the velocity vector with components ($U_1$, $U_2$ and $U_3$) in the $x_1$, $x_2$ and $x_3$ (vertical) directions respectively, ${P}$ is the pressure deviation from the hydrostatic background state, $f$ is the Coriolis's frequency, $\nu$ ($\mathrm{m}^2\, \mathrm{s}^{-1}$) is the kinematic viscosity and $\kappa$ ($\mathrm{m}^2\, \mathrm{s}^{-1}$) is the diffusion coefficient.\\

We decompose a variable into its mean $\langle B \rangle _H$ and fluctuating component $b$ following Reynolds decomposition:
\begin{equation}
\label{reynolds decomposition}
 B=\langle B \rangle _H + b,
\end{equation}
where, $\langle B \rangle _H$ is the average in the $x_1$ and $x_2$ horizontal directions, and $\langle b \rangle _H=0$. We consider periodic boundary conditions in both $x_1$ and $x_2$ directions and assume the quantities are statistically invariant in these directions. Therefore, we can define the horizontal average for any quantity, say $B(\boldsymbol{x},t)$ as follows:
 \begin{equation}
     \label{average}
     \langle B \rangle _H({x_3},t)=\lim_{\substack{(l_{x_1},l_{x_2}) \to (\infty,\infty) }} \: \frac{1}{l_{x_1} l_{x_2}} \int_{{-l_{x_1}}/{2}}^{{+l_{x_1}}/{2}} \int_{{-l_{x_2}}/{2}}^{{+l_{x_2}}/{2}} B (\boldsymbol{x},t)\mathrm{d}x_1\mathrm{d}x_2.
 \end{equation}
Substituting Reynolds decomposition (\ref{reynolds decomposition}) into the governing equations ({\ref{continuity}}), ({\ref{momentum}}) and ({\ref{concentration}}), with mean velocity field $\langle \boldsymbol{U} \rangle _H =0$ and $\langle \boldsymbol{u} \rangle _H=\langle c \rangle _H=\langle p \rangle _H=0$, we obtain the equations for fluctuating velocity $\boldsymbol{u}(\boldsymbol{x},t)$ and concentration $c(\boldsymbol{x},t)$ fields as follows:
\begin{subequations}
\begin{equation}
\label{continuity fluc}
\frac{\partial u_i}{\partial x_i}=0,
\end{equation}
\begin{equation}
\label{velocity fluc}
\frac{\partial u_i}{\partial t} + u_j \frac{\partial u_i}{\partial x_j} =  - \frac{\partial p}{\partial x_i}  + f\epsilon_{ij3} u_j\hat{e}_3 - 2\mathcal{A}g(t)c\delta_{i3} + \frac{\partial \langle u_i u_3 \rangle_H}{\partial x_3}+ \nu {\nabla}^2 u_i, 
\end{equation}
\begin{equation}
\label{concentration fluc}
\frac{\partial c}{\partial t} + u_j \frac{\partial c}{\partial x_j} = - u_3 \frac{\partial \langle C \rangle_H}{\partial x_3} +  \frac{\partial \langle u_3 c \rangle_H}{\partial x_3} + \kappa\, {\nabla}^2 c. 
\end{equation}
\end{subequations}

We will compute the mixing zone width $L(t)$ from the mean concentration profile \citep{andrews1990simple} as:
 \begin{equation}
 \label{mz width}
 L(t)=6\int_{-\infty}^{+\infty} \langle C \rangle _H (x_3,t) \left(1-\langle C \rangle _H (x_3,t)\right)\mathrm{d}x_3.
 \end{equation}
 
 We define the Brunt V\"{a}is\"{a}l\"{a} (or stratification) frequency as follows:
 \begin{equation}
      \label{strat fre}
      N=\sqrt{-2\mathcal{A} g_0\frac{\partial \langle C \rangle_H}{\partial x_3}} = \sqrt{-2\mathcal{A} g_0 \varGamma}.
 \end{equation}
 
Here, $\varGamma = {\partial \langle C \rangle_H}/{\partial x_3}$ is the vertical gradient of mean concentration, which can be approximated as ${\partial \langle C \rangle_H}/{\partial x_3}=-1/L$.
 
\subsection{Linear theory for the inhomogeneous inviscid rotating system under parametric oscillations}
In this section, we develop a theoretical model to investigate the effects of initial stratification along with Coriolis force on the triggering and development of the instability. We define the initial stratification by Atwood number ($\mathcal{A}$) and initial mixing zone size ($L_0$). We consider a piecewise concentration profile, constant in the bottom and upper pure (unmixed) fluids and linear in the mixing layer, with a vertical gradient of concentration $\varGamma=0$, in the unmixed region and $\varGamma=\text{constant}$, in the mixed region. We also assume that the {\color{blue} fluid is inviscid fluids} and the fluctuations in velocity and concentration are small. After neglecting the non-linear, viscous and diffusion terms from equations (\ref{velocity fluc}) and (\ref{concentration fluc}) we get:
\begin{subequations}
     \begin{equation}
     \label{velocity fluc1}
     \frac{\partial u_i}{\partial t} =  - \frac{\partial p}{\partial x_i}  + f\epsilon_{ij3} u_j\hat{e}_3 - 2\mathcal{A}g(t)c\delta_{i3} , 
    \end{equation}

     \begin{equation}
     \label{concentration fluc1}
     \frac{\partial c}{\partial t} = - u_3\varGamma . 
     \end{equation}
\end{subequations}
Differentiating equation (\ref{continuity fluc}) with respect to time and using the equation (\ref{velocity fluc1}) yields:
 \begin{equation}
 \label{pressure eq}
 \nabla^2_\mathrm{H} p = f\left(\frac{\partial u_2}{\partial x_1} - \frac{\partial u_1}{\partial x_2} \right) + \frac{\partial^2 u_3}{\partial x_3 \partial t},
 \end{equation}
where $\nabla^2_\mathrm{H} \equiv \partial^2/\partial x_1^2 + \partial^2/\partial x_2^2 $ is the horizontal Laplacian operator. Now, eliminating the pressure term from equation (\ref{pressure eq}) by taking the $\nabla^2_\mathrm{H}$ of equation (\ref{velocity fluc1}) for $\emph{i}=3$ and using equation (\ref{pressure eq}) we obtain:
 \begin{equation}
     \label{u3 fluc1}
     \frac{\partial}{\partial t}\left(\nabla^2 u_3\right) = -\frac{\partial}{\partial x_3}\left[f \left(\frac{\partial u_2}{\partial x_1}-\frac{\partial u_1}{\partial x_2}\right)\right] -  2 \mathcal{A} g(t) \nabla^2_\mathrm{H} c,
 \end{equation}
Next, we eliminate $u_1$ and $u_2$ from equation (\ref{u3 fluc1}) by differentiating it with respect to `$t$', and using equations (\ref{velocity fluc1}) and (\ref{concentration fluc1}):
 \begin{equation}
     \label{u3 fluc2}
     \frac{\partial^2}{\partial t^2}\left(\nabla^2 u_3\right) =
     -f^2\frac{\partial^2 u_3}{\partial x_3^2} + 2 \mathcal{A} g(t) \varGamma \nabla^2_\mathrm{H} u_3.
  \end{equation}
To solve \ref{u3 fluc2} we first consider a case constant forcing, $g(t)=g_0$ and assume a solution for $u_3$ of the form:  
 \begin{equation}
     \label{u3 sol1}
     u_3(x_1,x_2,x_3,t) = \phi(x_3) e^{i (kx_1+lx_2-\varOmega t)},
 \end{equation}
where $\phi(x_3)$ is the amplitude, $\varOmega$ is the temporal frequency or characteristic frequency, and $k$ and $l$ are the components of the wavevector in the horizontal directions $x_1$ and $x_2$, respectively. Substitution of (\ref{u3 sol1}) in equation (\ref{u3 fluc2}) yields (see details in appendix \ref{derivation}): 

 \begin{equation}
     \label{phi eqn}
     \frac{\partial^2\phi}{\partial x_3^2} + {K}^2 \frac{\left( N^2 -\varOmega^2 \right)}{\left(\varOmega^2 - f^2 \right)}  \phi = 0.
 \end{equation}
 To solve equation (\ref{phi eqn}) we consider the walls of the domain at $x_3 = \pm H/2$. Therefore, the boundary conditions at $x_3 = \pm H/2$ are $\phi=0$. As we are solving equation (\ref{phi eqn}) for the without periodic forcing case, we assume constant $\phi={u_3}_{top}$ at $x_3=L_0/2$ (interface between upper lighter fluid and mixed fluid) and constant $\phi={u_3}_{bot}$ at $x_3=-L_0/2$ (interface between mixed fluid and bottom denser fluid). Here $L_0$ represents the initial mixing zone size, and $H$ represents the domain height, as shown in the figure (\ref{fig:domain}). The equation (\ref{phi eqn}) has a finite number of solutions ($\phi_i$) corresponding to the characteristic frequency $\varOmega_i$ for a given horizontal wavenumber ($K$). The solutions of equation (\ref{phi eqn}) (see Appendix \ref{appA1} for the solution steps) in the upper pure fluid region ($x_3 \geq {L_0}/{2},\;\varGamma=0$), bottom pure fluid region ($x_3 \leq {-L_0}/{2},\;\varGamma=0$), and mixed fluid region ($\lvert x_3 \rvert \leq {L_0}/{2},\;\varGamma=-N^2/\left(2\mathcal{A}g_0\right)$) respectively are:
\begin{figure*}
 \centering
 \includegraphics[width=0.9\textwidth,trim={0cm 0cm 0 0cm},clip]{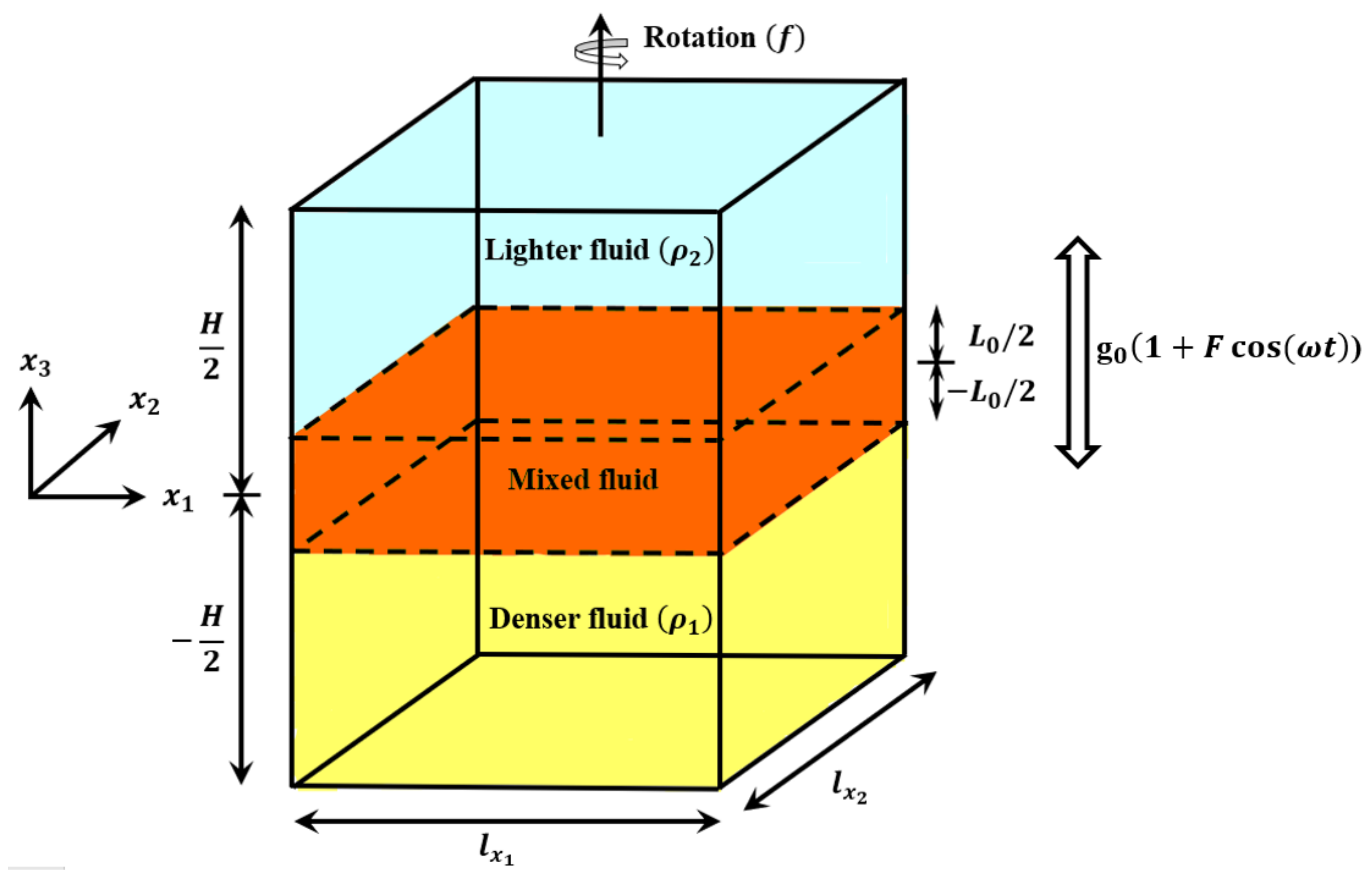}
 \caption{Schematic of the domain containing two miscible liquids subjected to vertically periodic forcing.}
 \label{fig:domain}
 \end{figure*}
 
\begin{subequations}
\begin{equation}
\label{phi sol1}
    \phi_i \left( x_3 \geq \frac{L_0}{2} \right)  =  -{u_3}_{top}\frac{\sinh{\left( K Y_i \left(x_3-\frac{H}{2} \right) \right)}}{\sinh{\left( K Y_i \left(\frac{H-L_0}{2} \right) \right)} },
\end{equation}

\begin{equation}
\label{phi sol2}
\phi_i \left( x_3 \leq \frac{L_0}{2} \right) =  +{u_3}_{bot}\frac{\sinh{\left( K Y_i \left(x_3+\frac{H}{2} \right) \right)}}{\sinh{\left( K Y_i \left(\frac{H-L_0}{2} \right) \right)} },
\end{equation}

\begin{equation}
\label{phi sol3}
\phi_i \left( \lvert x_3 \rvert \leq \frac{L_0}{2} \right)   =  \frac{{u_3}_{top}-{u_3}_{bot}}{2\sin{\left(K X_i \frac{L_0}{2} \right)} } \sin{\left(K X_i x_3 \right)} +\, \frac{{u_3}_{top}+{u_3}_{bot}}{2\cos{\left(K X_i\frac{L_0}{2} \right)} } \cos{\left(K X_i x_3 \right)},
\end{equation}
\end{subequations}

where,
\begin{equation}
\label{XiYi eq}
    X_i=\sqrt{\frac{N^2 - \varOmega_i^2}{\varOmega_i^2 -f^2}}\:, \quad  \quad Y_i=\sqrt{\frac{\varOmega_i^2}{\varOmega_i^2 -f^2}}\:. 
\end{equation}
Both $\phi_i$ and $\partial_{x_3}\phi$ are continuous at ${x_3}=\pm L_0/2$. Therefore, we differentiate equation (\ref{phi sol3}) with respect to $x_3$ at ${x_3}=+L_0/2$ and ${x_3}=-L_0/2$, and equate it to the derivative of equation (\ref{phi sol1}) at ${x_3}=+L_0/2$ and derivative of equation (\ref{phi sol2}) at ${x_3}=-L_0/2$, respectively. After, comparing the coefficients of ${u_3}_{top}$ and ${u_3}_{bot}$ we obtain the following equations (see appendix \ref{DEO}):  
\begin{subeqnarray}
\slabel{omega eq1}
\tan\left(K \sqrt{\frac{N^2 - \varOmega_i^2}{\varOmega_i^2 -f^2}}\frac{L_0}{2} \right) & = &-\, \sqrt{\frac{N^2-\varOmega_i^2}{\varOmega_i^2}}  \tanh{\left(K \sqrt{\frac{\varOmega_i^2}{\varOmega_i^2 -f^2}} \left(\frac{H-{L}_0}{2} \right) \right)},\\[3pt]
\slabel{omega eq2}
\tan\left(K \sqrt{\frac{N^2 - \varOmega_i^2}{\varOmega_i^2 -f^2}}\frac{L_0}{2} \right) & = & \sqrt{\frac{\varOmega_i^2}{N^2-\varOmega_i^2}} \frac{1}{ \tanh{\left(K \sqrt{\frac{\varOmega_i^2}{\varOmega_i^2 -f^2}} \left(\frac{H-{L}_0}{2} \right) \right)}}.
\end{subeqnarray}

\begin{figure*}
 \centering
 \includegraphics[width=0.6\textwidth,trim={0cm 6.2cm 0 7cm},clip]{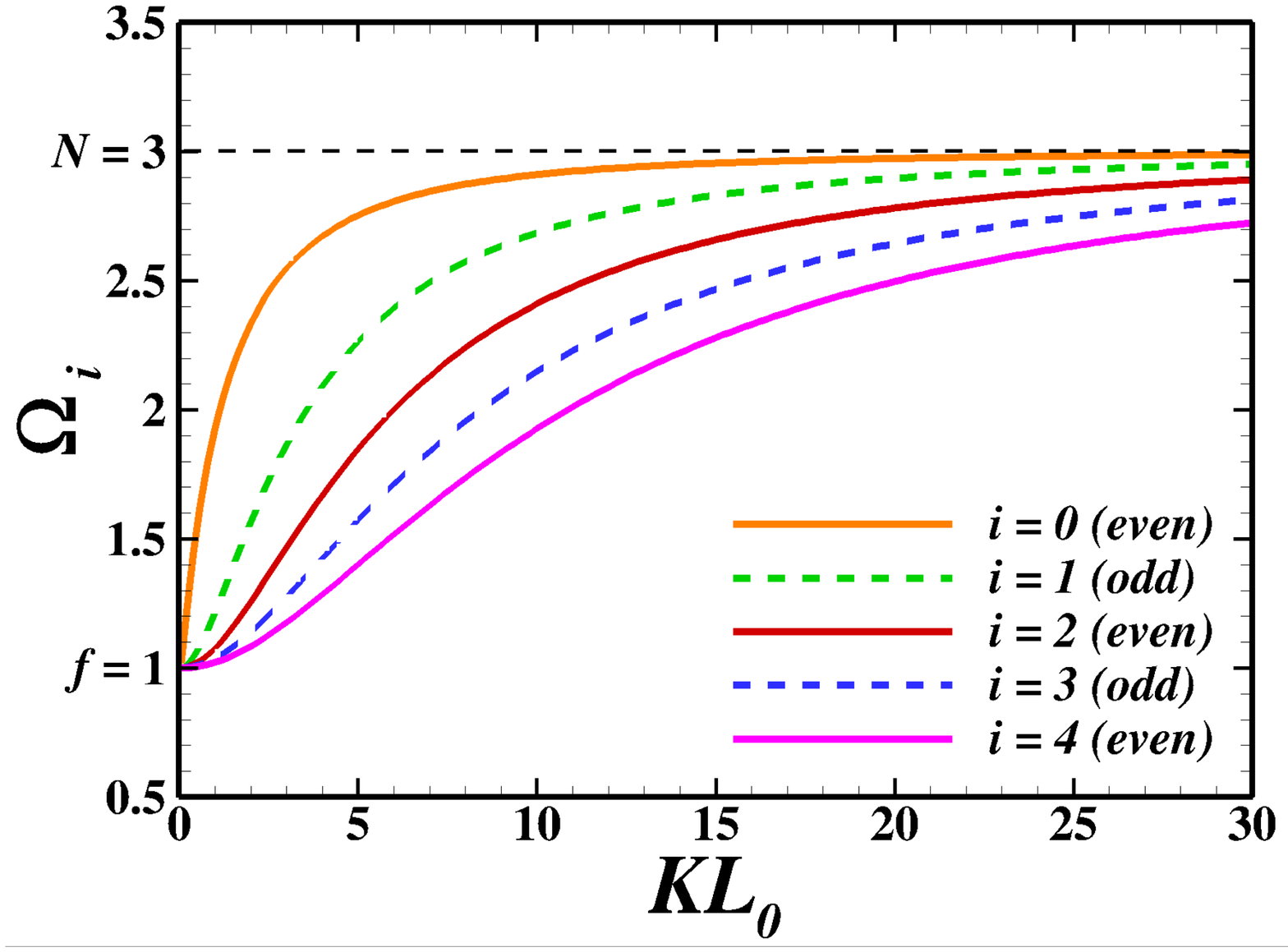}
 \caption{Possible values of characteristic frequency $\varOmega_i$ calculated from the solutions of equations (\ref{omega eq1}) and (\ref{omega eq2}) with $N=3$ and $f=1$, for different horizontal wavenumber $K$ and initial mixing zone width $L_0$. Solutions of equation (\ref{omega eq1}) are odd and of (\ref{omega eq2}) are even.}
 \label{fig:charterstic fre}
 \end{figure*}

If we assume the vertical height of the domain to be infinite, the terms with $\tanh{()}$ can be replaced by 1. We are interested in the solutions of equations (\ref{omega eq1}) and (\ref{omega eq2}) in terms of $\varOmega$ for a given value of $N$, $f$ and $KL_0$. In general, the solutions of equations (\ref{omega eq1}) and (\ref{omega eq2}) show that for given $KL_0$, the possible values of characteristic frequency $\varOmega_i$ fall between $f$ and $N$. To demonstrate this, we assume $N=3$ and $f=1$ and the solutions show that $1<\varOmega_i<3$ (as illustrated in figure \ref{fig:charterstic fre}). All the $\varOmega_i$ becomes asymptotic for $K L_0\gg 1 $, corresponding to the fully developed mixing zone case \citep{briard2020turbulent}.\\

Now we focus on solving the case with vertically periodic forcing (equation \ref{u3 fluc2}). The horizontal directions $x_1$ and $x_2$ are assumed to be periodic. Therefore, we apply Fourier transform on $u_3(x_1,x_2,x_3,t)$ with respect to $x_1$ and $x_2$ such that 
\begin{equation}
    \label{u3 sol2a}
    u_3(x_1,x_2,x_3,t) = \hat{u}_3(k,l,x_3,t)  e^{i (kx_1+lx_2)}.
\end{equation}
Substituting the solution (\ref{u3 sol2a}) in equation (\ref{u3 fluc2}),we get:
\begin{equation}
     \label{u3 hat}
   \frac{\partial^2}{\partial t^2} \left( \frac{\partial^2 \hat{u}_3 }{\partial x_3^2} - K^2  \hat{u}_3 \right) = -f^2 \frac{\partial^2 \hat{u}_3}{\partial x_3^2} - 2\mathcal{A} g(t) \varGamma K^2 \hat{u}_3.
\end{equation}
We further decompose the amplitude $\hat{u}_3(k,l,x_3,t)$ on the basis of previously calculated solutions $\phi_i$ of the stationary system (\ref{phi eqn}), as follows:
  \begin{equation}
    \label{u3 sol2b}
    \hat{u}_3(k,l,x_3,t) =\sum_{i} A_i(k,l,t)\phi_i(k,l,x_3).
  \end{equation} 
Here $A_i(k,l,t)$ represents the time-dependent part of the amplitude of vertical velocity field modes evolving as a result of periodic forcing and $\phi_i(k,l,x_3)$ corresponds to the spatially ($x_3$) dependent amplitude of the vertical velocity field modes. Substituting the decomposition (\ref{u3 sol2b}) in the equation (\ref{u3 hat}), we get 
  \begin{equation}
  \label{phiAi eq1}
  \sum_{i} \frac{\partial^2}{\partial t^2} \left( A_i\frac{\partial^2 \phi_i }{\partial x_3^2} - K^2 \phi_i A_i \right) = - \sum_{i}f^2 A_i \frac{\partial^2 \phi_i }{\partial {x_3}^2} -  \sum_{i}2\mathcal{A} g(t) \varGamma K^2 A_i \phi_i.
 \end{equation}
% \end{subequations}
 
Since we are interested in the temporally evolving amplitude ($A_i$) with periodic forcing, the elimination of $\phi_i$ and $K$ from equation (\ref{phiAi eq1}) by substituting equation \ref{phi eqn} to obtain (see appendix \ref{phiAi eq31A}):

 \begin{equation}
     \label{phiAi eq31}
     \sum_{i} \frac{\partial^2 A_i}{\partial t^2} =  \sum_{i}\left( -\frac{f^2 \left( N^2 -\varOmega_i^2 \right)  - \left( \varOmega_i^2 - f^2 \right) 2 \mathcal{A}g(t)\varGamma}{N^2 - f^2} \right) A_i.
 \end{equation}
 
After substituting $g(t)=g_0\left(1+F\cos{\omega t} \right)$ and $\varGamma=-N^2/(2 \mathcal{A}g_0)$ (see equation (\ref{strat fre})) in equation (\ref{phiAi eq31}) we obtain
 \begin{equation}
     \label{phiAi eq33}
     \sum_{i} \frac{\partial^2 A_i}{\partial t^2} =  \sum_{i}\left( -\frac{f^2 \left( N^2 -\varOmega_i^2 \right)  + N^2\left( \varOmega_i^2 - f^2 \right) \left(1+F\cos{\omega t}\right) }{N^2 - f^2} \right) A_i.
 \end{equation}
% \end{subequations}
Now, we define the amplitude of the concentration field mode $a_i$ which is related to the amplitude of vertical velocity mode $A_i$ via ${\partial c}/{\partial t} = - u_3\varGamma$ (from equation (\ref{concentration fluc1})), therefore
 \begin{subequations}
 \begin{equation}
     \label{conc mode1}
      \sum_{i} \frac{\partial a_i}{\partial t} = \sum_{i} - A_i\varGamma,
 \end{equation} 
 \begin{equation}
     \label{conc mode2}
     \sum_{i} A_i= \sum_{i} -\frac{1}{\varGamma} \frac{\partial a_i}{\partial t}.
 \end{equation}
 \end{subequations}
 
Substituting $A_i$ from equation (\ref{conc mode2}) in equation (\ref{phiAi eq33}) we get 
 \begin{equation}
     \label{a0 eq1e}
    \sum_{i} \frac{\partial}{\partial t} \left(\frac{\partial^2 a_i}{\partial t^2} \right)=  \sum_{i}\left( -\varOmega_i^2-\frac{ N^2\left( \varOmega_i^2 - f^2 \right) F\cos{(\omega t)} }{N^2 - f^2} \right)  \frac{\partial a_i}{\partial t}.
 \end{equation}

After integrating the above equation (\ref{a0 eq1e}) with respect to $t$, yields:
 \begin{equation}
     \label{a0 eq1}
     \frac{\partial^2 a_i}{\partial t^2} + \left( \varOmega_i^2 + N^2 \left(\frac{\varOmega_i^2 - f^2}{N^2 - f^2}\right) F \cos(\omega t) \right) a_i = 0.
 \end{equation}

\subsection{Fully developed mixing zone case}
In this section, we study the fully developed mixing zone case ($KL_0\gg1$), where all $\varOmega_i$ values are constant and $f<\varOmega_i<N$ are possible. First we rewrite the equation (\ref{phi eqn}) as:
 \begin{equation}
 \label{mphi eq}
 \frac{\partial^2\phi_i}{\partial {x_3}^2} + m^2 \phi_i = 0,
 \end{equation}
 
 where,  
 \begin{equation}
 \label{m eq}
 m^2=\frac{K^2\left(N^2-\varOmega_i^2\right)}{\left(\varOmega_i^2-f^2\right)}.
 \end{equation}
 
According to the approximate WKB solution of equation (\ref{mphi eq}) (see Appendix \ref{appB} for WKB approximation), $m$  represents the local wavenumber in the $x_3$ (vertical) direction. So, equation (\ref{m eq}) represents the dispersion relation for our system. We introduce $\tan(\theta_i)=K/m$, where $\theta_i$ is the angle between the vertical axis and the wavevector $\mathbf{K}$, with components $(k,l,m)$ in the $x_1$, $x_2$ and $x_3$ (vertical) directions, respectively. After substituting the definition of angle  $\theta_i$ in equation (\ref{m eq}), we obtain the characteristic frequency as follows (see appendix \ref{omega eqAP}):

\begin{equation}
     \label{omega eq}
     \varOmega_i^2=N^2\sin^2{(\theta_i)}+f^2\cos^2{(\theta_i)}.
\end{equation}

Now, we substitute the characteristic frequency $\varOmega_i$ from equation (\ref{omega eq}) in equation (\ref{a0 eq1}) to get (see appendix \ref{a0 eq28A}): 

 \begin{equation}
     \label{a0 eq28}
     \frac{\partial^2 a_i}{\partial t^2} + \left( f^2 \cos^2{(\theta_i)} + N^2 \sin^2{(\theta_i)} \left( 1 + F \cos{(\omega t)} \right) \right) a_i = 0 .
 \end{equation}

We define non-dimensional time as $\tau=\omega t$ and replace $\theta=\theta_i$ and $a=a_i$ such that equation (\ref{a0 eq28}) becomes: 
 \begin{equation}
     \label{a0 eq3}
     \frac{\partial^2 a}{\partial \tau^2} + \left( \frac{f^2 \cos^2{(\theta)}}{\omega^2} +  \frac{N^2 \sin^2{(\theta)}}{\omega^2} \left(1+F\cos{(\tau)} \right) \right) a = 0.
 \end{equation} 
 
This system of equations is equivalent to a set of Mathieu equations \citep{kovacic2018mathieu}, which governs the stability of our inviscid rotating system under vertically periodic forcing. We solve this equation using the Floquet theory, which provides the solutions to linear differential equations with periodic coefficients. Details of solving the equation (\ref{a0 eq3}) using Floquet theory are in Appendix \ref{appA}.

 \section{\textbf{Numerical methodology and simulation parameters}}\label{sec:numerical methods}
 %\subsection{Numerical methodology and simulation parameters}
 The governing equations (\ref{continuity}), (\ref{momentum}) and (\ref{concentration}) are discretized using the finite-difference method on a staggered grid arrangement. We store the velocity fields at the cell faces and the pressure and concentration fields at the cell centers. A second-order central finite-difference scheme is employed to compute all the spatial derivatives. For the advancement in time, we use an explicit third-order Runge–Kutta method except for the diffusion terms, which are solved implicitly using the Crank–Nicolson method \citep{pal2020evolution, pal2020deep}. The pressure Poisson equation is employed to project a velocity field into a divergence-free space and is solved using a parallel multigrid iterative solver to obtain the dynamic pressure. This numerical solver has been extensively validated and used for several DNS of stratified free-shear and wall-bounded turbulent flows \citep{brucker2010comparative, pal2013spatial, pal2015effect, pal2020evolution, pal2020deep, naskar2022direct, naskar2022effects}. To control the spurious reflections from the disturbances propagating out of the domain, we use sponge regions near the top and bottom boundaries, where damping functions gradually relax the values of the variables to their corresponding values at the boundary. These damping functions are added on the right-hand side of the momentum equation (\ref{momentum}) as explained \citep{brucker2010comparative}. The sponge region is far away from the mixing zone such that it does not affect the dynamics of the mixing of fluids. Periodic boundary conditions are used in the $x_1$ and $x_2$ (horizontal) directions for all the variables. In the $x_3$ (vertical) direction, the top and bottom boundaries are walls, with Dirichlet and Neumann boundary conditions as follows:
\begin{equation}
\label{BC1}
     U_1=U_2=U_3=0, \quad \frac{\partial P}{\partial x_3}=\frac{\partial C}{\partial x_3}=0, \quad \text{at} \quad x_3=\pm H/2.
 \end{equation}
 The initial concentration profile (\citet{briard2019harmonic,briard2020turbulent}) in the present simulations is given as:
 \begin{equation}
\label{conc_profile}
     C\left(\boldsymbol{x},t=0 \right)=\frac{1}{2}\left(1-\tanh{\left(\frac{2x_3}{\delta}\right)}\right),
 \end{equation}
where the parameter $\delta$ is used to change the initial mixing zone size-$L_0$. All of the cases simulated here have the same $\delta=0.035$ and corresponding $L_0=0.096$.
Broadband fluctuations \citep{pal2015effect} are imposed on the initial concentration profile (\ref{conc_profile}), with an initial energy spectrum as follows:
 \begin{equation}
\label{initial pert}
     E(k)=\left(k/k_0\right)^4 \mathrm{e}^{-2\left(k/k_0\right)^2},
 \end{equation}
where, $k_0=30$. The fluctuations are localized at the centerline and are damped exponentially away from the centerline by multiplying the fluctuating concentration $c'$ with a damping function given by
\begin{equation}
\label{crop fluc}
     g(x_3)=a\mathrm{e}^{-\frac{1}{2}\left({x_3}/{\delta}\right)^2},
     %g(r)=a\left[1+\left(r/r_0\right)^2\right] \mathrm{e}^{-\frac{1}{2}\left(\frac{r}{r_0}\right)^2},
 \end{equation}
where $a=0.7$ is the maximum intensity of fluctuation at the centerline. The size of the computational domain is $l_{x_1}=2\pi \:\mathrm{m}$ and $l_{x_2}=2\pi \:\mathrm{m}$ in the horizontal directions. The height of the domain is $l_{x_3}=2H=3.5\pi \:\mathrm{m}$ which includes sponge region of thickness $H_s=0.64\pi \:\mathrm{m}$ at top and bottom boundaries. We have used uniform grids $N_{x_1}=N_{x_2}=512$ in the horizontal direction, while in the vertical direction $x_3$, non-uniform grids $N_{x_3}=512$ are used. The grid is clustered at the vertical centre region of thickness $1.53\pi \:\mathrm{m}$ with $\Delta x_{3_{min}}=\Delta x_1=\Delta x_2$, which is fine enough to resolve the mixing zone. We perform simulations with different parameters listed in table \ref{tab:parameters} to verify our theoretical solutions. The forcing frequency $\omega$ is chosen according to the saturation criterion (\ref{Lsat}) for all simulation cases with Coriolis frequency $f=0$, such that $L_{sat}\simeq2.45$ at all forcing amplitudes $F$. We use the same $\omega$ for rotation cases at the corresponding $F$ to compare non-rotation and rotation cases. We choose a fixed time step for each case such that $dt=T/1120$ where $T$ is the period of vertical oscillations corresponding to $\omega$.
 
 \begin{table}
  \begin{center}
\def~{\hphantom{0}}
\setlength{\tabcolsep}{8pt} % Default value: 6pt
\renewcommand{\arraystretch}{1.2} % Default value: 1
  \begin{tabular}{lcccccccc}
       Case        & $F$      & $\omega \:(\mathrm{rad\:s^{-1}})$    & $f\:(\mathrm{s^{-1}})$   & $f/\omega$ & $\left(f/\omega \right)^2$ & $L_{end} \:(\mathrm{m})$  \\[3pt]
       F075f/$\omega0$  & 0.75 & 0.67   & 0     & 0  &   0   & $\simeq$3.1\\
       F075f/$\omega$48      & 0.75 & 0.67   & 0.322 & 0.481   & 0.23   & $\simeq$3.25\\
       F075f/$\omega$59      & 0.75 & 0.67   & 0.396 & 0.591   & 0.35   & never saturate\\
       F1f/$\omega0$  & 1.0 & 0.7   & 0     & 0  &   0   & $\simeq$3.22\\
      % F1f/$\omega$24      & 1.0 & 0.7   & 0.168 & 0.24   & 0.058   & $\simeq$3.22\\
    %   F1f/$\omega$32      & 1.0 & 0.7   & 0.225 & 0.32   & 0.103   & -\\
       F1f/$\omega$48      & 1.0 & 0.7   & 0.338 & 0.482   & 0.23   & $\simeq$3.22\\
       F1f/$\omega$59      & 1.0 & 0.7   & 0.414 & 0.591   & 0.35   & never saturate\\
       F2f/$\omega0$  & 2.0 & 0.8   & 0     & 0  &   0   & $\simeq$3.22\\
    %   F2f/$\omega$32      & 2.0 & 0.8   & 0.257 & 0.32   & 0.103   & -\\
       F2f/$\omega$48      & 2.0 & 0.8   & 0.384 & 0.48   & 0.23   & $\simeq$3.16\\
       F2f/$\omega$59      & 2.0 & 0.8   & 0.473 & 0.592   & 0.35   & never saturate\\
       F3f/$\omega0$  & 3.0 & 0.9   &   0   & 0  &   0   & $\simeq$3.0\\
    %   F3f/$\omega$32      & 3.0 & 0.9   & 0.29  & 0.32   & 0.103   & -\\
       F3f/$\omega$48      & 3.0 & 0.9   & 0.432 & 0.48   & 0.23   & $\simeq$3.0\\
       F3f/$\omega$59      & 3.0 & 0.9   & 0.532 & 0.592   & 0.35   & never saturate\\
    %   F4f/$\omega0$  & 4.0 & 0.98  &   0   & 0  &   0   & -\\
    %   F4f/$\omega$32      & 4.0 & 0.98  & 0.315 & 0.32   & 0.103   & -\\
    %   F4f/$\omega$48      & 4.0 & 0.98  & 0.47 & 0.48   & 0.23   & -\\
    %   F4f/$\omega$59      & 4.0 & 0.98  & 0.58 & 0.592   & 0.35   & -\\
      
  \end{tabular}
  \caption{Parameters for the simulation: here $F$ is the forcing amplitude, $\omega$ is the forcing frequency, $f$ is the Coriolis frequency and $L_{end}$ is the final mixing zone size. The Atwood number $\mathcal{A}=0.01$, initial mixing zone width $L_0=0.096 \:\mathrm{m}$, gravitational acceleration $g_0=10\:\mathrm{m\:s^{-2}}$, kinematic viscosity $\nu=1\times10^{-4}\:\mathrm{m^2\:s^{-1}} $ and diffusion coefficient $\kappa=1\times10^{-4}\:\mathrm{m^2\:s^{-1}}$ is used for all of the cases ( \citet{briard2019harmonic}).}
  \label{tab:parameters}
  \end{center}
\end{table}

\begin{figure*}
\centering
\includegraphics[width=1.0\textwidth,trim={0cm 0cm 0 0cm},clip]{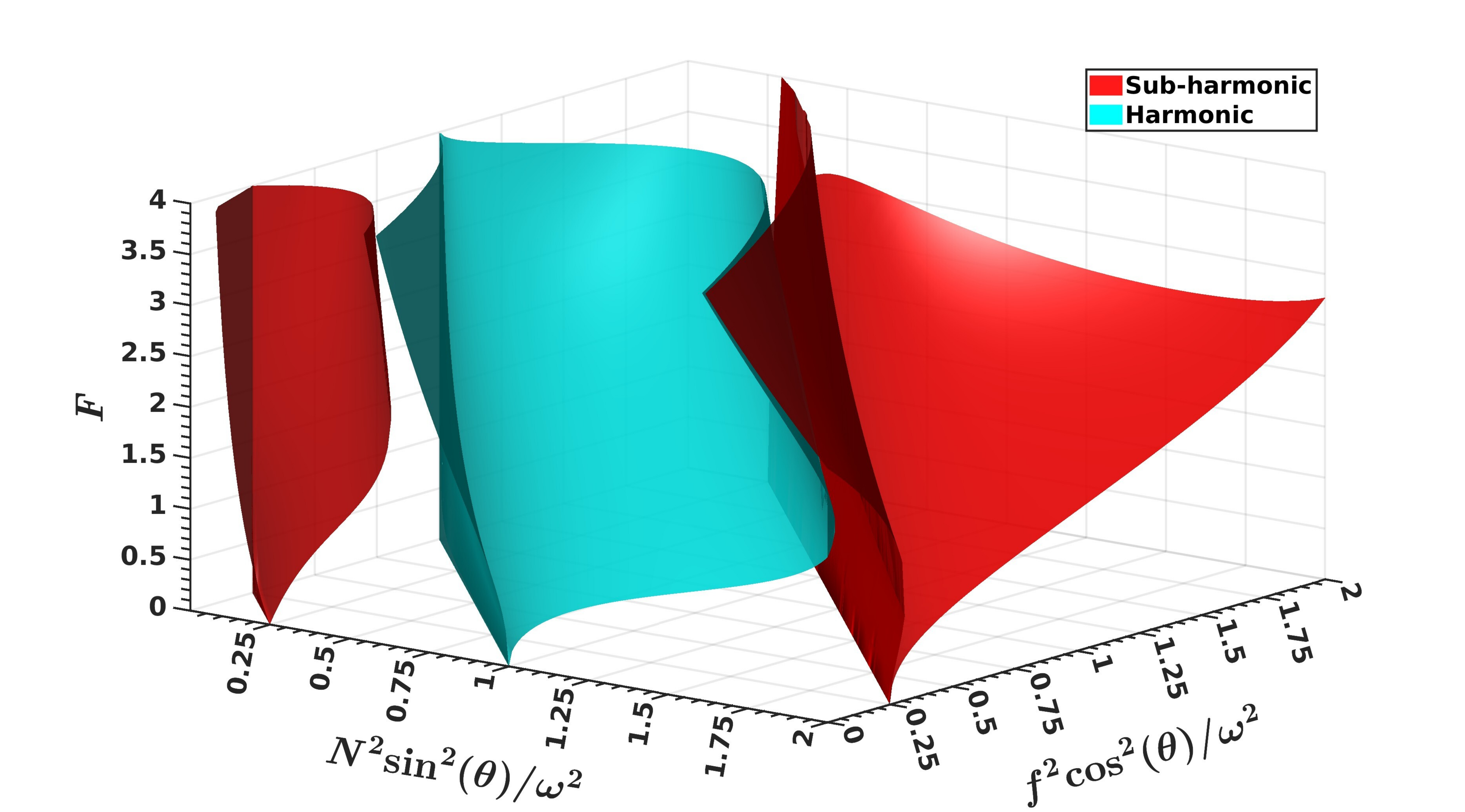}
\caption{A three-dimensional stability diagram obtained by solving the Mathieu equation (\ref{a0 eq3}) in the parameter space $\left(N^2 \sin^2(\theta)/\omega^2,f^2 \cos^2(\theta)/\omega^2,F \right)$. All solutions inside the red and cyan coloured regions are unstable in nature with sub-harmonic and harmonic responses, respectively. Solutions are stable outside the unstable regions.}
\label{fig:3d}
\end{figure*} 

\section{Results}\label{sec:results}
The three-dimensional stability diagram of our Mathieu equation (\ref{a0 eq3}) is shown in the figure (\ref{fig:3d}). The stability diagram shows the effect of the Coriolis force on the onset, development, and saturation of the instability through representative examples (line segments of different colors). The coloured curves correspond to the neutral stability curves, which separate the $\left(N^2 \sin^2(\theta)/\omega^2,f^2 \cos^2(\theta)/\omega^2 \right)$-plane into regions of stable and unstable (sub-harmonic or harmonic) solutions. Inside the red-colored sub-harmonic instability tongues, the possible frequencies of the growing solutions are only odd multiple of the $\omega/2$, which is half of the external forcing frequency. The solutions inside the cyan-colored harmonic instability tongues can have frequencies that are integer multiples of $\omega$. Other solutions that lie outside the instability tongues are stable.

\subsection{Exploring the stability diagram in the presence of  the Coriolis force}

\begin{figure}
	\centering
	\begin{subfigure}{0.85\textwidth}
		\centering
		\includegraphics[width=0.85\textwidth,trim={0cm 0.2cm 0cm 0cm},clip]{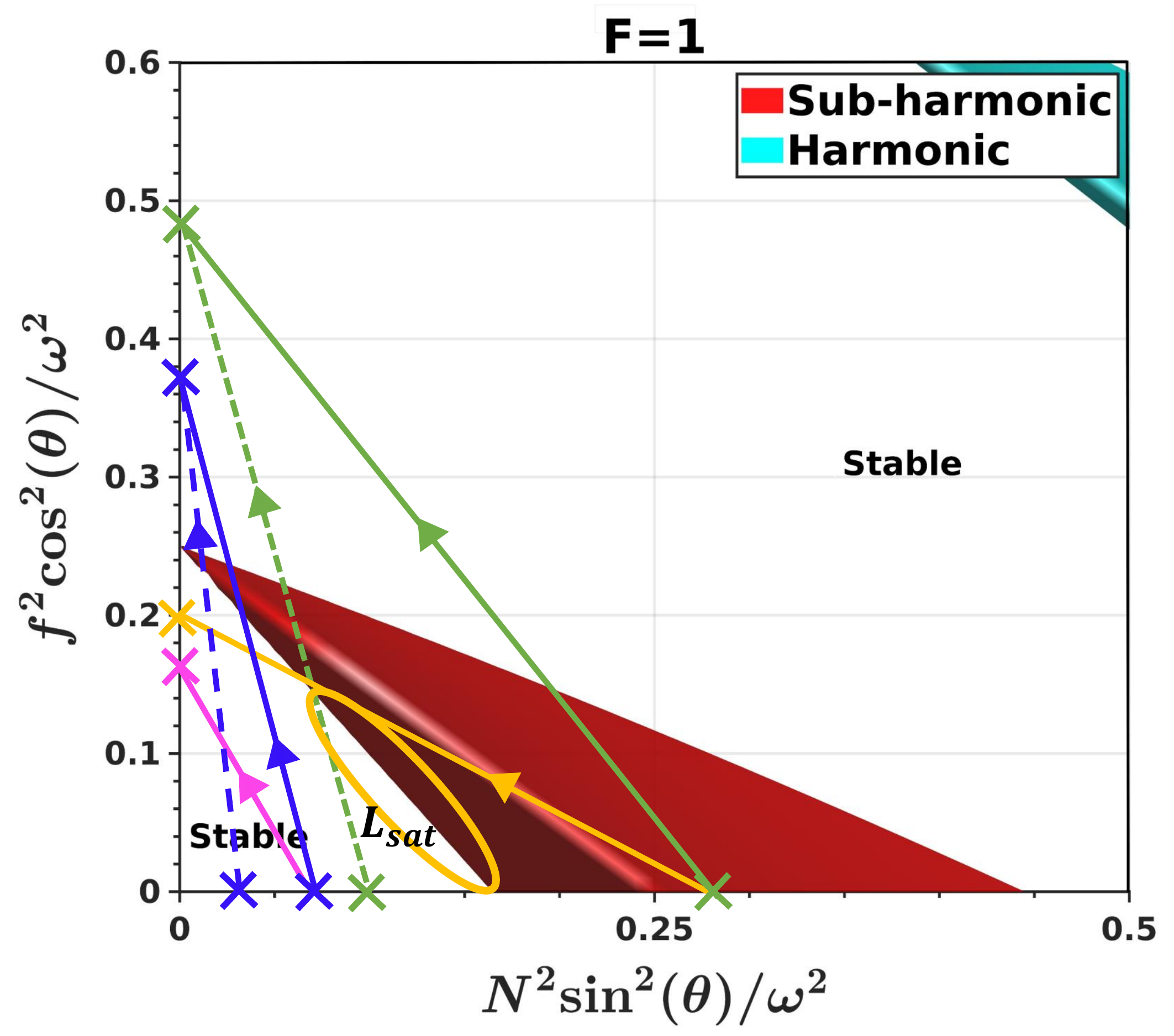}
		\caption{}
		\label{subfig:3dF1_1}
	\end{subfigure}
	\hfill
	\begin{subfigure}{0.85\textwidth}
		\centering
		\includegraphics[width=0.85\textwidth,trim={0cm 0.2cm 0cm 0cm},clip]{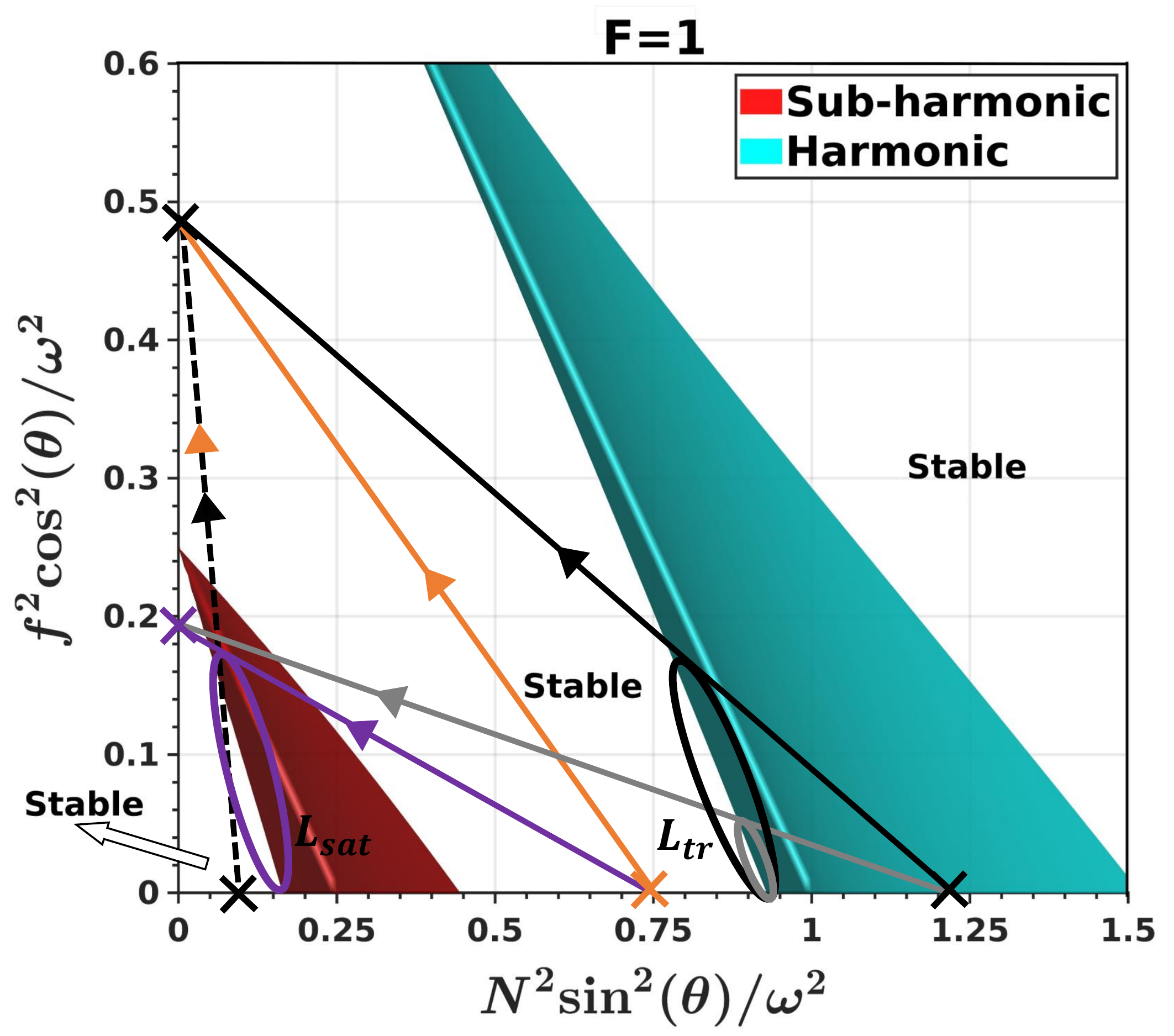}
		\caption{}
		\label{subfig:3dF1_2}
	\end{subfigure}
	\caption{Top view of the three-dimensional stability curve (figure \ref{fig:3d}) at forcing amplitude $F = 1$. For different cases the initially excited inclined segment (solid line) with the left extreme end (shown by cross $\times$) corresponding to a given Coriolis frequency $f$ and the right end (cross $\times$) corresponding to the initial stratification condition $N_0$ , (\textit{a}) in the first stable region and first unstable sub-harmonic (red) tongue, and (\textit{b}) in the first unstable sub-harmonic (red) and harmonic (cyan) tongues. }
	\label{fig:3dFplane_F1}
\end{figure}

Figure \ref{fig:3dFplane_F1} shows the top view of the three-dimensional stability diagram at forcing amplitude $F=1$. In all the cases discussed below, for the given Coriolis frequency $f$ and initial mixing zone width-$L_0$ or initial buoyancy frequency $N_0$, some $\theta$-modes are initially excited by a periodic forcing of a fixed amplitude $F$ and forcing frequency $\omega$, which excites the inclined segment delineated by two crosses ($\times)$ (as shown in figure \ref{fig:3dFplane_F1}). The right end of the inclined segment (at $\theta=\pi/2$) corresponds to the initial stratification condition $\left(N_0/\omega \right)^2$, while the left end (at $\theta=0$) corresponds to the given Coriolis frequency $f$. Further, at each time the inclined segment is excited by some $\theta$-modes where the right end $\left(N/\omega \right)^2$ of the segment represents the mixing zone size-$L$ because $N=\sqrt{2\mathcal{A}g_0/L}$ (from equation \ref{strat fre}) therefore $\left(N/\omega \right)^2=2\mathcal{A}g_0/\left(L \omega^2\right)$. Each segment ranges from $\left[\left(N/\omega \right)^2,0\right]$ to $\left[\left(N/\omega \right)^2,\left(f/\omega \right)^2\right]$ for a given value of Coriolis frequency ($f$). Four cases at $F=1$ are possible to explore the effect of rotation on the stability diagram.  \\
 \captionsetup[subfigure]{textfont=normalfont,singlelinecheck=off,justification=raggedright}

 \begin {enumerate}
 \item The condition for the occurrence of instabilities (similar to \citet{grea2018final}) is  
         \begin{equation}
          \label{inst condition}
            \left(\frac{N}{\omega} \right)^2>\mathcal{G}(f,F),
         \end{equation}
where $\mathcal{G}$ represents the extreme left curve (or boundary) of the first sub-harmonic tongue. Beyond $\mathcal{G}$ Faraday instability is triggered, resulting in the growth of the mixing zone size-$L$. If the initial stratification $N_0$ %(right end of the initially excited segment) 
is such that $\left(N_0/\omega \right)^2\leq\mathcal{G}(f,F)$, then two scenarios are possible. \\
 
     \begin{itemize} \label{sec:MathNot}
         \item $\left(f/\omega \right)^2<0.25$:  For these conditions, in the initial phase of the periodic forcing, some $\theta$-modes are excited on the inclined pink line segment, as shown in the figure \ref{subfig:3dF1_1}. The instability is not triggered because all the $\theta$-modes are in the stable region. Therefore, solutions within this regime will not grow, and the mixing zone size will remain at its initial value of $L(t=0)$. We define the condition for instability based on the fact that if at least one $\theta$-modes fall in any unstable tongues solution will grow, resulting in instability and mixing. 
         We can define the saturation state of the Faraday instability when the instability condition (\ref{inst condition}) is no longer fulfilled and $\left(f/\omega \right)^2<0.25$. The estimation of the saturated mixing zone size in this situation will be as follows:
        \begin{equation}
         \label{sat condition}
            \left(\frac{N}{\omega}\right)^2=\mathcal{G}(f,F)\quad \mathrm{or} \quad L_{sat}=\frac{2\mathcal{A}g_0}{\omega^2}\frac{1}{\mathcal{G}(f,F)}\;,\quad (\mathrm{when} \left(f/\omega \right)^2<0.25).
        \end{equation}\\
        
        \item $\left(f/\omega \right)^2 \geq 0.25$: This condition represents the second scenario, where the dark blue line segment (figure \ref{subfig:3dF1_1}) is excited initially. Some of the excited $\theta$-modes are in the first unstable sub-harmonic tongue, which triggers the instability leading to the growth of the mixing zone size-$L$. With the growth in $L$, $\left(N/\omega \right)^2$ starts decreasing (because $\left(N/\omega \right)^2=2\mathcal{A}g_0/\left(L \omega^2\right) \propto 1/L$) and moves toward the left end of the segment, as indicated by the arrow. Surprisingly, the mixing zone size-$L$ will never reach a saturation state because unstable $\theta$-modes (indicated by the dark dashed blue line segment in figure \ref{subfig:3dF1_1}) are always excited, and the instability condition (\ref{inst condition}) is always satisfied. Therefore, $L$ will continue to grow. However, the growth rate of $L$ might be slow at later stages because $\theta$-modes that fall in the first unstable tongue decrease as $\left(N/\omega \right)^2$ decreases. As a result, we can conclude that the saturation will never be achieved for $\left(N_0/\omega \right)^2\leq\mathcal{G}(f,F)$ and $\left(f/\omega \right)^2 \geq 0.25$.\\
\end{itemize}
     
\item When the initial condition is chosen to be in the first unstable sub-harmonic (red) region (figure \ref{subfig:3dF1_1}) then the instability condition (\ref{inst condition}) is always satisfied and the mixing zone size-$L$ grows. The following two configurations are possible for this condition.\\
 
    \begin{itemize} \label{sec:MathNot}
        \item $\left(f/\omega \right)^2<0.25$: %For this case the instability condition (\ref{inst condition}) is satisfied. 
        The most unstable sub-harmonic $\theta$-modes are initially excited, and the corresponding, the solid yellow line segment is shown in figure \ref{subfig:3dF1_1}. The instability is triggered quickly, which grows the mixing zone size-$L$ and decreases $\left(N/\omega \right)^2$. The instability will achieve a state of saturation when no sub-harmonic (unstable) $\theta$-modes are excited, i.e., instability condition (\ref{inst condition}) is no longer satisfied. The corresponding mixing zone size $L_{sat}$ is given by the saturation condition (\ref{sat condition}).
$L_{sat}$ may lie anywhere on the left stability curve of the first sub-harmonic tongue, as shown by the yellow ellipse (see figure \ref{subfig:3dF1_1}).\\
        
        \item $\left(f/\omega \right)^2 \geq 0.25$: Similar to the previous case, the excited unstable sub-harmonic $\theta$-modes trigger the instability, as shown by the green line segment (see figure \ref{subfig:3dF1_1}). Therefore, the mixing zone width-$L$ begins to grow. However, the mixing zone size-$L$ will never reach the saturated state because, at every instant, all the exciting segments will cross the first sub-harmonic tongue, as demonstrated by the dashed green line segment. Initially, $L$ increases quickly, but at a later stage, its growth rate will decrease. \\ 
\end{itemize}

\item In the third case, the initial condition is in the second stable region between the first sub-harmonic and first harmonic tongues, as shown in figure \ref{subfig:3dF1_2}. The instability condition (\ref{inst condition}) is satisfied, and the discussion of the related two cases is as follows.\\
 
    \begin{itemize} \label{sec:MathNot}
        \item $\left(f/\omega \right)^2<0.25$: In this case, the purple line segment is excited initially, which lies in both the stable and unstable sub-harmonic regions (see figure \ref{subfig:3dF1_2}). If the initially excited $\theta$-modes are in a stable region, then the mixing zone size will grow due to molecular diffusion. However, the instability is triggered when the excited $\theta$-mode lies in the first unstable sub-harmonic region, and the mixing zone size will grow. At a later time, when no more unstable $\theta$-modes are excited, the instability saturation condition (\ref{sat condition}) is satisfied. At this instant, the mixing zone size will saturate at $L_{sat}$ on the left stability curve of the first sub-harmonic region (shown by a purple ellipse, see figure \ref{subfig:3dF1_2}).\\
    
        \item $\left(f/\omega \right)^2 \geq 0.25$: Initially, the entire orange line segment is excited in the stable region and will remain in the stable region until any of the $\theta$-modes are in the unstable region. In the stable region, the mixing zone size-$L$ grows due to molecular diffusion. However, at a later time, it will increase due to the instability caused by unstable $\theta$-modes in the sub-harmonic region. The $L$ continues to grow for all time because the $\theta$-modes excited over the segment will always pass through the unstable sub-harmonic region thus instability condition (\ref{inst condition}) is always satisfied. The dashed orange line overlapping the dashed black line segment demonstrates this scenario. The growth rate of $L$ will decrease when a small number of the excited $\theta$-modes are in the unstable region. \\
    \end{itemize}
    
\item When the initial stratification condition $\left(N_0/\omega \right)^2$ is in the first harmonic tongue or far away (see figure \ref{subfig:3dF1_2}), both harmonic and sub-harmonic instabilities are triggered by the unstable $\theta$-modes due to the instability condition (\ref{inst condition}). We discuss the corresponding two cases below.\\
 
\begin{itemize} \label{sec:MathNot}
\item $\left(f/\omega \right)^2<0.25$: For these initial conditions, the grey line segment (figure \ref{subfig:3dF1_2}) is excited initially. For some initial time, the excited $\theta$-modes fall in the harmonic region, so the harmonic region is more visible, which grows the mixing zone size-$L$ and thus $\left(N/\omega \right)^2$ decreases. The transition from harmonic to sub-harmonic occurs when no more harmonic $\theta$-modes are excited, such that $\left(N/\omega \right)^2$ will cross the left stability curve of the first harmonic tongue. The corresponding mixing zone size is defined by $L_{tr}$, which is shown by the gray ellipse (see figure \ref{subfig:3dF1_2}). We will also demonstrate this transition using numerical simulations. Finally, the mixing zone width saturates ($L_{sat}$) due to the fulfillment of the saturation condition (\ref{sat condition}). The corresponding $L_{sat}$ will be on the left stability curve of the first sub-harmonic region (shown by a purple ellipse, see figure \ref{subfig:3dF1_2}).\\
    
\item $\left(f/\omega \right)^2 \geq 0.25$: In this case, the instability criteria and harmonic to sub-harmonic transition are the same as in the former case, except for the saturation state. After some time, during the growth of the mixing zone width-$L$, the initially excited solid black segment can be in the first stable region, as shown by the dashed black line (figure \ref{subfig:3dF1_2}). The mixing zone size never achieves a state of saturation because, at all times, the excited $\theta$-modes over the segments will pass through the first unstable sub-harmonic region (dashed black line segment). Therefore, the instability condition (\ref{inst condition}) is always satisfied, which causes the continuous growth of the mixing zone. \\
\end{itemize}
    
 \end {enumerate}

 \begin{figure*}
 \centering
 \includegraphics[width=0.9\textwidth,trim={0cm 0.0cm 0 0cm},clip]{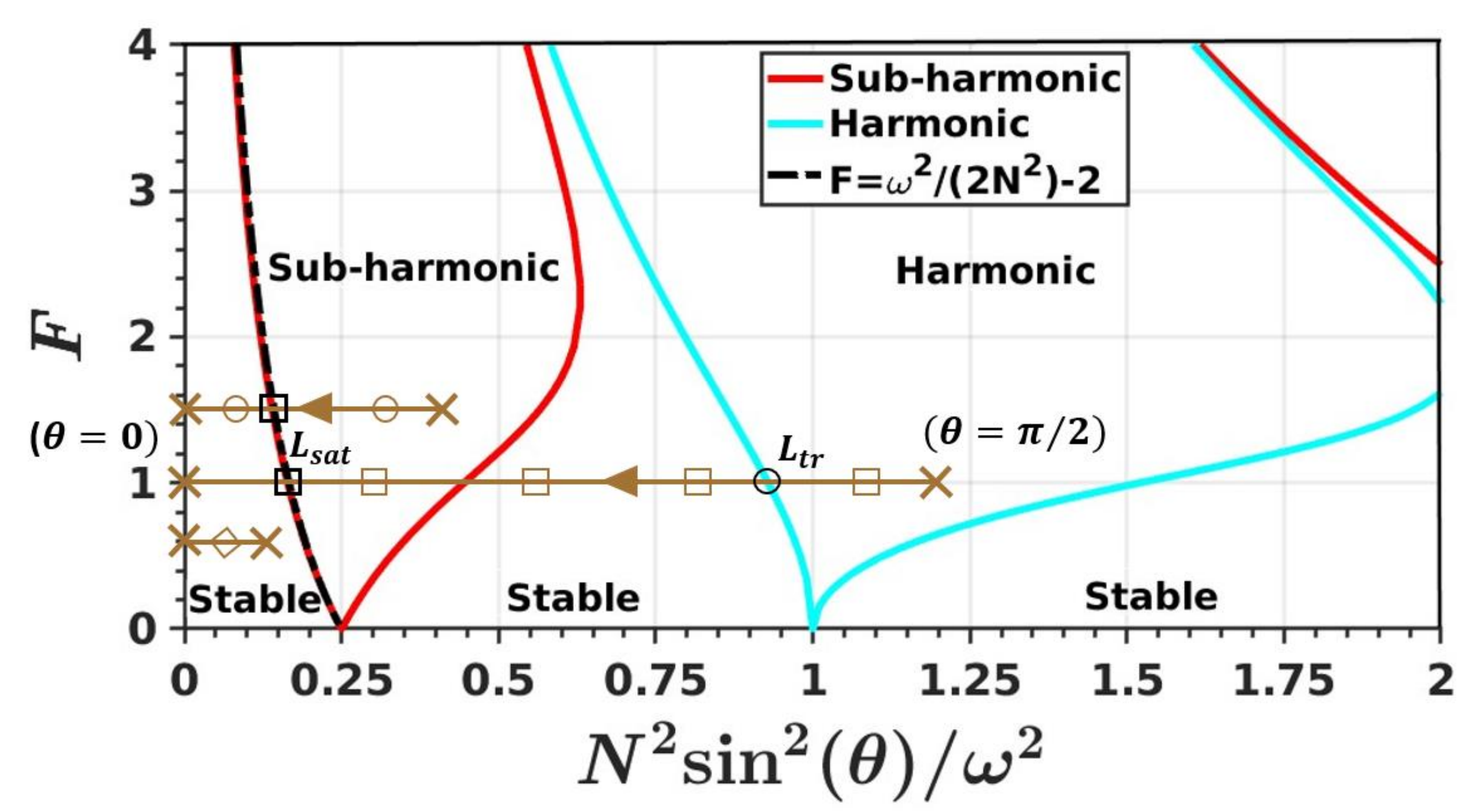}
 \caption{Stability diagram of Mathieu equation (\ref{a0 eq3}) for stratification ($N$) initial condition in the absence of rotation (Coriolis frequency, $f=0$).  The red and cyan-colored stability curves separate the stable and unstable (sub-harmonic and harmonic) regions. The brown line segment, delineated by two crosses (right end ($\times$) at $\theta=\pi/2$ and left end ($\times$) at $\theta=0$), is excited corresponding to the different initial conditions in the different tongue-like zones.}
 \label{fig:3d_f0}
 \end{figure*} 

To summarize, the mixing zone saturates at $L_{sat}$ for $\left(f/\omega \right)^2<0.25$, whereas for $\left(f/\omega \right)^2 \geq 0.25$, the mixing zone never saturates and continues to grow due to the unstable sub-harmonic $\theta$-modes at all times.

 \captionsetup[subfigure]{textfont=normalfont,singlelinecheck=off,justification=raggedright}
 \begin{figure}
 \centering
 	\begin{subfigure}{0.71\textwidth}
 		\centering
 		\includegraphics[width=0.71\textwidth,trim={0cm 0.2cm 0cm 0cm},clip]{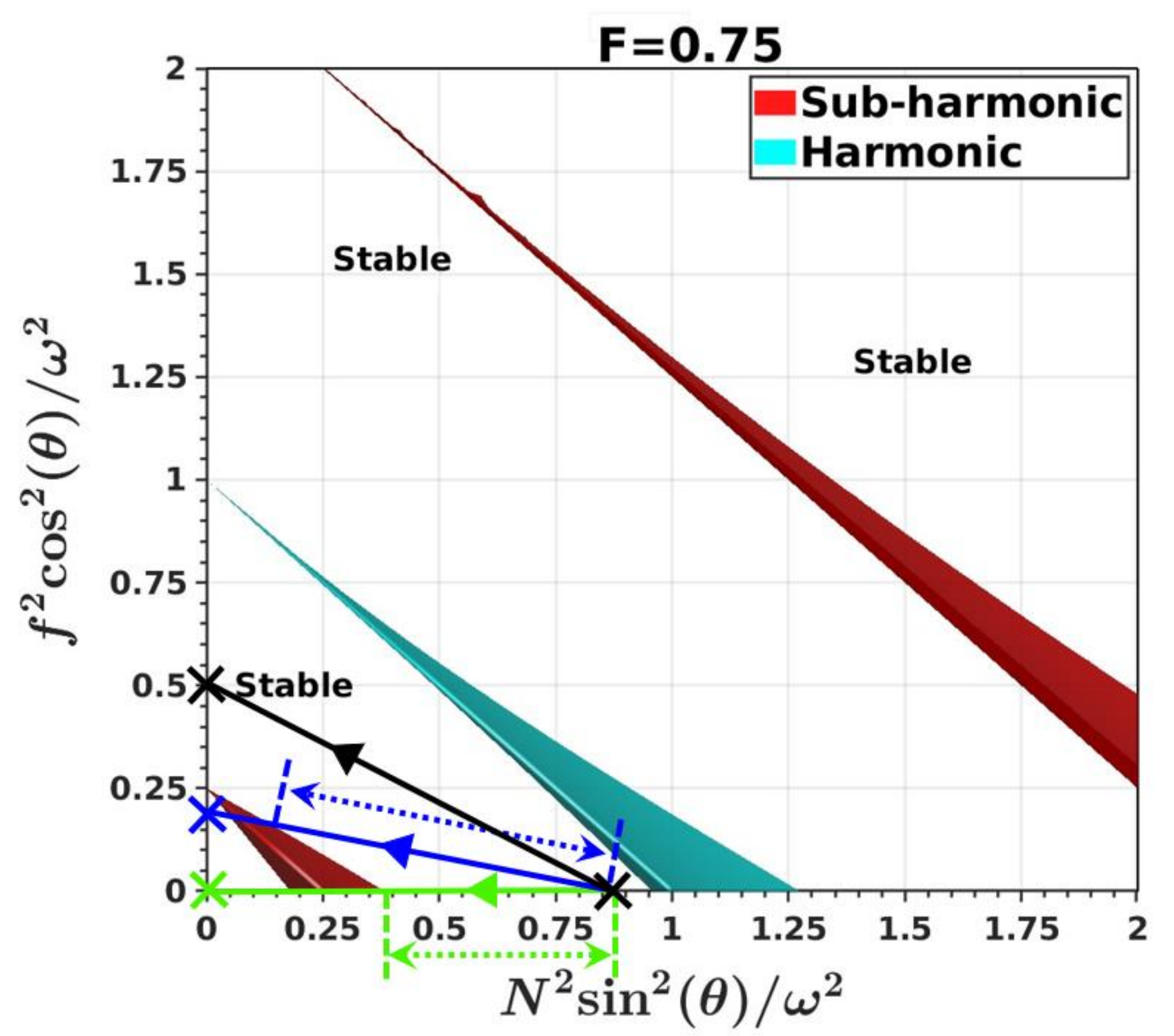}
 		\caption{}
 		\label{subfig:3dF075}
 	\end{subfigure}
%  	\hfill
% 	\centering
% 	\begin{subfigure}{0.485\textwidth}
% 		\centering
% 		\includegraphics[width=1.0\textwidth,trim={0cm 0.2cm 0cm 0cm},clip]{figures/3d_F1.pdf}
% 		\caption{}
% 		\label{subfig:3dF1}
% 	\end{subfigure}
	%\hfill
	\begin{subfigure}{0.71\textwidth}
		\centering
		\includegraphics[width=0.71\textwidth,trim={0cm 0.2cm 0cm 0cm},clip]{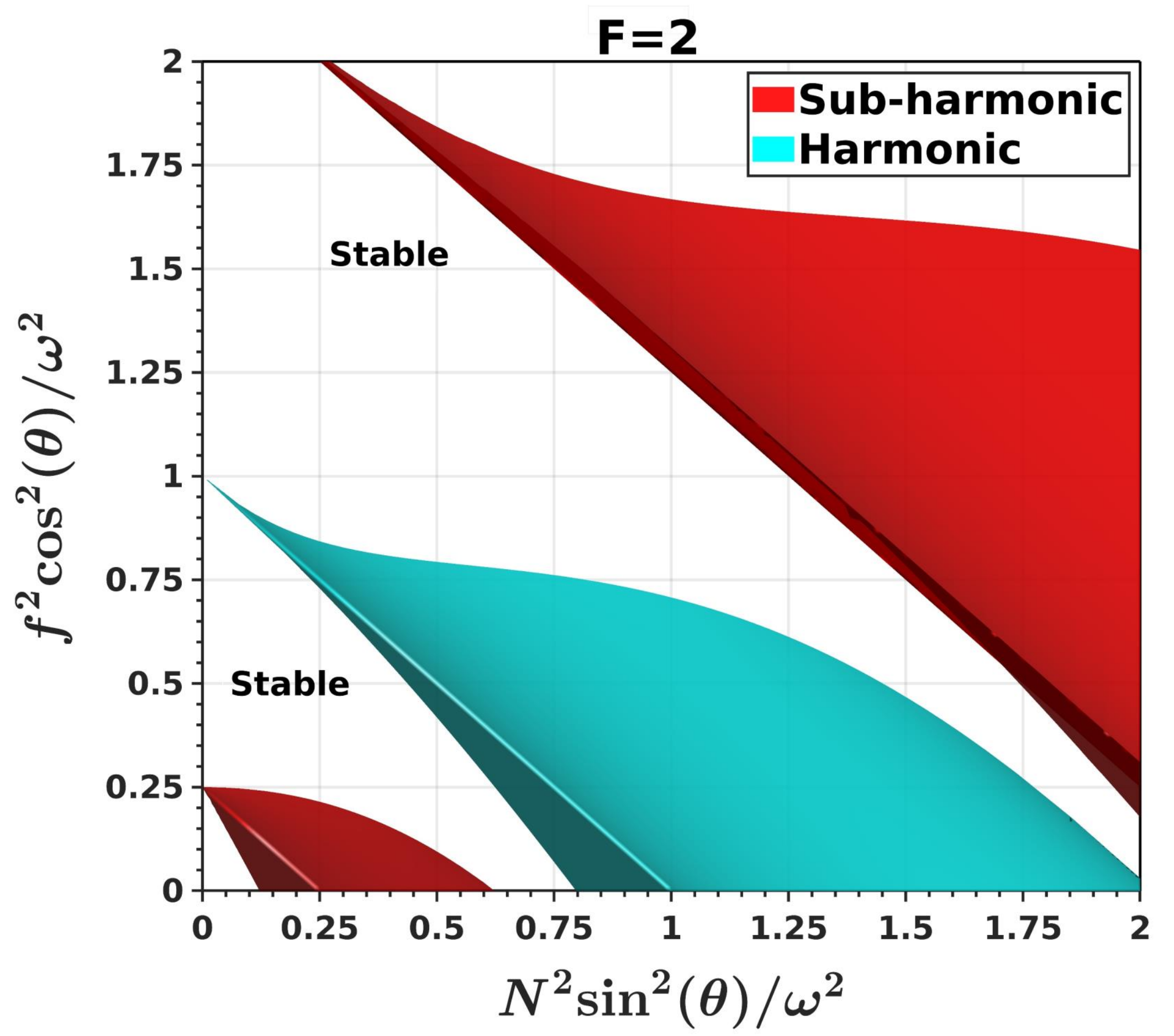}
		\caption{}
		\label{subfig:3dF2}
	\end{subfigure}
%	\hfill
	\begin{subfigure}{0.71\textwidth}
		\centering
		\includegraphics[width=0.71\textwidth,trim={0cm 0.2cm 0cm 0cm},clip]{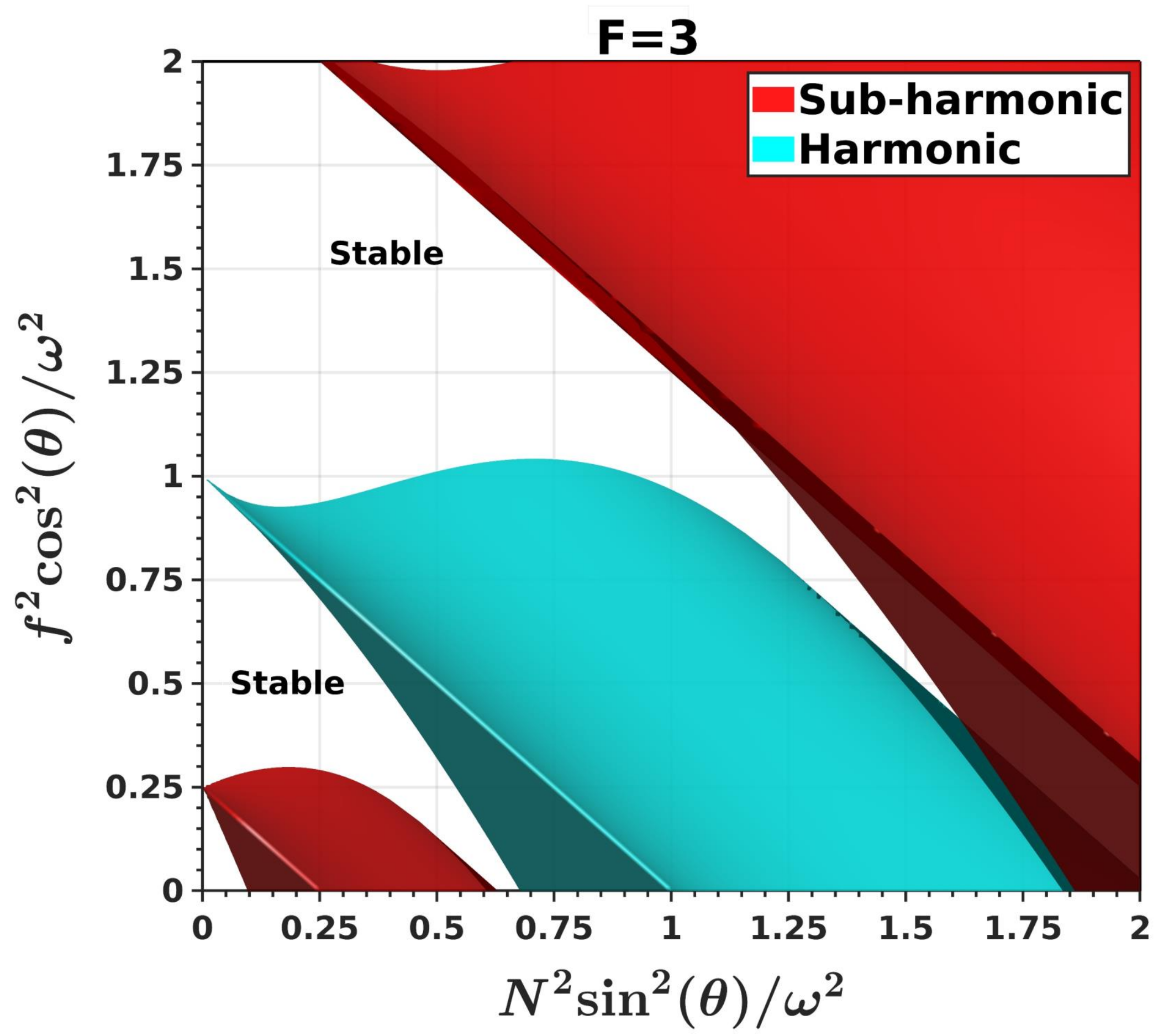}
		\caption{}
		\label{subfig:3dF3}
	\end{subfigure}
% 	\hfill
%  	\begin{subfigure}{0.485\textwidth}
%  		\centering
%  		\includegraphics[width=1.0\textwidth,trim={0cm 0.2cm 0cm 0cm},clip]{figures/3d_F4.pdf}
%  		\caption{}
%   		\label{subfig:3dF4}
%  	\end{subfigure}
	\caption{Stability diagrams corresponding to three-dimensional stability curve (figure \ref{fig:3d}) at four different values of the forcing amplitude (\textit{a}) $F=0.75$, (\textit{b}) $F=2$ and (\textit{c}) $F=3$. The stable (white) regions are in between the unstable red (sub-harmonic) and cyan (harmonic) colored tongue-like zones. }
	\label{fig:3dFplane}
\end{figure}

\subsection{Exploration of the stability diagram without Coriolis force}

Figure (\ref{fig:3d_f0}) shows the stability curves only for initial stratification conditions in the absence of the Coriolis force (f = 0) i.e., in ($\left(N/\omega \right)^2-F$) parameter plane. We obtained this stability curve similar to that of \citet{grea2018final}. The stable regions are separated by the unstable sub-harmonic (inside the red stability curve) and harmonic (inside the red stability curve) regions. The horizontal line segment (delineated by two crosses, $\times$), as shown by brown line in figure (\ref{fig:3d_f0}), is excited for various initial conditions of $\left(N_0/\omega \right)^2$. The segment ranges from $0$ (left end; $\theta=0$) to $\left(N/\omega \right)^2$ (right end; $\theta=\pi/2$). The instability condition (\ref{inst condition}) for $f=0$ was calculated by \citet{grea2018final} and provides the expression for $\mathcal{G}(0,F)=1/(2F+4)$. When the initial condition is in the first stable region (shown by the brown line with a diamond, see figure (\ref{fig:3d_f0})), the instability does not trigger. However, if the initial conditions are in the first sub-harmonic tongue (brown line segments with circles) the instability is triggered immediately by the unstable sub-harmonic $\theta$-modes, resulting in the rapid growth of $L$. However, for the initial condition in the first harmonic tongue (brown line segments with squares) or far away, the mixing zone size grows due to both sub-harmonic and harmonic instabilities. Hence, the mixing zone size-$L$ grows and saturates ($L_{sat}$) after crossing the first marginal stability curve (as shown by black dashed curve in figure (\ref{fig:3d_f0})), defined by $\left(N/\omega \right)^2=\mathcal{G}(0,F)=1/(2F+4)$. The black squares indicate the $L_{sat}$ on the marginal stability curve. Therefore, the mixing zone size-$L$ saturates for all cases without rotation. \\

\subsection{Does Coriolis force delay the onset of Faraday instability and turbulent mixing?}     \label{subsec:instability delays}

In the presence of the Coriolis force, the blue (left end ($\times$) at $f^2/\omega^2=0.2$) and black (left end ($\times$) at $f^2/\omega^2=0.5$) the $\theta$-modes are excited over the inclined segments at each time, as shown in the figure (\ref{subfig:3dF075}). In contrast, the $\theta$-modes are excited over horizontal segments in the absence of the Coriolis force, as shown by the green line segment (left end ($\times$) at $f^2/\omega^2=0$) in the figure (\ref{subfig:3dF075}). The right end ($\times$) of each segment corresponding to the initial stratification condition is considered the same, which lies in the stable region between the first sub-harmonic (red) and harmonic (cyan) tongue.
We know that the unstable $\theta$-modes which lie in unstable tongues trigger the instability. Before the onset of the instability, the stable $\theta$-modes will lie between the right end ($\times$) of the segment, and the first crossing point of the line segment on the right boundary of the first sub-harmonic tongue, as shown by the dotted lines of the corresponding color in figure (\ref{subfig:3dF075}). The length of the line segment over which the stable $\theta$-modes lie progressively increases with increasing $f^2/\omega^2$ as shown by the dotted green line and the dotted blue and solid black line segments in figure \ref{subfig:3dF075}. This broadening of the stable region owing to rotation delays the onset of the sub-harmonic instability, and the turbulent mixing will start when the right end ($N^2/\omega^2$ corresponding to $L$) of the $\theta$-modes excited over the segment enters the first sub-harmonic (red) tongue \citep{grea2018final,briard2019harmonic}. Therefore, turbulent mixing also delays with increasing Coriolis frequency ($f$).\\

We can observe that the first sub-harmonic (red) and harmonic (cyan) tongues becomes wider with increasing forcing amplitude $F$ (see comparison among figures \ref{subfig:3dF075}, \ref{subfig:3dF2}, and \ref{subfig:3dF3}). Therefore, the right boundary of the first sub-harmonic tongue (red) shifts towards the right-hand side, and the left boundary of the first harmonic tongue (cyan) moves towards the left-hand side. As a result, the stable region between these tongues shrinks. Consequently, the onset of the turbulent mixing zone and Faraday instability occurs much earlier at a higher forcing amplitude for all the initial stratification conditions that satisfy the instability condition (\ref{inst condition}). At higher forcing amplitudes, the flow is energetic enough to overcome the stabilizing effect of rotation. Therefore, the effect of rotation decreases with an increase in the forcing amplitude. The numerical simulations presented in section \cref{subsec: Numerical results} will further support these findings. \\

\subsection{Numerical results} \label{subsec:Numerical results}

 \captionsetup[subfigure]{textfont=normalfont,singlelinecheck=off,justification=raggedright}
 \begin{figure}
 \centering
 	\begin{subfigure}{0.83\textwidth}
 		\centering
 		\includegraphics[width=0.83\textwidth,trim={0cm 0.2cm 0.07cm 0.0cm},clip]{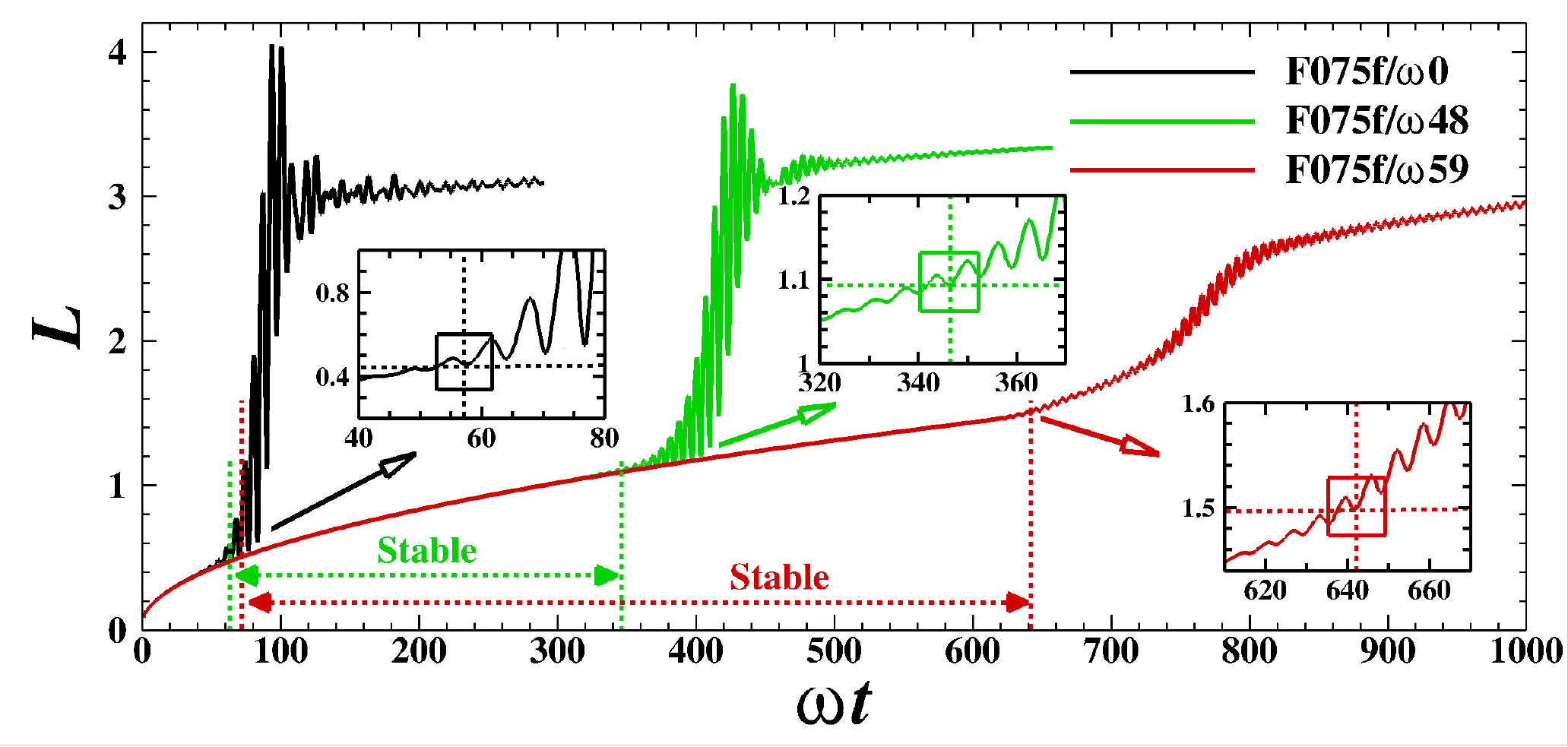}
 		\caption{}
 		\label{subfig:L_F075}
 	\end{subfigure}
 	\hfill
	\begin{subfigure}{0.83\textwidth}
		\centering
		\includegraphics[width=0.83\textwidth,trim={0cm 0.2cm 0.07cm 0.0cm},clip]{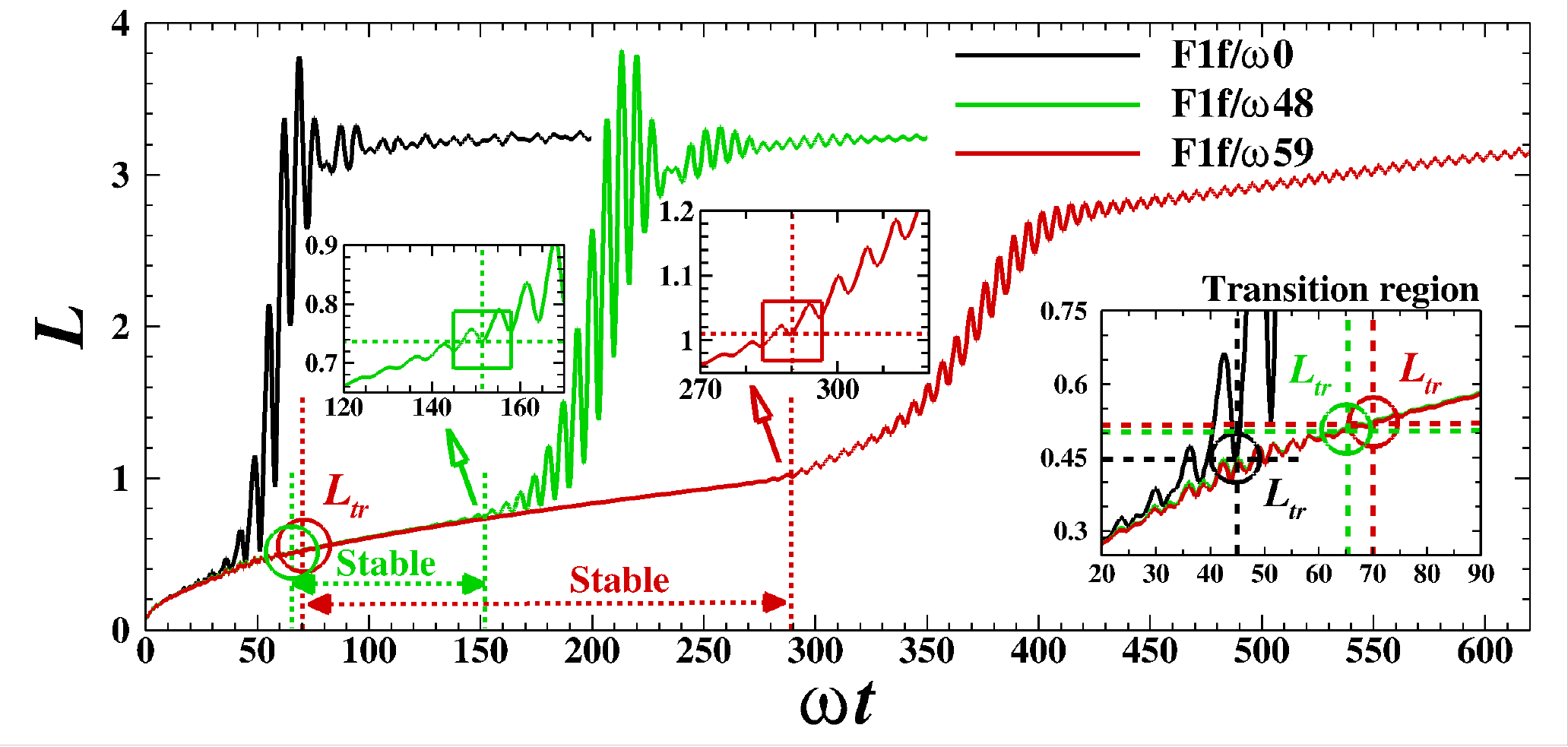}
		\caption{}
		\label{subfig:L_F1}
	\end{subfigure}
	\hfill
	\begin{subfigure}{0.83\textwidth}
		\centering
		\includegraphics[width=0.83\textwidth,trim={0cm 0.2cm 0.07cm 0.0cm},clip]{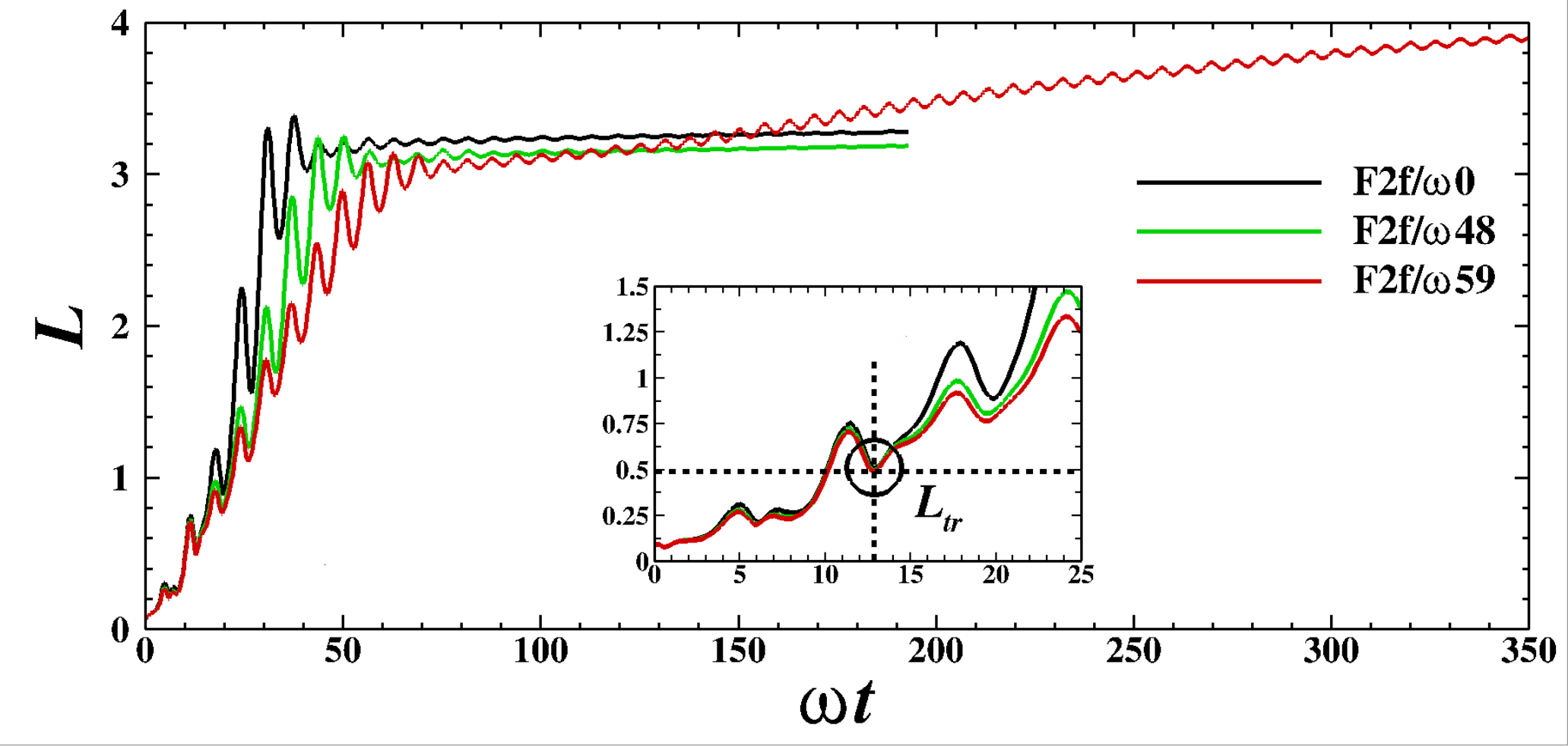}
		\caption{}
		\label{subfig:L_F2}
	\end{subfigure}
	\hfill
	\begin{subfigure}{0.83\textwidth}
		\centering
		\includegraphics[width=0.83\textwidth,trim={0cm 0.2cm 0.07cm 0.0cm},clip]{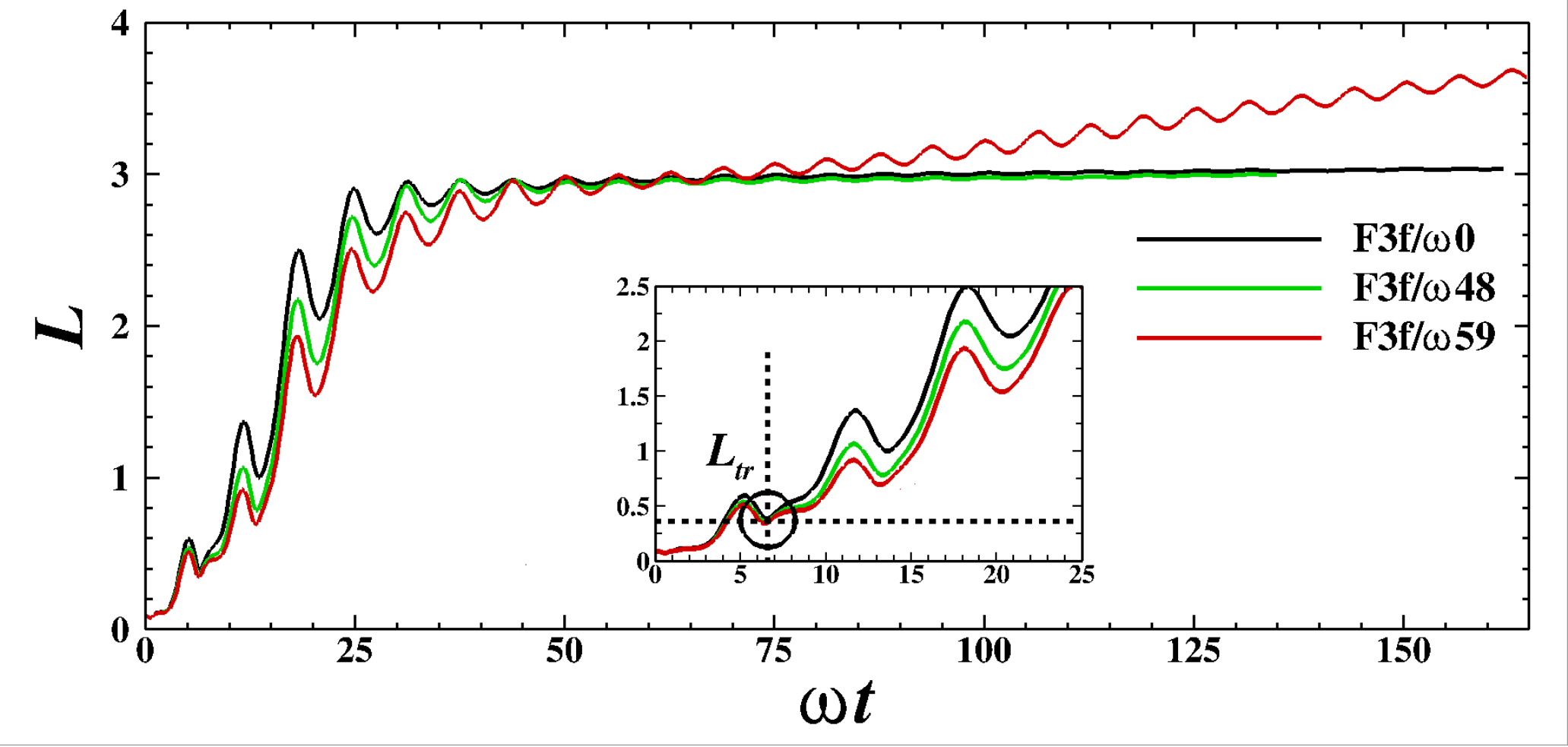}
		\caption{}
		\label{subfig:L_F3}
	\end{subfigure}
	\caption{Evolution of the mixing zone size-$L$ with non-dimensional time $\omega t$ for without rotation and rotation cases, (\textit{a}) at $F=0.75$, (\textit{b}) at $F=1$, (\textit{c}) at $F=2$ and (\textit{d}) at $F=3$. Inset labeled by transition region: the enlarged view of small oscillation before the harmonic to sub-harmonic transition $L_{tr}$. Insets indicated by arrows: corresponds to the onset of main sub-harmonic instability (shown by square box).}
	\label{fig:L}
\end{figure}

In the analytical model, the instability of the interface under the influence of rotation is presented in terms of the mixing zone size-$L$ $\left( \propto \omega^2/N^2\right)$. We present the time evolution of the mixing zone size-$L$  (equation \ref{mz width}) for forcing amplitude $F=0.75$ in figure \ref{subfig:L_F075} obtained from our simulations. The initial stratification condition corresponding to the initial mixing zone size-$L_0=0.096$ is $N_0^2/\omega^2=4.64$ that satisfies the instability condition \ref{inst condition} and ensures the onset of the Faraday instability. For the case without rotation (F075f/$\omega$0) at $\omega t\simeq 58$, $L\simeq0.48$ and begins to oscillate with a small amplitude that rapidly grows (shown in the inset of figure \ref{subfig:L_F075} indicated by black arrow) to triggers the instability resulting in turbulent mixing. According to the stability diagram in figure \ref{subfig:3dF075} (green line segment), the $N^2/\omega^2\simeq0.93$ corresponding to $L\simeq0.48$ lies in the stable region between the first sub-harmonic and harmonic tongues. When $L$ increases, $N^2/\omega^2$ decreases and enters the first sub-harmonic tongue resulting in the rapid growth of $L$ and turbulent mixing similar to the simulations. After $\omega t\simeq130$, an asymptotic state is achieved, where the mixing zone size oscillates without any further increase, signifying the saturation of instability. The final size of the mixing zone is defined by $L_{end}$ and the value of $L_{end}\simeq3.1$ for F075f/$\omega$0. In the stability diagram (\ref{subfig:3dF075}, green line segment), $N^2/\omega^2\simeq0.144$ corresponding to $L_{end}\simeq3.1$ lies in the first stable region and signifies the saturation state of the instability.\\ 

The presence of rotation at $F = 0.75$ stabilizes the flow by impeding the growth of the instabilities. $L$ initially grows slowly due to the delay in harmonic to sub-harmonic transition, followed by a quick growth because of the sub-harmonic instability. For the case with $f/\omega=0.48$ the sub-harmonic instability is triggered at $\omega t\simeq340-355$, $L\simeq1.08-1.12$, $N^2/\omega^2\simeq0.412-0.398$, as illustrated by the case F075f/$\omega$48 in figure \ref{subfig:L_F075} (see the inset indicated by green arrow). When the Coriolis frequency is increased to $f/\omega=0.59$, the sub-harmonic instability begins at $\omega t\simeq635-650$, $L\simeq1.48-1.52$, $N^2/\omega^2\simeq0.3-0.293$, as depicted by the case F075f/$\omega$59 in figure \ref{subfig:L_F075} (see the inset indicated by red arrow). Therefore, the onset of the sub-harmonic instability causing turbulent mixing is significantly delayed owing to rotation. The large Coriolis frequency ($f/\omega=0.59$) also suppresses the amplitude of oscillations in $L$ in the sub-harmonic instability phase. For $f/\omega=0.48$ ($f^2/\omega^2=0.23$) the instability saturates ($L_{end}\simeq3.25$) and small oscillations in the asymptotic state are damped out by diffusion, whereas for $f/\omega=0.59$ ($f^2/\omega^2=0.35$) the instability never saturates and hence the mixing zone size-$L$ continues to evolve with oscillations. This result is consistent with our theoretical prediction that the instability saturates for $f^2/\omega^2<0.25$ but never saturates for $f^2/\omega^2\geq0.25$. \\

At forcing amplitude $F=1$, the time evolution of the $L$ for without rotation (F1f/$\omega$0) and with rotation (F1f/$\omega$48 and F1f/$\omega$59) cases are illustrated in figure \ref{subfig:L_F1}. For all the cases, the instability condition \ref{inst condition} is satisfied by the initial stratification condition $N_0^2/\omega^2=4.25$ corresponding to the initial mixing zone size-$L_0=0.096$ that ensures the onset of the Faraday instability. Before the rapid growth of $L$ small oscillations of varying time periods (shown in the inset labeled by transition region of figure \ref{subfig:L_F1}) are observed for both the without rotation ($f/\omega=0$) and with rotation ($f/\omega$=0.48 and $f/\omega=0.59$) cases. These small oscillations represent the harmonic to sub-harmonic transition \citep{briard2019harmonic}. For $f/\omega =0.48$ and $0.59$ this transition occurs at $\omega t\simeq 60-70$ and $65-75$ (see inset of figure \ref{subfig:L_F1} labeled by transition region) followed by the stable region till $\omega t\simeq150-160$ and $290-300$ respectively. No harmonic or sub-harmonic modes are excited in the stable region, and $L$ grows due to diffusion. Similar stable regions are also be observed for F075f/$\omega$48 and F075f/$\omega$59 in figure \ref{subfig:L_F075} (see the supplementary movies $1$ and $2$ of the concentration field for $F = 0.75$ and $1$). The presence of this stable region delays the onset of the sub-harmonic instability and is also present in our theoretical analysis in \cref{subsec:instability delays}. $L$ rapidly grows after the triggering of the sub-harmonic instability resulting in turbulent mixing. Similar to $F = 0.75$, we observe that $L$ saturates for F1f/$\omega$48, but continues to grow for F1f/$\omega$59.\\ 

The effect of rotation on the evolution of $L$ at higher forcing amplitudes $F=2$ and $F=3$ is demonstrated in figures \ref{subfig:L_F2} and \ref{subfig:L_F3} respectively. The initial stratification condition is $N_0^2/\omega^2=3.26$ for $F=2$ and $N_0^2/\omega^2=2.6$ for $F=3$ corresponding to the initial mixing zone size-$L_0=0.096$ which ensures the onset of Faraday instability. At $F=2$, the harmonic to sub-harmonic transition ($L_{tr}\simeq0.5$) occurs at $\omega t\simeq13$ for F2f/$\omega$0, F2f/$\omega$48 and F2f/$\omega$59 as indicated by the black circle in the inset of figure \ref{subfig:L_F2}. Further increasing the forcing amplitude to $F=3$, the transition ($L_{tr}\simeq0.375$) occurs at $\omega t\simeq6.5$ for F3f/$\omega$0, F3f/$\omega$48 and F3f/$\omega$59 cases, as shown by the black circle in the inset of figure \ref{subfig:L_F3}. The effect of rotation in stabilizing the flow is subdued as the forcing amplitude increases. This observation is consistent with our analytical model that predicts the shrinkage of the stable region between the unstable harmonic and sub-harmonic tongues with increasing forcing amplitudes (see stability diagrams at F = 2,3 in figures \ref{subfig:3dF2} and \ref{subfig:3dF3} respectively). For cases with $f^2/\omega^2=0$ and $f^2/\omega^2=0.23$, the instability saturates, and the oscillations in the asymptotic state are damped. However, the instability never saturates for cases with $f^2/\omega^2=0.35$ and $L$ continues to evolve with oscillations. \\ 

 \captionsetup[subfigure]{textfont=normalfont,singlelinecheck=off,justification=raggedright}
 \begin{figure}
%	\centering
	\begin{subfigure}{0.263\textwidth}
		\centering
		\caption{}
		\includegraphics[width=1.0\textwidth,trim={0cm 0.1cm 1.4 0cm},clip]{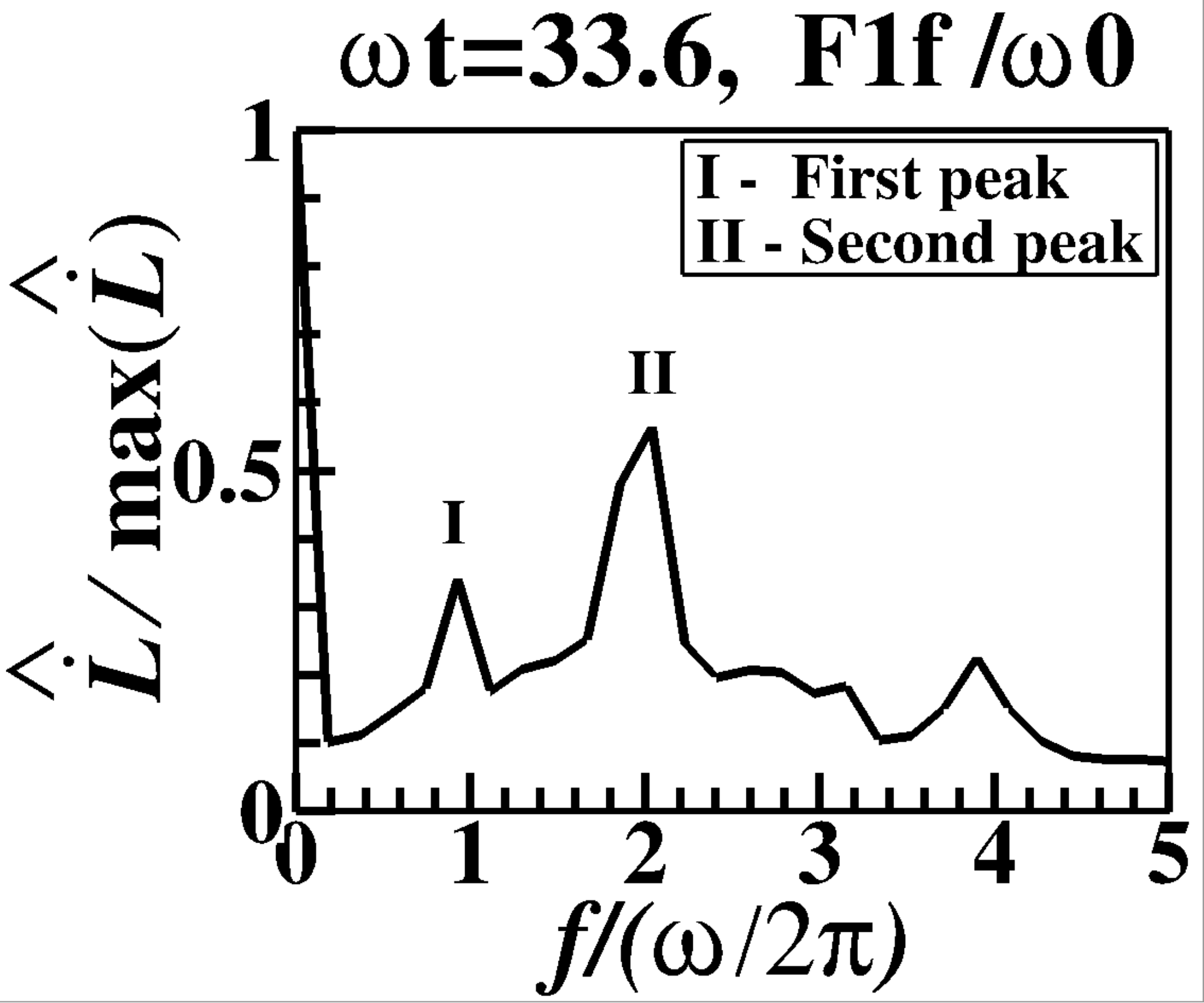}
		\label{subfig:FFT_F1fw0_a}
	\end{subfigure}
%	\hfill
	\begin{subfigure}{0.24\textwidth}
		\centering
		\caption{}
		\includegraphics[width=1.0\textwidth,trim={1.8cm 0.1cm 1.4 0cm},clip]{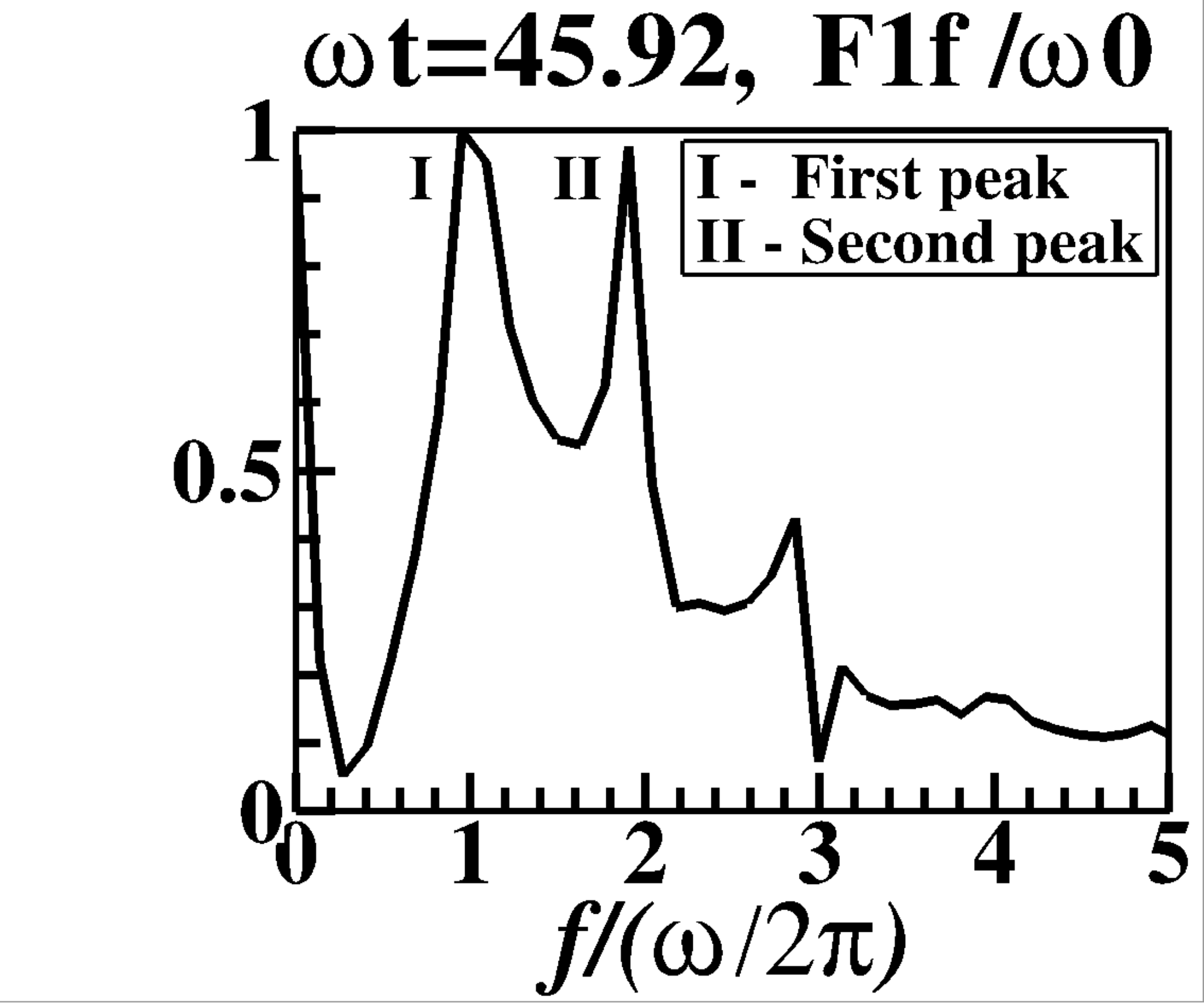}
		\label{subfig:FFT_F1fw0_b}
	\end{subfigure}
%	\hfill
	\begin{subfigure}{0.24\textwidth}
		\centering
		\caption{}
		\includegraphics[width=1.0\textwidth,trim={1.8cm 0.1cm 1.4 0cm},clip]{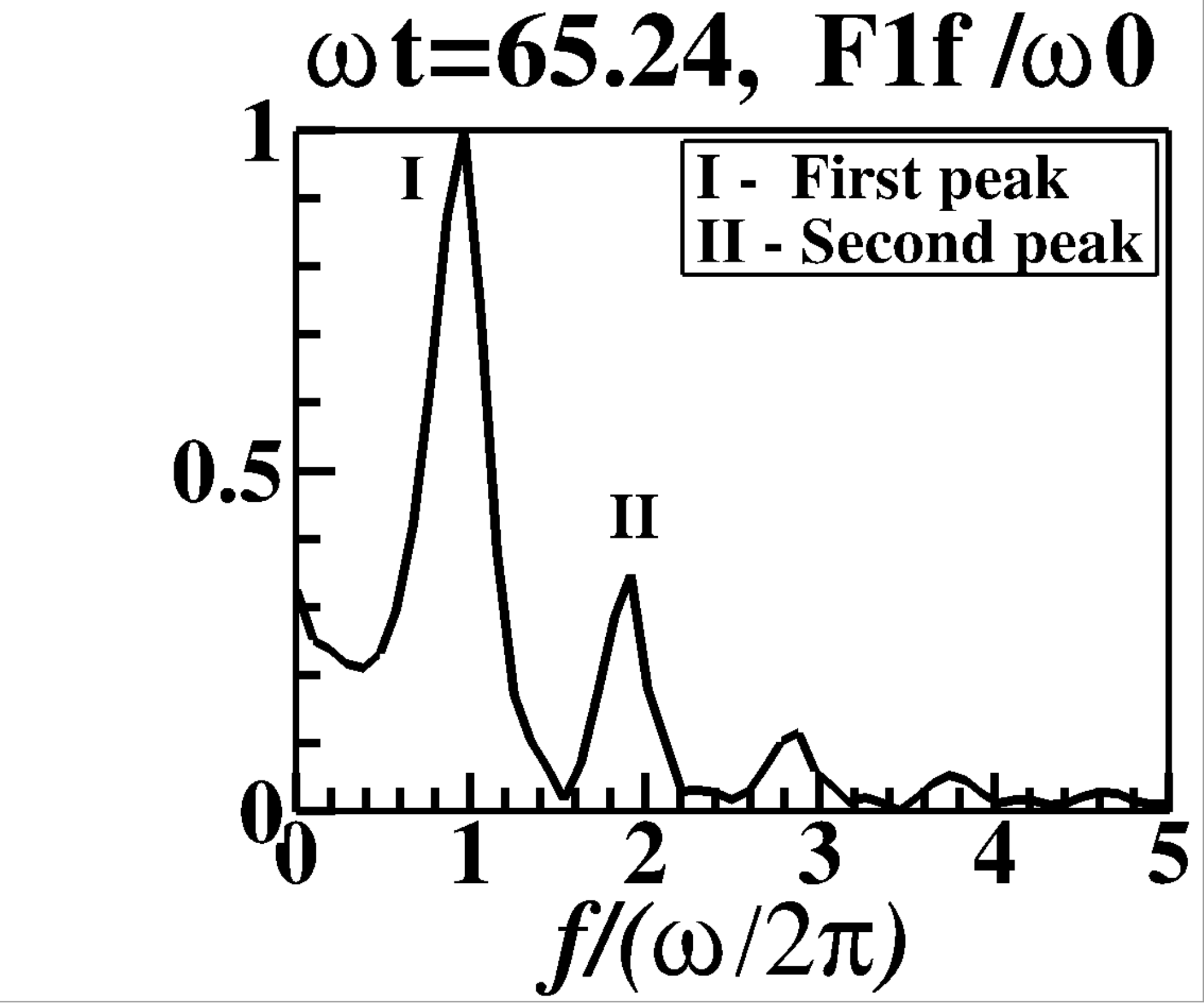}
		\label{subfig:FFT_F1fw0_c}
	\end{subfigure}
%	\hfill
    \begin{subfigure}{0.24\textwidth}
		\centering
		\caption{}
		\includegraphics[width=1.0\textwidth,trim={1.8cm 0.1cm 1.4 0cm},clip]{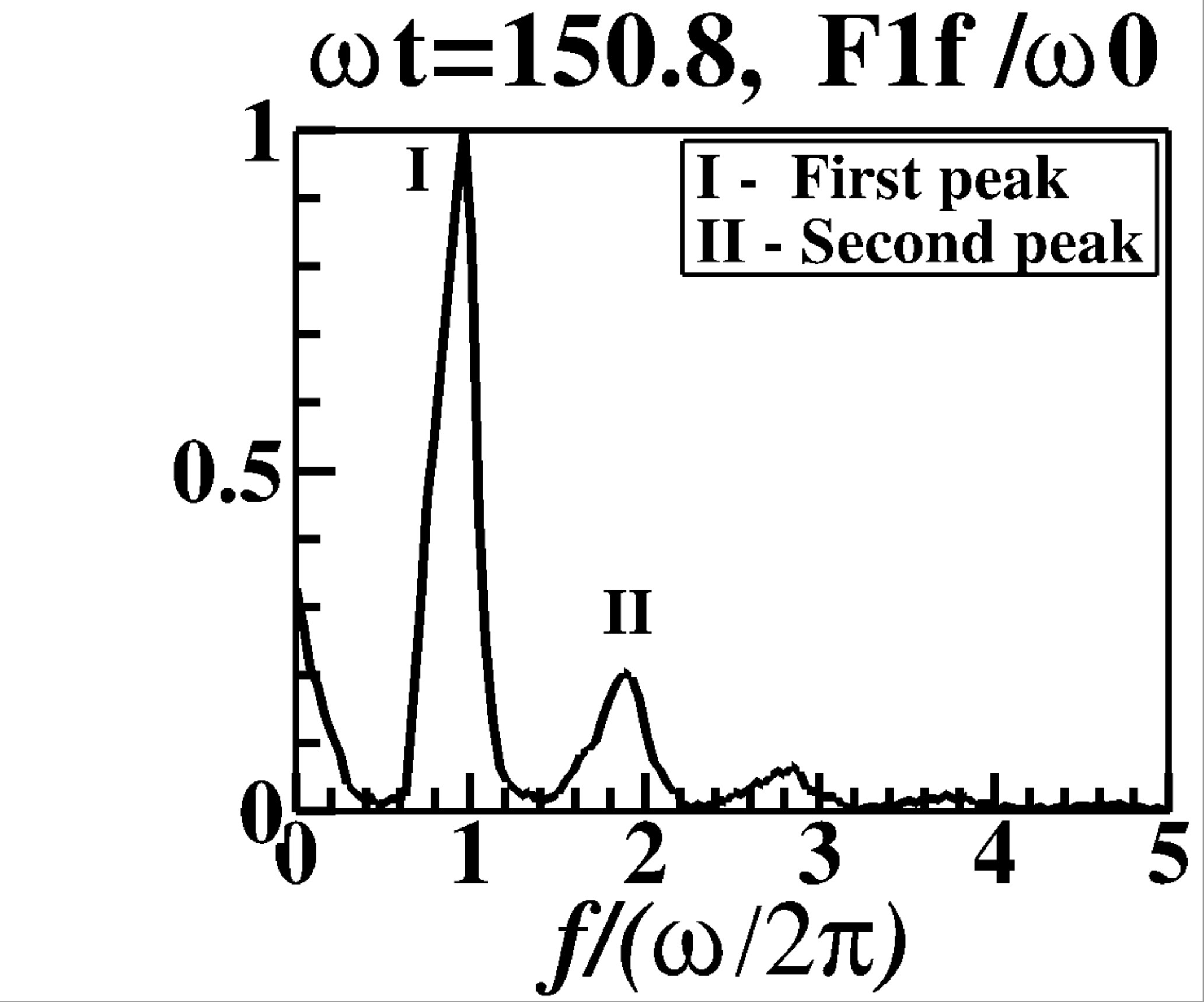}
		\label{subfig:FFT_F1fw0_d}
	\end{subfigure}
	\centering
%	\centering
	\begin{subfigure}{0.263\textwidth}
		\centering
		\caption{}
		\includegraphics[width=1.0\textwidth,trim={0cm 0.1cm 1.4 0cm},clip]{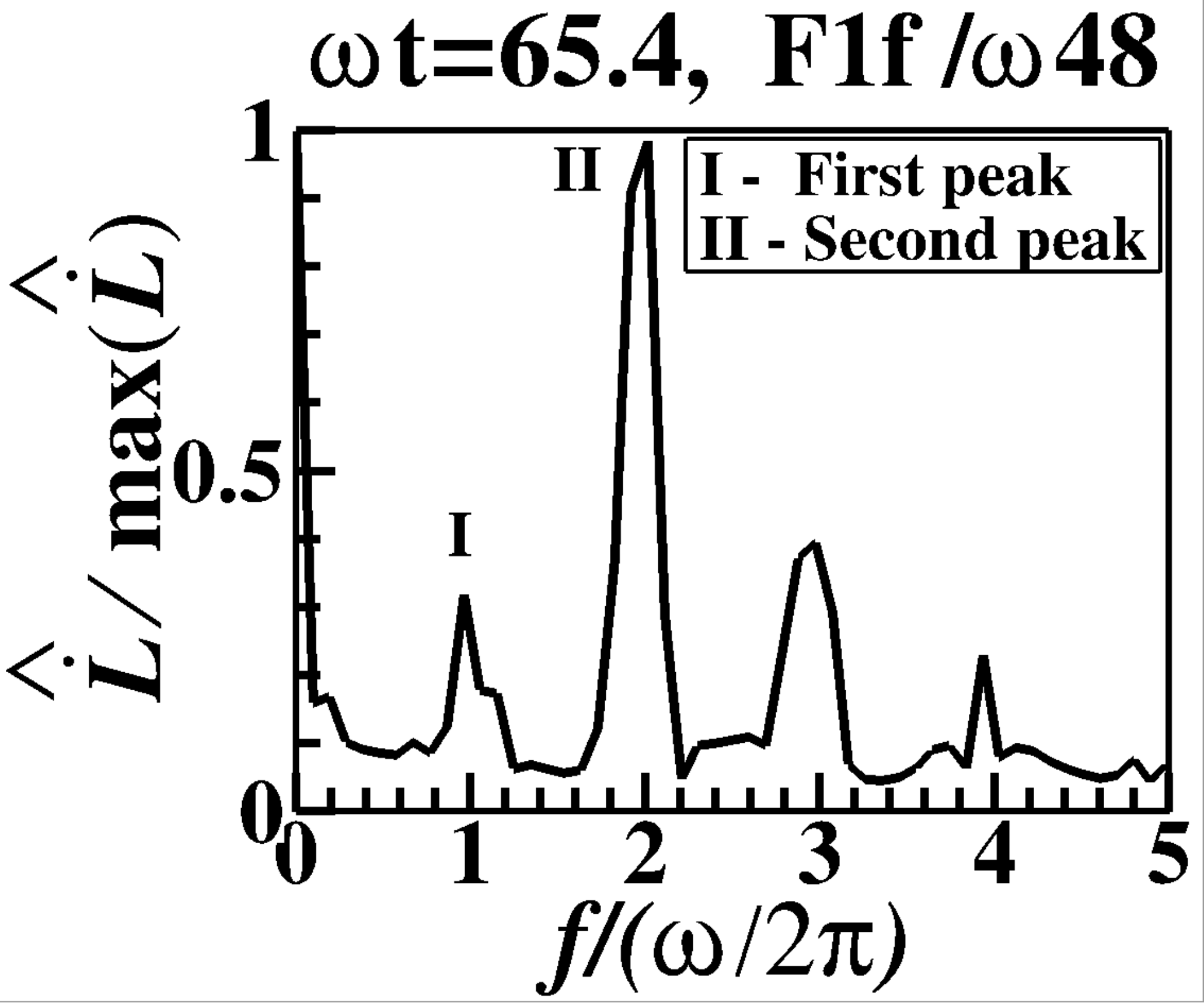}
		\label{subfig:FFT_F1fw48_a}
	\end{subfigure}
%	\hfill
	\begin{subfigure}{0.24\textwidth}
		\centering
		\caption{}
		\includegraphics[width=1.0\textwidth,trim={1.8cm 0.1cm 1.4 0cm},clip]{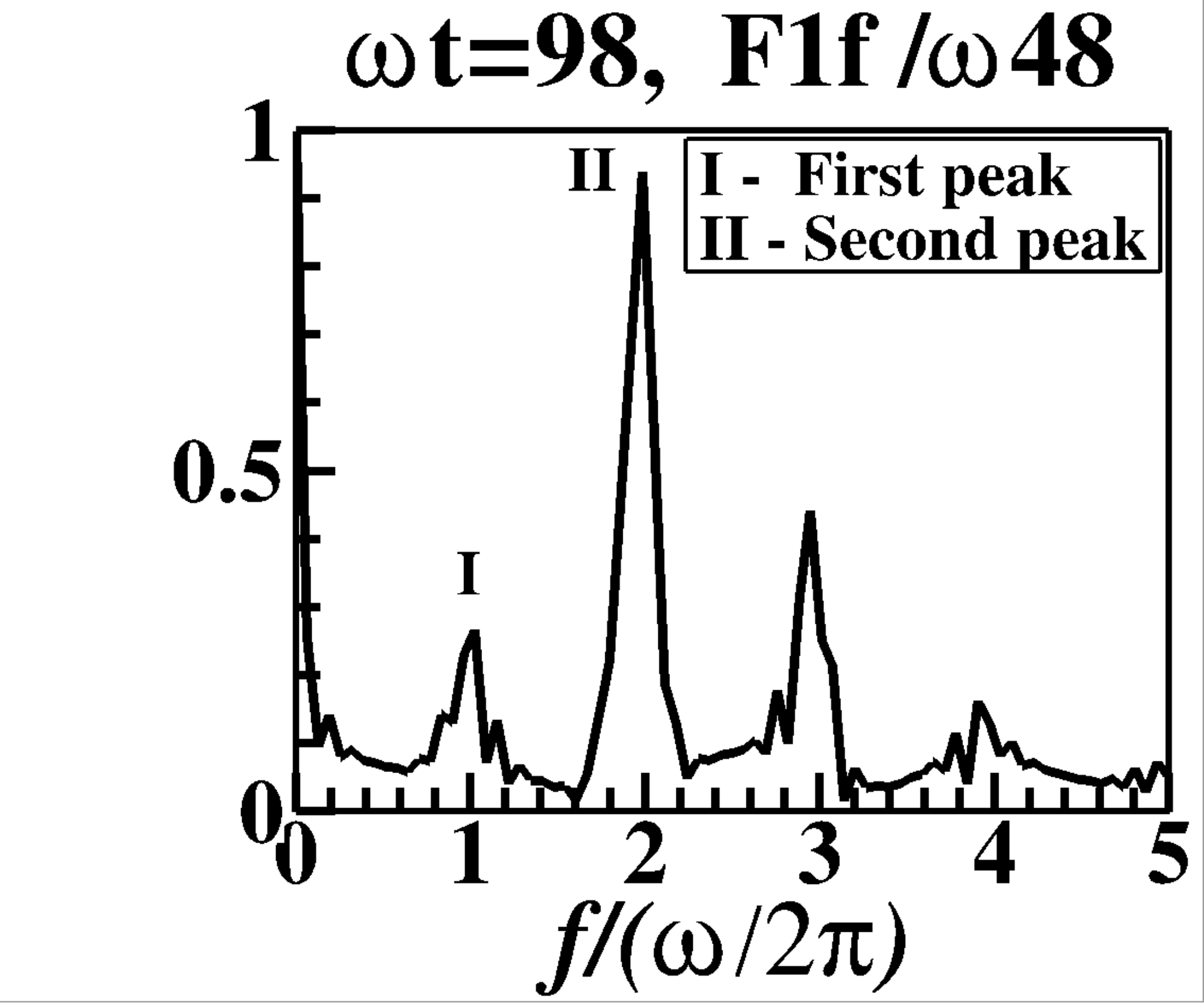}
		\label{subfig:FFT_F1fw48_b}
	\end{subfigure}
%	\hfill
	\begin{subfigure}{0.24\textwidth}
		\centering
		\caption{}
		\includegraphics[width=1.0\textwidth,trim={1.8cm 0.1cm 1.4 0cm},clip]{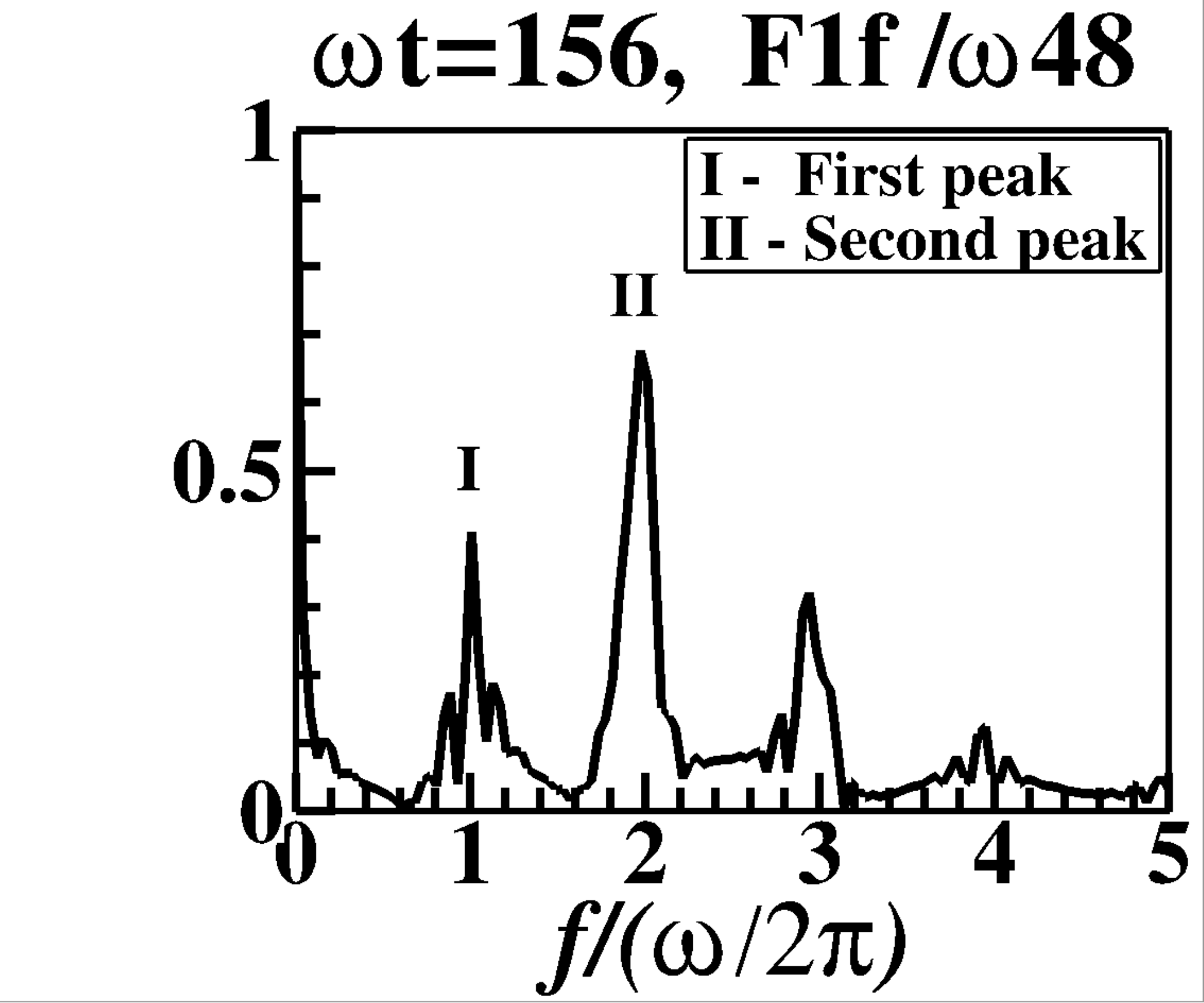}
		\label{subfig:FFT_F1fw48_c}
	\end{subfigure}
%	\hfill
    \begin{subfigure}{0.24\textwidth}
		\centering
		\caption{}
		\includegraphics[width=1.0\textwidth,trim={1.8cm 0.1cm 1.4 0cm},clip]{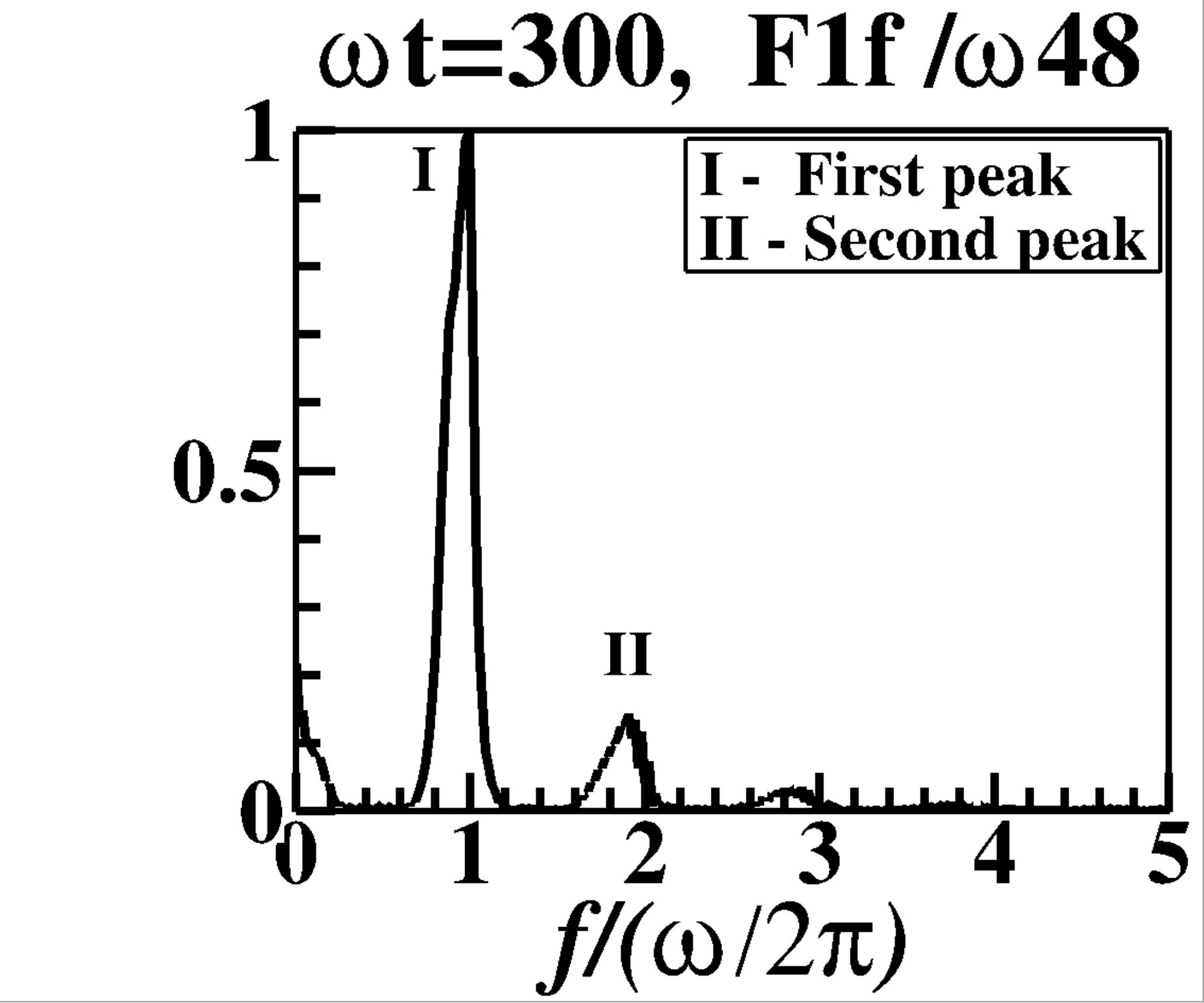}
		\label{subfig:FFT_F1fw48_d}
	\end{subfigure}
	\caption{Normalized Fourier transform of ${\dot L}$ as a function of normalized frequency, (\textit{a,b,c,d}) for F1f/$\omega$0 case, (\textit{e,f,g,h}) for F1f/$\omega$48 case, for different time intervals each starting from $\omega t=0 $ to $\omega t$ mentioned on the top of each figure. First peak at $f /\left(\omega/2 \pi \right)\simeq1$ corresponds to the sub-harmonic regime and second peak at $f /\left(\omega/2 \pi \right)\simeq2$ corresponds to the harmonic regime \citep{briard2019harmonic}. }
	\label{fig:FFT_peaks}
 \end{figure}

Fourier transform of the derivatives of $L$ for different time intervals, each starting from $t=0$, are performed to capture the harmonic to sub-harmonic transition in our simulations. For case F1f/$\omega$0, the normalized Fourier transform of $dL/dt$ as a function of the normalized frequency $f/(\omega2\pi)$ is depicted in figure \ref{subfig:FFT_F1fw0_d}. Two peaks of different frequencies are present in $L$. The first and the second peaks represent the sub-harmonic and harmonic regimes, respectively. We can observe that for $\omega t \leq 45$, the second peak is higher (see figure \ref{subfig:FFT_F1fw0_a}), denoting the dominance of the harmonic regime. At $\omega t = 45.92$ (figure \ref{subfig:FFT_F1fw0_b}) the amplitude of both the peaks are similar (see figure \ref{subfig:FFT_F1fw0_b}) indicating the transition from harmonic to sub-harmonic regime. Note that the small oscillations in $L$ also rapidly grow from $\omega t\simeq 45$ (see figure \ref{subfig:L_F1}) owing to the onset of the sub-harmonic instability. When $\omega t > 45$ the first peak becomes the dominating peak as illustrated in the figures \ref{subfig:FFT_F1fw0_c} and \ref{subfig:FFT_F1fw0_d}, signifying the primary role of the sub-harmonic instabilities in turbulent mixing. The Fourier transform of $dL/dt$ for F1f/$\omega$48 demonstrates the harmonic to sub-harmonic transition, stable region, and the onset of the sub-harmonic instability. The dominating second peak at $\omega t = 65.4$ in figure \ref{subfig:FFT_F1fw48_a} shows small oscillations of $L$ (see inset of figure \ref{subfig:L_F1} indicated by green arrow) are in the harmonic region. The amplitude of the first and second peak remains approximately unchanged till $\omega t\simeq 150$ (figures \ref{subfig:FFT_F1fw48_b} and \ref{subfig:FFT_F1fw48_c}) which represents the stable region. At $\omega t\simeq 150-160$, the first peak start increasing, and the second peak start decreasing. Eventually the first peak becomes the dominant (figure \ref{subfig:FFT_F1fw48_d}) marking the onset of the sub-harmonic instability at $\omega t\simeq 150-160$. An increase in the Coriolis frequency to $f/\omega=0.59$ impedes the harmonic to sub-harmonic transition, expands the stable region, and delays the onset of the main sub-harmonic instability (figures not shown). \\

\begin{figure}
%	\centering
	\begin{subfigure}{0.23\textwidth}
		\centering
		\caption{}
		\includegraphics[width=1.0\textwidth,trim={0cm 0.1cm 0.1cm 0cm},clip]{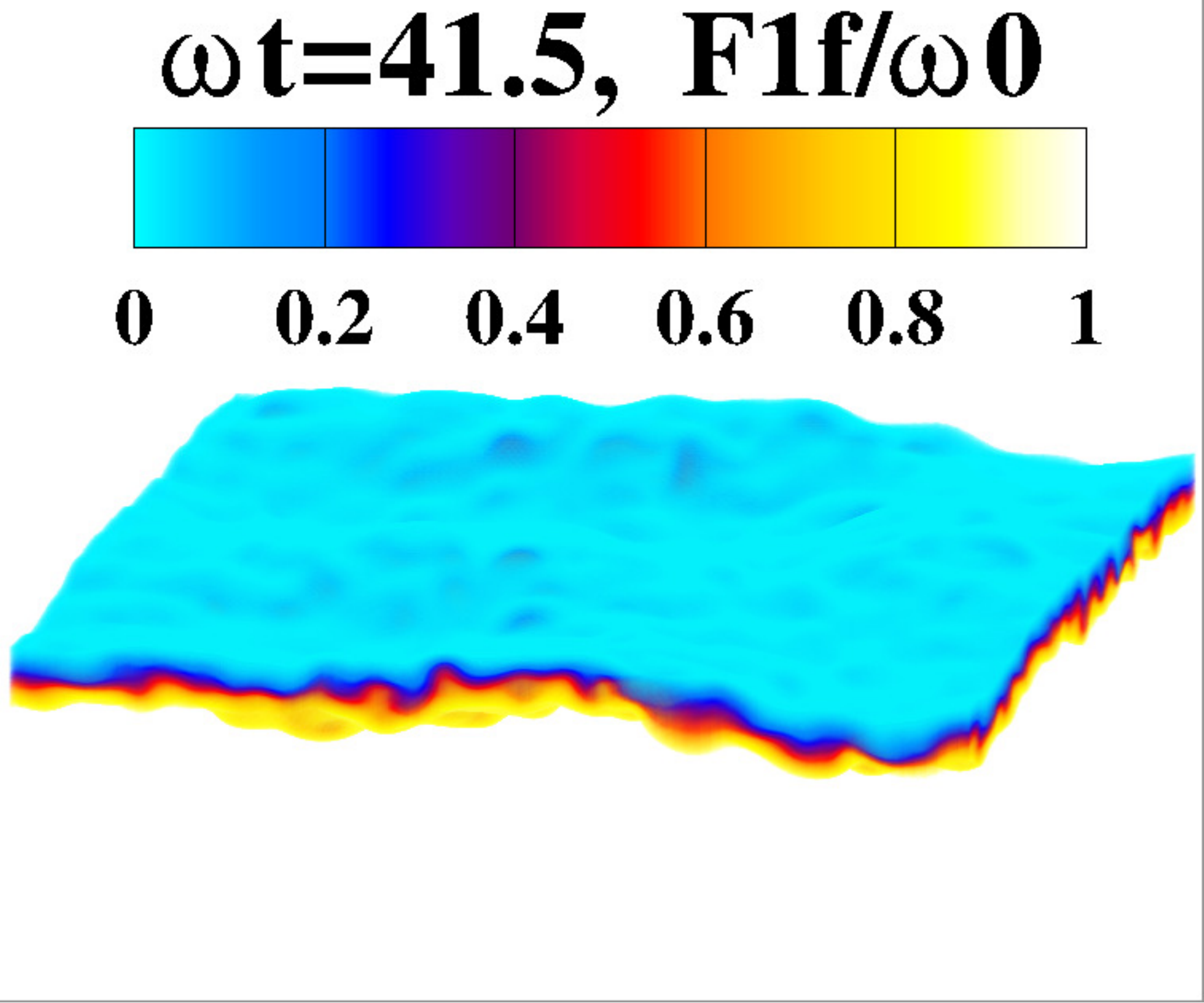}
		\label{subfig:Fig_c9}
	\end{subfigure}
%	\hfill
	\begin{subfigure}{0.23\textwidth}
		\centering
		\caption{}
		\includegraphics[width=1.0\textwidth,trim={0cm 0.1cm 0.1cm 0cm},clip]{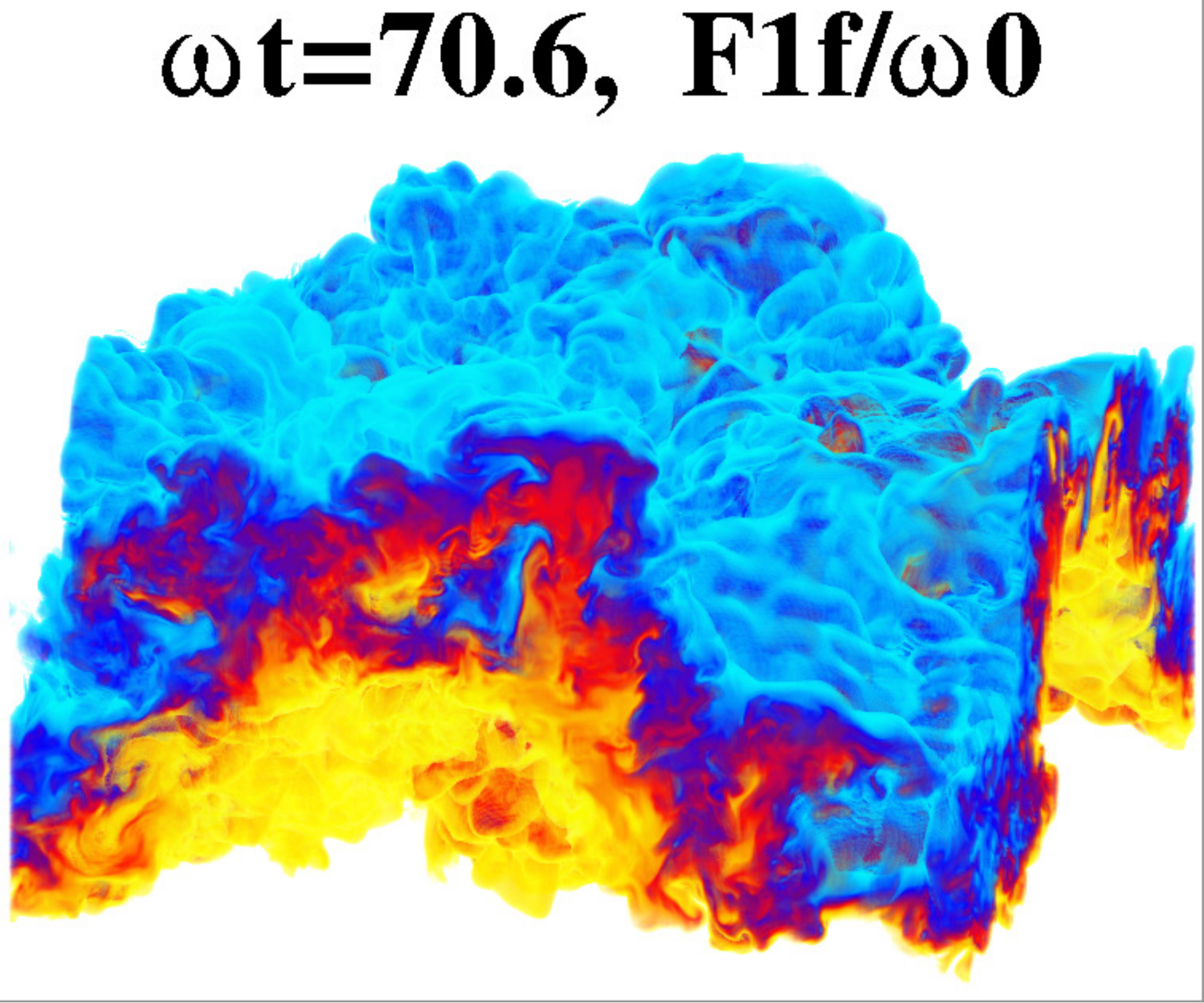}
		\label{subfig:Fig_c10}
	\end{subfigure}
%	\hfill
	\begin{subfigure}{0.23\textwidth}
		\centering
		\caption{}
		\includegraphics[width=1.0\textwidth,trim={0cm 0.1cm 0.1cm 0cm},clip]{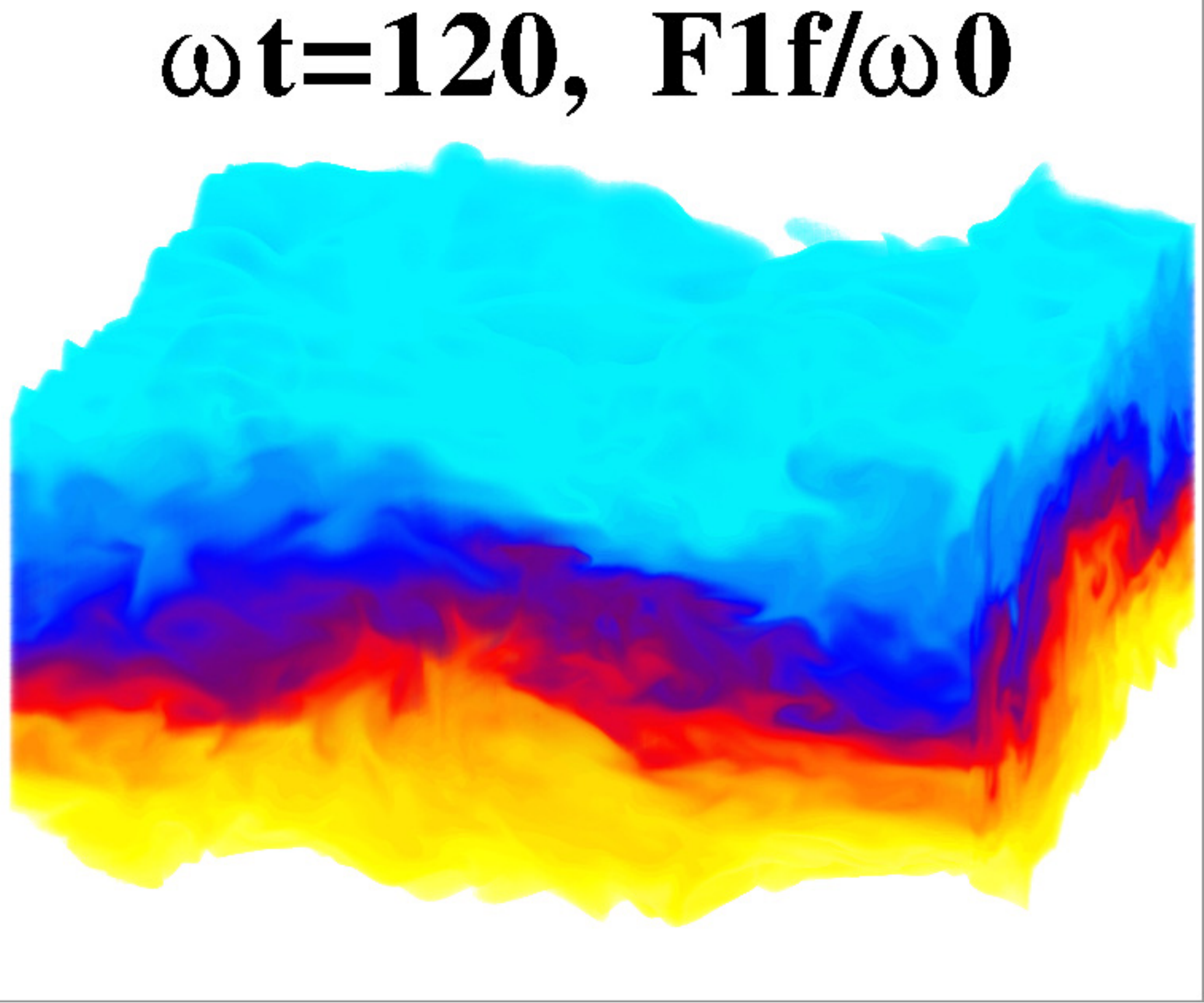}
		\label{subfig:Fig_c11}
	\end{subfigure}
%	\hfill
    \begin{subfigure}{0.23\textwidth}
		\centering
		\caption{}
		\includegraphics[width=1.0\textwidth,trim={0cm 0.1cm 0.1cm 0cm},clip]{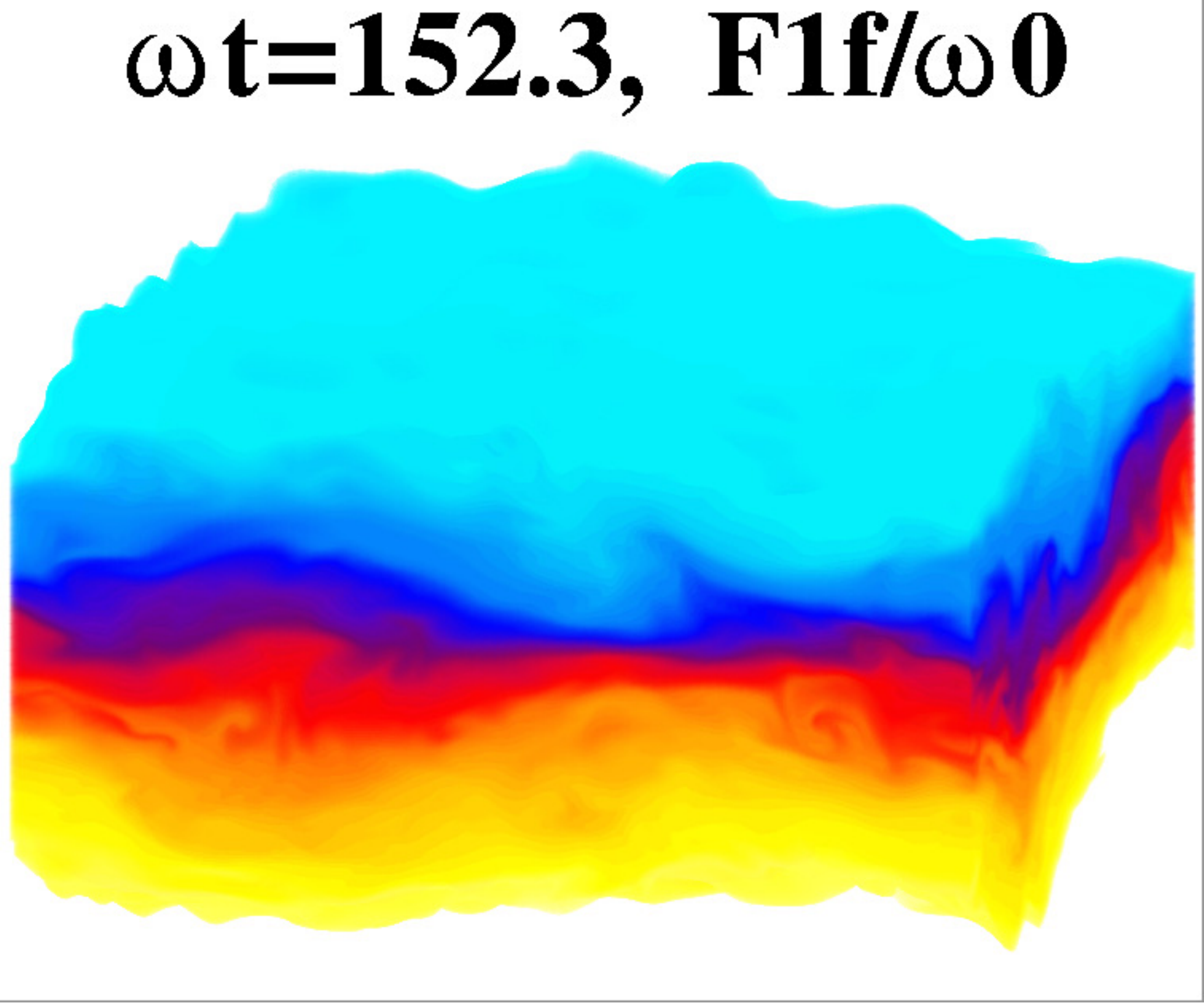}
		\label{subfig:Fig_c12}
	\end{subfigure}
	\centering
	%\hfill
 	\begin{subfigure}{0.23\textwidth}
 		\centering
 		\caption{}
 		\includegraphics[width=1.0\textwidth,trim={0cm 0.1cm 0.1cm 0cm},clip]{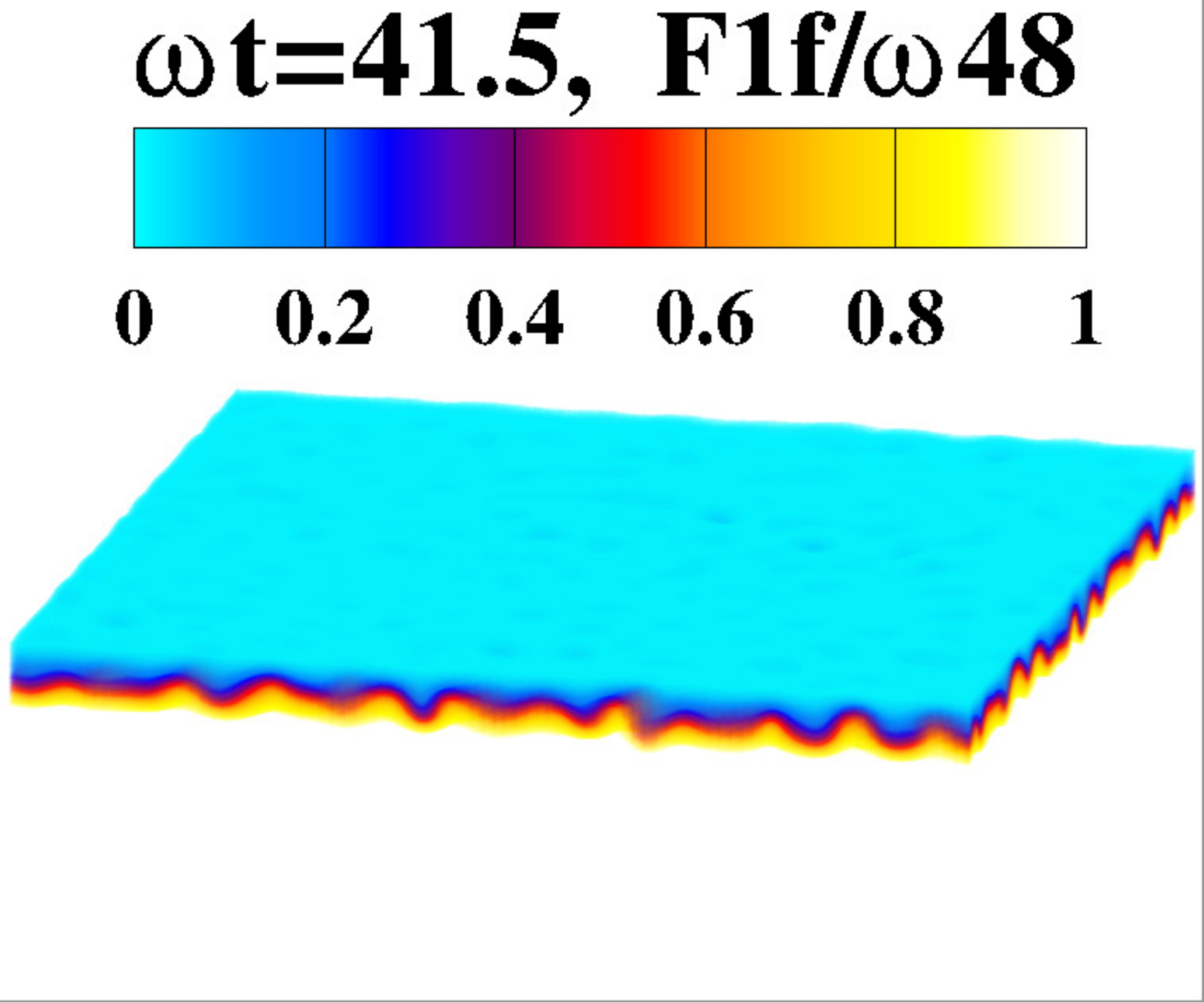}
 		\label{subfig:Fig_c13}
 	\end{subfigure}
 	%\hfill
 	\begin{subfigure}{0.23\textwidth}
 		\centering
 		\caption{}
 		\includegraphics[width=1.0\textwidth,trim={0cm 0.1cm 0.1cm 0cm},clip]{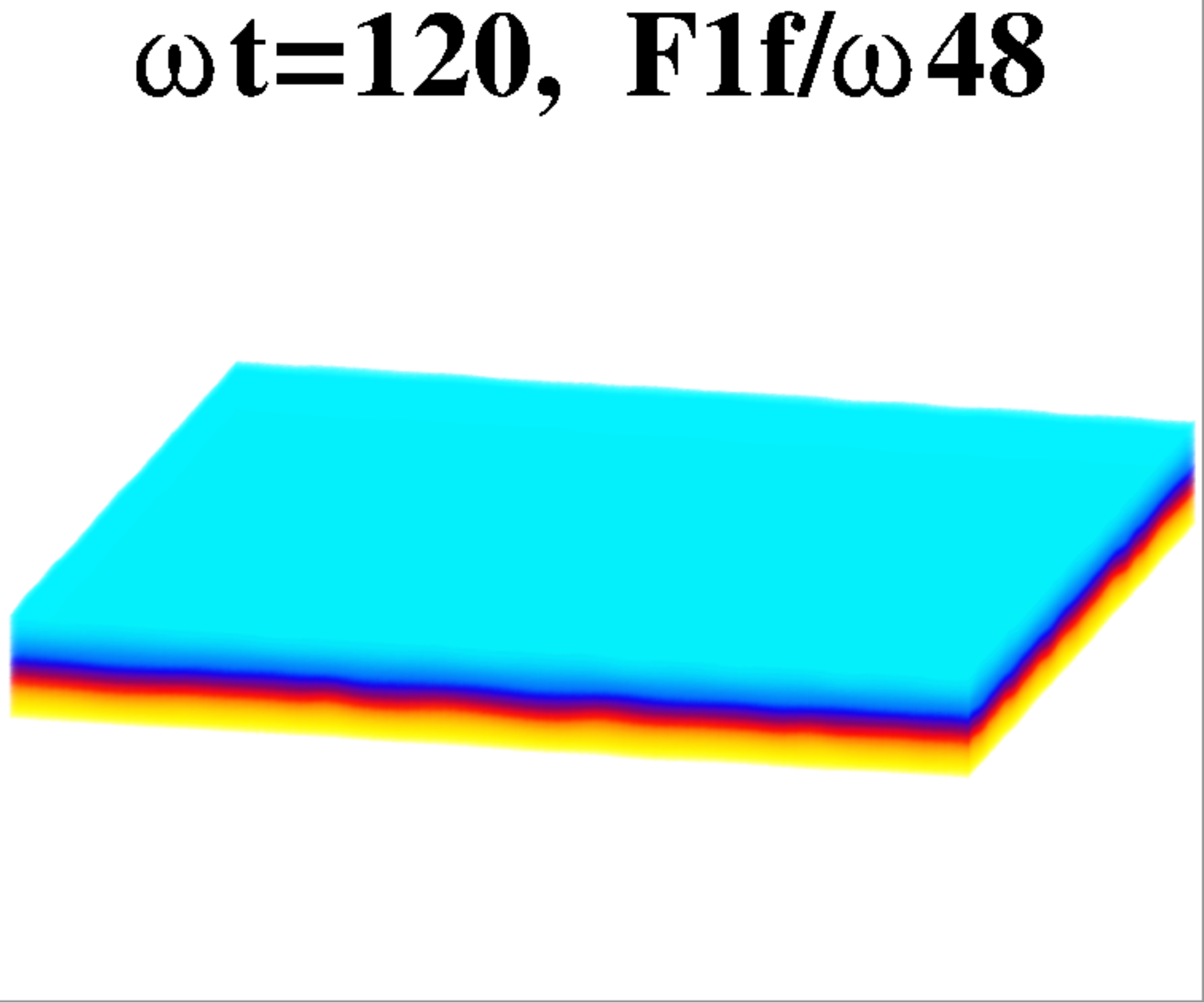}
 		\label{subfig:Fig_c14}
 	\end{subfigure}
 	%\hfill
 	\begin{subfigure}{0.23\textwidth}
 		\centering
 		\caption{}
 		\includegraphics[width=1.0\textwidth,trim={0cm 0.1cm 0.1cm 0cm},clip]{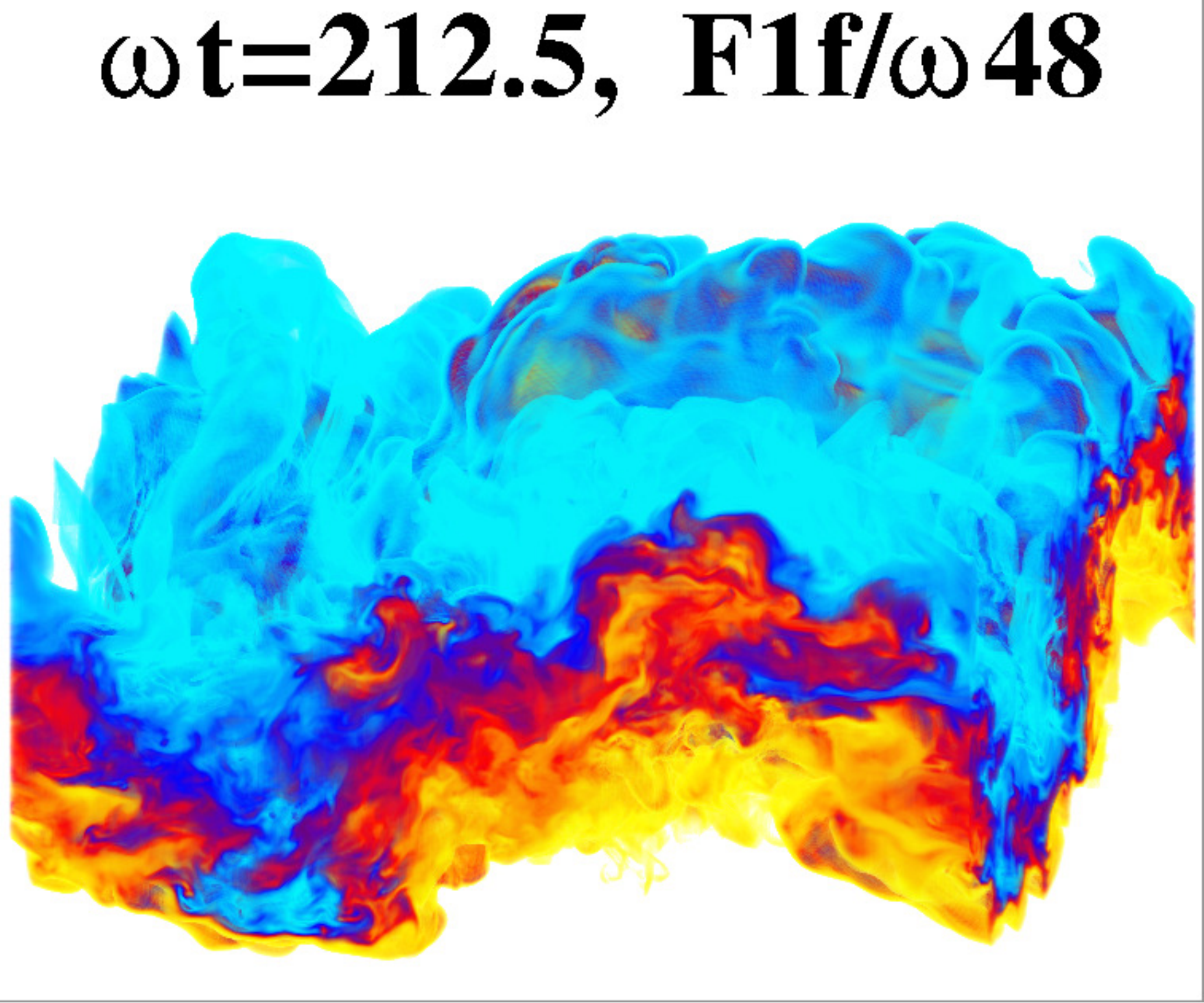}
 		\label{subfig:Fig_c15}
 	\end{subfigure}
 	%\hfill
     \begin{subfigure}{0.23\textwidth}
 		\centering
 		\caption{}
 		\includegraphics[width=1.0\textwidth,trim={0cm 0.1cm 0.1cm 0cm},clip]{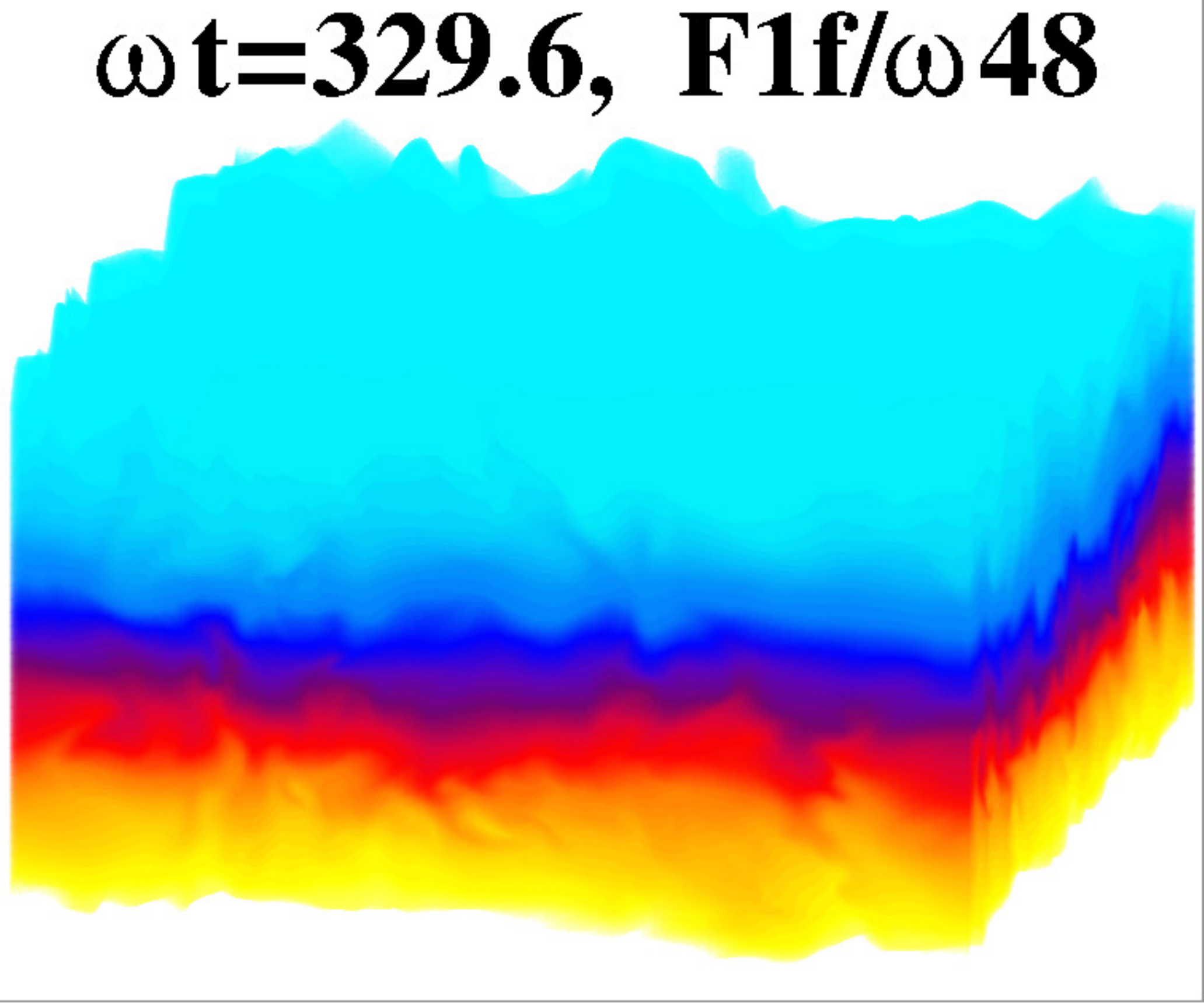}
 		\label{subfig:Fig_c16}
 	\end{subfigure}
	\caption{
	Visualization of three-dimensional concentration field, (\textit{a,b,c,d}) for F1f/$\omega$0 case, (\textit{e,f,g,h}) for F1f/$\omega$48 case, at different time instants. Color map: blue for $C=0$ (lighter fluid) and red for $C=1$ (denser fluid). For better visualisation of the mixing of fluids, the pure fluids are made transparent.  }
	\label{fig:conc_field_F1}
 \end{figure}

Figure \ref{fig:conc_field_F1} depicts the three-dimensional concentration field in the mixing zone for without rotation F1f/$\omega$0 and with rotation F1f/$\omega$48 cases. The random wave patterns on the interface at $\omega t=41.5$ for F1f/$\omega$0 indicate the inception of the Faraday instability as shown in figure \ref{subfig:Fig_c9}. The turbulent mixing caused by the sub-harmonic instability is demonstrated at $\omega t=70.6$ for F1f/$\omega$0 in \ref{subfig:Fig_c10} followed by the beginning of instability saturation and fully saturated instability in figures \ref{subfig:Fig_c11} and \ref{subfig:Fig_c12} at $\omega t=120$ and $\omega t\sim152$ respectively. For F1f/$\omega$48, the stable region after the harmonic to sub-harmonic transition is depicted by figure \ref{subfig:Fig_c14} at $\omega t=120$. The subsequent figures \ref{subfig:Fig_c15} and \ref{subfig:Fig_c16} for F1f/$\omega$48 at $\omega t \sim 212$ and $329$ show a significant delay in turbulent mixing and saturation of the instability with respect to F1f/$\omega$0. Movies $(1-4)$ of the concentration field at $x_2 = 0$ plane for different forcing amplitudes are included as supplementary data, to demonstrate the influence of rotation on the evolution of the mixing zone.\\

 \captionsetup[subfigure]{textfont=normalfont,singlelinecheck=off,justification=raggedright}

\section{Conclusions}\label{sec:conclusions}

The effect of rotation on the onset and saturation of Faraday instability in two miscible fluids subjected to vertical oscillations is investigated  by means of theoretical analysis and DNS. The analytical model utilizes  Floquet theory to solve a set of Mathieu equations derived from linear stability analysis. On solving the set of Mathieu equations, a three-dimensional stability diagram is obtained that consists of stable and unstable harmonic and sub-harmonic regions. It was observed that the onset of the sub-harmonic instabilities occurring after the harmonic to sub-harmonic transition triggers turbulent mixing. We found that rotation stabilizes the flow and delays the onset of the sub-harmonic instability at lower forcing amplitudes ($F$). In contrast, the effect of rotation diminishes with an increase in $F$. The reason for this occurrence is the increase in the area of all the unstable tongue in the 3D stability diagram resulting the majority of excited $\theta$-modes to be unstable.\\
 
Further, we have performed simulations at different forcing amplitudes and rotation rates to verify the findings of our theoretical model. The harmonic to sub-harmonic transition followed by the onset of the sub-harmonic instability, and its saturation is shown in terms of the time evolution of the mixing zone size $L$. For $F = 0.75$  and $1$, the evolution of the mixing zone size manifests a significantly slow growth under the influence of rotation. This slow growth of $L$ represents the stable region during the harmonic to sub-harmonic transition. After this transition, the sub-harmonic instability is triggered, and $L$ rapidly grows, resulting in turbulent mixing. At forcing amplitudes of $2$ and $3$, the vertical vibrations are strong enough to diminish the stabilizing effect of rotation. Therefore, the mixing zone size for the cases with $F = 2$ and $3$ with rotation grows similar to the non-rotating cases.  We also perform the Fourier transform of the rate of change of $L$ to capture the harmonic to sub-harmonic transition and demonstrate the delay in this transition in terms of the peak frequencies.\\
 
Finally, we have compared the simulations for $f/\omega=0.48$ and $f/\omega=0.59$, which correspond to $\left(f/\omega \right)^2 < 0.25$ and $\left(f/\omega \right)^2 \geq 0.25$ respectively on the 3D stability diagram, at different forcing amplitudes. We find that for $f/\omega=0.48$, the instability saturates, and the mixing zone size asymptotes. However, for $f/\omega$=0.59, the instability does not saturate, and the mixing zone size keeps growing. This confirms our theoretical observation that for $\left(f/\omega \right)^2 < 0.25$ the instability saturates, whereas it never saturates for $\left(f/\omega \right)^2 \geq 0.25$ because all the excited segments cross the first instability tongue at each instant. \\

The turbulent mixing, attributed to the onset of the sub-harmonic instability, is associated with exchanges among the total potential energy (TPE), the background potential energy (BPE), and the available potential energy (APE). We aim to extend the present investigation to explore the influence of rotation on these energy exchanges through reversible buoyancy flux, irreversible diapycnal mixing, and irreversible kinetic energy dissipation to quantify the irreversible mixing efficiency. \\

\backsection[Supplementary data]{\label{SupMov}  \\https://doi.org/**.****/jfm.***...}

\backsection[Acknowledgements]{We gratefully acknowledge the support of the Science and Engineering Research Board, Government of India grant no. SERB/ME/2020318. We also want to thank the Office of Research and Development, Indian Institute of Technology Kanpur for the financial support through grant no. IITK/ME/2019194. The support and the resources provided by PARAM Sanganak under the National Supercomputing Mission, Government of India at the Indian Institute of Technology, Kanpur are gratefully acknowledged.}

%\backsection[Funding]{This research received no specific grant from any funding agency, commercial or not-for-profit sectors.}

\backsection[Declaration of interests]{ The authors report no conflict of interest.}

%\backsection[Data availability statement]{The data that support the findings of this study are openly available in [repository name] at http://doi.org/[doi], reference number [reference number].}

%\backsection[Author ORCID]{Souvik Naskar, https://orcid.org/0000-0003-0445-8417; Anikesh Pal, https://orcid.org/****-****-****-****}

\backsection[Author contributions]{N.S. and A.P. designed research; N.S. and A.P. performed research; N.S. analyzed data; N.S. and A.P. wrote the paper.}

%\backsection[Author ORCID]{Narinder Singh,  https://orcid.org/
%0000-0002-1529-1061; Anikesh Pal,  https://orcid.org/0000-0003-2085-7231 }

\appendix
\section{Steps to derive \ref{phi eqn}}
\label{derivation}
 \begin{equation}
 \begin{aligned}
     \label{phi eqn1}
     \frac{\partial^2}{\partial t^2} \left( \left(\phi \left(i^2k^2+i^2l^2\right)+\frac{\partial^2\phi}{\partial x_3^2}\right) e^{i (kx_1+lx_2-\varOmega t)} \right) = & -f^2\frac{\partial^2 \phi}{\partial x_3^2}e^{i (kx_1+lx_2-\varOmega t)} \\
     & + 2 \mathcal{A} g_0 \varGamma \phi \left(i^2k^2+i^2l^2\right) e^{i (kx_1+lx_2-\varOmega t)} ,
 \end{aligned}
 \end{equation}
 
 where $i^2=-1$. Substituting $N^2=-2\mathcal{A}g_0 \varGamma$ (from equation (\ref{strat fre})) and horizontal wavenumber $K$ which is defined as $K=\sqrt{k^2 + l^2}$, in equation (\ref{phi eqn1}) we get
 \begin{subequations}
 \begin{equation}
 \begin{aligned}
     \label{phi eqn2}
     \left(-i\varOmega\right)^2 \left(-\phi K^2 +\frac{\partial^2\phi}{\partial x_3^2}\right) e^{i (kx_1+lx_2-\varOmega t)}  = & -f^2\frac{\partial^2 \phi}{\partial x_3^2}e^{i (kx_1+lx_2-\varOmega t)} \\
     & +N^2 \phi K^2 e^{i (kx_1+lx_2-\varOmega t)} ,
 \end{aligned}
 \end{equation}
 
 \begin{equation}
     \label{phi eqn3}
     -\varOmega^2 \left(-\phi K^2 +\frac{\partial^2\phi}{\partial x_3^2}\right)  =  -f^2\frac{\partial^2 \phi}{\partial x_3^2} +N^2 \phi K^2 .
 \end{equation}
 \end{subequations}

Rearranging equation \ref{phi eqn3} we get equation \ref{phi eqn}.

\section{Steps for solving \ref{phi eqn}}\label{appA1}
We show the steps to solve equation (\ref{phi eqn}) which represents the second order linear homogeneous equation with constant coefficients. First we solve this equation in the upper pure fluid regime where vertical gradient of mean concentration $\varGamma=0$ and thus $N=\sqrt{-2\mathcal{A} g_0 \varGamma}=0$, according to our assumption of piecewise background concentration profile. So equation (\ref{phi eqn}) becomes 
\begin{equation}
\label{A13}
\frac{\partial^2\phi}{\partial x_3^2} - \frac{{K}^2 \varOmega^2 }{\left(\varOmega^2 - f^2 \right)}  \phi = 0 .
\end{equation}

Let $q^2=K^2\varOmega^2/\left(\varOmega^2 - f^2 \right)$ and thus the above equation can be written as 
\begin{equation}
\label{A12}
\frac{\partial^2\phi}{\partial x_3^2} - q^2  \phi = 0 .
\end{equation}

Let $\phi=e^{rx_3}$ be a solution to this equation, for some as-yet-unknown constant $r$. After substituting this assumed solution in equation (\ref{A12}) we get, $e^{rx_3}\left(r^2-q^2 \right)=0$. Since $e^{rx_3}$ is never zero, the above equation is satisfied if and only if $r^2-q^2=0$ and roots of this characteristic equation are $r=\pm q$, which are real and distinct. Therefore, the general solution of equation (\ref{A12}) is 
\begin{equation}
\label{A13}
\phi= {Ae^{qx_3}+Be^{-qx_3}} .
\end{equation}

To find out the constants $A$ and $B$, we consider the boundary condition $\phi=0$ at $x_3=H/2$ and $\phi={u_3}_{top}$ at $x_3=L_0/2$. Substituting the boundary conditions in equation (\ref{A13}) and solving for the constants $A$ and $B$ we get:

\begin{subeqnarray}
\slabel{A14a}
A & = & {{-u_3}_{top}\frac{ \mathrm{exp}\left({-q H/{2}}\right)}{\mathrm{exp}\left({q (H-L_0)/{2}}\right)-\mathrm{exp}\left({-q (H-L_0)/{2}}\right)} },\\[3pt]
\slabel{A14b}
B & = & {{u_3}_{top}\frac{ \mathrm{exp}\left({q H/{2}}\right)}{\mathrm{exp}\left({q (H-L_0)/{2}}\right)-\mathrm{exp}\left({-q (H-L_0)/{2}}\right)} } .
\end{subeqnarray}

After substituting the constants $A$ and $B$ from equations (\ref{A14a}) and (\ref{A14b}) respectively into equation (\ref{A13}), we get:
\begin{equation}
\label{A15}
\phi\left(x_3\geq\frac{L_0}{2}\right) = -{u_3}_{top}\frac{\mathrm{exp}\left({q (x_3-H/2)}\right)-\mathrm{exp}\left({-q (x_3-H/2)}\right)}{\mathrm{exp}\left({q (H-L_0)/{2}}\right)-\mathrm{exp}\left({-q (H-L_0)/{2}}\right)} .
\end{equation}

We can write the exponential terms in the form of hyperbolic trigonometric functions as:

\begin{equation}
\label{A16}
\mathrm{sinh}(a)=\frac{\mathrm{exp}(a)-\mathrm{exp}(-a)}{2} .
\end{equation}

Therefore, we can write,
\begin{equation}
\label{A17}
\phi \left( x_3 \geq \frac{L_0}{2} \right)  =  -{u_3}_{top}\frac{\sinh{\left( q \left(x_3-\frac{H}{2} \right) \right)}}{\sinh{\left(q \frac{\left( H-L_0 \right)}{2} \right)} } .
\end{equation}

Now, substituting the value of $q$ in above equation (\ref{A17}) we get the final solution as follows:
\begin{equation}
\label{A18}
\phi \left( x_3 \geq \frac{L_0}{2} \right)  =  -{u_3}_{top} \frac{ \sinh{\left( K\sqrt{\frac{\varOmega^2}{\varOmega^2 - f^2}} \left(x_3-\frac{H}{2} \right) \right)}} {\sinh{\left( K\sqrt{\frac{\varOmega^2}{\varOmega^2 - f^2}} \left(\frac{H-L_0}{2} \right) \right)}}.
\end{equation}

We follow the above same procedure to find out the solution for equation (\ref{phi eqn}) in the bottom pure fluid region from $x_3=-L_0/2$ to $x_3=-H/2$. In this region, we consider the vertical gradient of mean concentration $\varGamma=0$, thus $N=\sqrt{-2\mathcal{A} g_0 \varGamma}=0$ and with boundary condition $\phi={u_3}_{bot}$ at $x_3=-L_0/2$ and $\phi=0$ at $x_3=-H/2$, we get the solution of equation (\ref{phi eqn}) as follows:

\begin{equation}
\label{A19}
\phi \left( x_3 \leq -\frac{L_0}{2} \right)  =  {u_3}_{bot} \frac{ \sinh{\left( K\sqrt{\frac{\varOmega^2}{\varOmega^2 - f^2}} \left(x_3+\frac{H}{2} \right) \right)}} {\sinh{\left( K\sqrt{\frac{\varOmega^2}{\varOmega^2 - f^2}} \left(\frac{H-L_0}{2} \right) \right)}} .
\end{equation}

Next, we see the solution of equation (\ref{phi eqn}) in the mixed fluid region between $x_3=L_0/2$ to $x_3=-L_0/2$ where the vertical gradient of mean concentration is $\varGamma=-N^2/2\mathcal{A}g_0$. Let $q^2={K}^2 \frac{\left( N^2 -\varOmega^2 \right)}{\left(\varOmega^2 - f^2 \right)}$ then equation (\ref{phi eqn}) can be written as 
\begin{equation}
\label{A20}
\frac{\partial^2\phi}{\partial x_3^2} + q^2  \phi = 0 .
\end{equation}

Again we assume the solution of above equation of the form $\phi=e^{rx_3}$ where $r$ is unknown constant. Substituting this solution in equation (\ref{A20}) we get $e^{rx_3}\left(r^2+q^2 \right)=0$, where $e^{rx_3}\neq0$, therefore we get the characteristic equation $r^2+q^2=0$ which has two complex roots $r=\pm iq$. Thus the general solution of equation (\ref{A20}) is
\begin{equation}
\label{A21}
\phi= {Ae^{iqx_3}+Be^{-iqx_3}} .
\end{equation}

We use the Euler's formula $e^{ia}=\mathrm{cos}(a)+i\mathrm{sin}(a)$ to write the solution in form of trigonometric terms and we get,
\begin{subequations}
\begin{equation}
\label{A22a}
\phi= A\mathrm{cos}(qx_3)+iA\mathrm{sin}(qx_3)+B\mathrm{cos}(-qx_3)+iB\mathrm{sin}(-qx_3),
\end{equation}
\begin{equation}
\label{A22b}
\phi= (A+B)\mathrm{cos}(qx_3)+i(A-B)\mathrm{sin}(qx_3) ,
\end{equation}
\begin{equation}
\label{A22c}
\phi= C\mathrm{cos}(qx_3)+iD\mathrm{sin}(qx_3) ,
\end{equation}
\end{subequations}

where $C$ and $D$ are the unknown constants. Now we apply the boundary conditions $\phi={u_3}_{top}$ at $x_3=+L_0/2$ and $\phi={u_3}_{bot}$ at $x_3=-L_0/2$ to the equation (\ref{A22c}) and solving for constants $C$ and $D$, we get
\begin{equation}
\label{A23}
C=\frac{{u_3}_{top}+{u_3}_{bot}}{2\mathrm{cos}\left(\frac{qL_0}{2}\right)};\quad D=\frac{{u_3}_{top}-{u_3}_{bot}}{2i\mathrm{sin}\left(\frac{qL_0}{2}\right)} .
\end{equation}

Substituting values of these constants in equation (\ref{A22c}) we obtain,
\begin{subeqnarray}
\slabel{A24a}
\phi_i \left( \lvert x_3 \rvert \leq \frac{L_0}{2} \right)&=&\frac{{u_3}_{top}+{u_3}_{bot}}{2\mathrm{cos}\left(\frac{qL_0}{2}\right)}\mathrm{cos}\left(qx_3\right)+i\frac{{u_3}_{top}-{u_3}_{bot}}{2i\mathrm{sin}\left(\frac{qL_0}{2}\right)}\mathrm{sin}\left(qx_3\right) ,\\
\phi_i \left( \lvert x_3 \rvert \leq \frac{L_0}{2} \right)&=&\frac{{u_3}_{top}+{u_3}_{bot}}{2\mathrm{cos}\left(\frac{qL_0}{2}\right)}\mathrm{cos}\left(qx_3\right)+\frac{{u_3}_{top}-{u_3}_{bot}}{2\mathrm{sin}\left(\frac{qL_0}{2}\right)}\mathrm{sin}\left(qx_3\right) .
\end{subeqnarray}

Substituting the value of $q=K \sqrt{\frac{N^2 -\varOmega^2}{\varOmega^2 - f^2}}$ we finally obtain,
\begin{equation}
\begin{aligned}
\label{A25}
\phi_i \left( \lvert x_3 \rvert \leq \frac{L_0}{2} \right)=& \frac{{u_3}_{top}+{u_3}_{bot}}{2\mathrm{cos}\left(K \sqrt{\frac{N^2 -\varOmega^2}{\varOmega^2 - f^2}}\frac{L_0}{2}\right)}\mathrm{cos}\left(K \sqrt{\frac{N^2 -\varOmega^2}{\varOmega^2 - f^2}}x_3\right) \\
&+ \frac{{u_3}_{top}-{u_3}_{bot}}{2\mathrm{sin}\left(K \sqrt{\frac{N^2 -\varOmega^2}{\varOmega^2 - f^2}}\frac{L_0}{2}\right)}\mathrm{sin}\left(K \sqrt{\frac{N^2 -\varOmega^2}{\varOmega^2 - f^2}}x_3\right) .
\end{aligned}
\end{equation}

\section{Derivation of the equation for $\varOmega_i$}
\label{DEO}
we differentiate equation (\ref{phi sol1}) with respect to $x_3$ which yields:
\begin{subequations}
\begin{equation}
\label{der phi1a}
    \frac{\partial{\phi_i}}{\partial x_3}\left( x_3 \geq \frac{L_0}{2} \right)=  -{u_3}_{top}\frac{K Y_i \cosh{\left( K Y_i \left(x_3-\frac{H}{2} \right) \right)}}{\sinh{\left( K Y_i \left(\frac{H-L_0}{2} \right) \right)} },
\end{equation}

\begin{equation}
\label{der phi1a1}
    \frac{\partial{\phi_i}}{\partial x_3}\left( x_3 \geq \frac{L_0}{2} \right)=  -{u_3}_{top}\frac{K Y_i \cosh{\left( K Y_i \left(\frac{H}{2}-x_3 \right) \right)}}{\sinh{\left( K Y_i \left(\frac{H-L_0}{2} \right) \right)} },
\end{equation}

here we have used the trigonometric hyperbolic function identity $\cosh{(-\theta_i)}=\cosh{(\theta_i)}$. Further calculating the above derivative of $\phi_i$ at ${x_3}=L_0/2$ we get
\begin{equation}
\label{der phi1b}
    \left(\frac{\partial{\phi_i}}{\partial x_3}\right)_{x_3=L_0/2}=  -{u_3}_{top}\frac{K Y_i }{\tanh{\left( K Y_i \left(\frac{H-L_0}{2} \right) \right)} } .
\end{equation}
\end{subequations}
 
 Differentiating the equation (\ref{phi sol3}) with respect to $x_3$ we get
 \begin{subequations}
 \begin{equation}
 \label{der phi3a}
 \frac{\partial{\phi_i}}{\partial x_3} \left( \lvert x_3 \rvert \leq \frac{L_0}{2} \right)   =  \frac{{u_3}_{top}-{u_3}_{bot}}{2\sin{\left(K X_i \frac{L_0}{2} \right)} } K X_i \cos{\left(K X_i x_3 \right)} + \frac{{u_3}_{top}+{u_3}_{bot}}{2\cos{\left(K X_i\frac{L_0}{2} \right)} } \left(-K X_i \right) \sin{\left(K X_i x_3 \right)} ,
 \end{equation}
 
 again calculating this derivative at $\phi_i$ at ${x_3}=L_0/2$ we get
 \begin{equation}
 \label{der phi3b}
 \left(\frac{\partial{\phi_i}}{\partial x_3}\right)_{x_3=L_0/2} = \left({u_3}_{top}-{u_3}_{bot}\right) \frac{K X_i}{2\tan{\left(K X_i \frac{L_0}{2} \right)}} - \left({u_3}_{top}+{u_3}_{bot}\right) \frac{K X_i}{2} \tan{\left(K X_i \frac{L_0}{2} \right)} .
 \end{equation}
 \end{subequations}

We equate the equations (\ref{der phi1b}) and (\ref{der phi3b}) for continuity of $\partial \phi_i/\partial x_3$ at $x_3=L_0/2$. First we compare the coefficients of ${u_3}_{top}$ and get:
\begin{subequations}
\begin{equation}
 \label{omega eq1a}
 \frac{-K Y_i}{\tanh{\left(K Y_i \frac{H-L_0}{2}\right)}} = \frac{K X_i}{2\tan\left(K X_i\frac{L_0}{2}\right)}-\frac{K X_i}{2} \tan\left(K X_i\frac{L_0}{2}\right) ,
 \end{equation}
 
 then we compare the coefficients of ${u_3}_{bot}$ and get:
 \begin{equation}
 \label{omega eq1b}
 0 = -\frac{K X_i}{2\tan\left(K X_i\frac{L_0}{2}\right)}-\frac{K X_i}{2} \tan\left(K X_i\frac{L_0}{2}\right) ,
 \end{equation}
 \begin{equation}
 \label{omega eq1c}
 \frac{1}{\tan\left(K X_i\frac{L_0}{2}\right)}=-\tan\left(K X_i\frac{L_0}{2}\right) .
 \end{equation}
 
 Substituting the above equation (\ref{omega eq1c}) in (\ref{omega eq1a}) we get
 \begin{equation}
 \label{omega eq1d}
 \frac{-Y_i}{\tanh{\left(K Y_i \frac{H-L_0}{2}\right)}} = \frac{-X_i }{2}\tan\left(K X_i\frac{L_0}{2}\right)-\frac{X_i}{2} \tan\left(K X_i\frac{L_0}{2}\right) ,
 \end{equation}
 
 \begin{equation}
 \label{omega eq1e}
 \frac{-Y_i}{\tanh{\left(K Y_i \frac{H-L_0}{2}\right)}} = - X_i \tan\left(K X_i\frac{L_0}{2}\right) ,
 \end{equation}
 
 rearranging the terms we get,
 \begin{equation}
 \label{omega eq1f}
 \tan\left(K X_i\frac{L_0}{2}\right) = \frac{Y_i}{X_i}\frac{1}{\tanh{\left(K Y_i \frac{H-L_0}{2}\right)}} .
 \end{equation}
 \end{subequations}
 
 Substituting the $X_i$ and $Y_i$ from equation (\ref{XiYi eq}) in above equation (\ref{omega eq1e}), we get
 
 \begin{equation}
 \label{omega eqA1}
 \tan\left(K \sqrt{\frac{N^2 - \varOmega_i^2}{\varOmega_i^2 -f^2}}\frac{L_0}{2} \right) =\sqrt{\frac{\varOmega_i^2}{N^2-\varOmega_i^2}} \frac{1}{ \tanh{\left(K \sqrt{\frac{\varOmega_i^2}{\varOmega_i^2 -f^2}} \left(\frac{H-{L}_0}{2} \right) \right)}} .
 \end{equation}
 
 Now we substitute 
 \begin{subequations}
  \begin{equation}
 \label{omega eq2a}
 -\tan\left(K X_i\frac{L_0}{2}\right)=\frac{1}{\tan\left(K X_i\frac{L_0}{2}\right)} ,
  \end{equation}
  
  from equation (\ref{omega eq1c}) in equation (\ref{omega eq1a}) and we get
 \begin{equation}
 \label{omega eq2b}
 \frac{-Y_i}{\tanh{\left(K Y_i \frac{H-L_0}{2}\right)}} = \frac{X_i}{2\tan\left(K X_i\frac{L_0}{2}\right)}+\frac{X_i}{2\tan\left(K X_i\frac{L_0}{2}\right)} ,
 \end{equation}
 
 rearranging the terms we get,
 \begin{equation}
 \label{omega eq2c}
 \tan\left(K X_i\frac{L_0}{2}\right) = -\frac{X_i}{Y_i} \tanh{\left(K Y_i \frac{H-L_0}{2}\right)} .
 \end{equation}
 \end{subequations}
 
 Further, substituting the $X_i$ and $Y_i$ from equation (\ref{XiYi eq}) in equation (\ref{omega eq2b}) we get
 \begin{equation}
 \label{omega eqA2}
 \tan\left(K \sqrt{\frac{N^2 - \varOmega_i^2}{\varOmega_i^2 -f^2}}\frac{L_0}{2} \right)  = -\sqrt{\frac{N^2-\varOmega_i^2}{\varOmega_i^2}}  \tanh{\left(K \sqrt{\frac{\varOmega_i^2}{\varOmega_i^2 -f^2}} \left(\frac{H-{L}_0}{2} \right) \right)} .
 \end{equation}

\section{Derivation of equation \ref{phiAi eq31}}
\label{phiAi eq31A}
 
Substituting equation (\ref{phi eqn}) into equation (\ref{phiAi eq1}) yields

 \begin{equation}
 \begin{aligned}
  \label{phiAi eq21}
  \sum_{i} \frac{\partial^2}{\partial t^2} \left( -A_i K^2 \left(\frac{ N^2 -\varOmega_i^2 }{\varOmega_i^2 - f^2}\right)\phi_i - K^2 \phi_i A_i \right) = & \sum_{i}f^2 A_i K^2 \left(\frac{ N^2 -\varOmega_i^2 }{\varOmega_i^2 - f^2}\right) \phi_i \\
  &-  \sum_{i}2\mathcal{A} g(t) \varGamma K^2 A_i \phi_i ,
 \end{aligned}
 \end{equation}
 
 rearranging the terms we get,
 \begin{equation}
 \begin{aligned}
  \label{phiAi eq22}
  \sum_{i} - K^2 \left(\frac{ N^2 -\varOmega_i^2 }{\varOmega_i^2 - f^2}+1\right)\phi_i  \frac{\partial^2 A_i}{\partial t^2} = & \sum_{i}f^2 A_i K^2 \left(\frac{ N^2 -\varOmega_i^2 }{\varOmega_i^2 - f^2}\right)\phi_i \\
  &-  \sum_{i}2\mathcal{A} g(t) \varGamma K^2 A_i \phi_i .
  \end{aligned}
 \end{equation}

 Taking the $K^2$ and $\phi_i$ terms common from both sides of the above equation (\ref{phiAi eq22}) which cancels out each other, and after arranging the terms, we get 
 \begin{equation}
 \begin{aligned}
  \label{phiAi eq23}
  \sum_{i} -\left(\frac{ N^2 -f^2 }{\varOmega_i^2 - f^2}\right)\frac{\partial^2 A_i}{\partial t^2}
  =  \sum_{i} \left( f^2 \left(\frac{ N^2 -\varOmega_i^2 }{\varOmega_i^2 - f^2}\right) - 2\mathcal{A} g(t) \varGamma \right) A_i .
  \end{aligned}
 \end{equation}
 
 Further rearranging the terms of the above equation (\ref{phiAi eq23}) and doing some simplifications, we get equation \ref{phiAi eq31}.

\section{Derivation of equation \ref{omega eq}}
\label{omega eqAP}
First we rearrange the equation (\ref{m eq}) as follows:
  \begin{equation}
    \label{m eq1}
    \varOmega_i^2-f^2 =\frac{K^2}{m^2}\left(N^2-\varOmega_i^2\right) .
 \end{equation}
 
Then we substitute the definition of angle $\theta_i$ in equation (\ref{m eq1}) and obtain the characteristic frequency $\varOmega_i$ in terms of the angle $\theta_i$ as follows:  
 \begin{subequations}
 \begin{equation}
    \label{m eq2}
    \varOmega_i^2-f^2 =\tan^2(\theta_i)\left(N^2-\varOmega_i^2\right) ,
 \end{equation}
 \begin{equation}
    \label{m eq3}
    \varOmega_i^2\left(1+\tan^2(\theta_i)\right) = N^2\tan^2(\theta_i)+f^2 ,
 \end{equation}
 
 using the trigonometric identity $1+\tan^2(\theta_i)=\sec^2(\theta_i)=1/\cos^2{(\theta_i)}$  in above equation (\ref{m eq3}) and rearranging the terms we get
 \begin{equation}
    \label{m eq4}
    \varOmega_i^2 = \frac{N^2\tan^2(\theta_i)+f^2}{1/\cos^2{(\theta_i)}} ,
 \end{equation}
 \begin{equation}
    \label{m eq5}
    \varOmega_i^2 = N^2\frac{\sin^2(\theta_i)}{\cos^2(\theta_i)}\cos^2(\theta_i) + f^2 \cos^2(\theta_i) .
 \end{equation}
 \end{subequations}
Rearranging equation \ref{m eq5} will give equation \ref{omega eq}. 

\section{Derivation of equation \ref{a0 eq28}}
\label{a0 eq28A}

 \begin{equation}
     \label{a0 eq21}
     \frac{\partial^2 a_i}{\partial t^2} + \left( N^2\sin^2{(\theta_i)}+f^2\cos^2{(\theta_i)} + \underbrace{ \frac{N^2\left( N^2\sin^2{(\theta_i)}+f^2\cos^2{(\theta_i)} - f^2\right)}{N^2 - f^2}}_{\displaystyle(\mathrm{I})} F \cos{(\omega t)} \right) a_i = 0 .
 \end{equation}
 
 Now we simplify the term $(\mathrm{I})$ of above equation (\ref{a0 eq21}) as follows:
\begin{subequations}
  \begin{equation}
     \label{a0 eq22}
    (\mathrm{I}) = \frac{N^2\left( N^2\sin^2{(\theta_i)}+f^2\cos^2{(\theta_i)} - f^2\right)}{N^2 - f^2} ,
 \end{equation}
 \begin{equation}
     \label{a0 eq23}
    (\mathrm{I}) = \frac{N^2\left( N^2\sin^2{(\theta_i)}+f^2\cos^2{(\theta_i)} + \left(f^2\sin^2{(\theta_i)}-f^2\sin^2{(\theta_i)} \right) - f^2\right)}{N^2 - f^2} ,
 \end{equation}
  \begin{equation}
     \label{a0 eq24}
    (\mathrm{I}) = \frac{N^2\left( N^2\sin^2{(\theta_i)}+f^2\left(\cos^2{(\theta_i)} + \sin^2{(\theta_i)}\right)-f^2\sin^2{(\theta_i)}- f^2\right)}{N^2 - f^2} ,
 \end{equation}
 
 using the Pythagorean trigonometric identity $\cos^2{(\theta_i)} + \sin^2{(\theta_i)} =1$ , we get
 \begin{equation}
     \label{a0 eq24}
    (\mathrm{I}) = \frac{N^2\left( N^2\sin^2{(\theta_i)}+f^2-f^2\sin^2{(\theta_i)}-f^2\right)}{N^2 - f^2} ,
 \end{equation}
 \begin{equation}
     \label{a0 eq25}
    (\mathrm{I}) = \frac{N^2\left( N^2\sin^2{(\theta_i)}-f^2\sin^2{(\theta_i)}\right)}{N^2 - f^2} .
 \end{equation}
\end{subequations}

Now we substitute the term $(\mathrm{I})$ from above equation (\ref{a0 eq25}) in equation (\ref{a0 eq21}), and we get

\begin{equation}
     \label{a0 eq26}
     \frac{\partial^2 a_i}{\partial t^2} + \left( N^2\sin^2{(\theta_i)}+f^2\cos^2{(\theta_i)} + \frac{N^2\left( N^2\sin^2{(\theta_i)}-f^2\sin^2{(\theta_i)}\right)}{N^2 - f^2} F \cos{(\omega t)} \right) a_i = 0 .
 \end{equation}
 
 After doing some simplifications, we get
 \begin{subequations}

 \begin{multline}
 \label{a0 eq27}
     \frac{\partial^2 a_i}{\partial t^2} + \left( \frac{f^2 \left(N^2-f^2\right)\cos^2{(\theta_i)}}{N^2 - f^2} \right.\\
  \left.
  +  \frac{\left(N^2\right)^2\sin^2{(\theta_i)}\left(1+F \cos{(\omega t)} \right) - f^2 N^2\sin^2{(\theta_i)}\left(1+F \cos{(\omega t)} \right)}{N^2 - f^2} \right) a_i = 0 ,
 \end{multline}
 
 \begin{equation}
 \label{a0 eq27}
     \frac{\partial^2 a_i}{\partial t^2} + \left( f^2 \cos^2{(\theta_i)} + \frac{N^2 \left(N^2-f^2\right)\sin^2{(\theta_i)}}{{N^2 - f^2}}\left(1+F \cos{(\omega t)}\right) \right) a_i = 0 .
 \end{equation}
 \end{subequations}
After rearrangement of equation \ref{a0 eq27} we obtain equation \ref{a0 eq28}.

\section{Floquet Analysis}\label{appA} 
In this appendix, we briefly discuss the theorems related to the Floquet theory, and provide the steps to solve the Mathieu equation (\ref{a0 eq3}) by using these theorems. Proofs of the theorems and solution steps are given in \citet[pp. 308-317]{jordan2007nonlinear}. 
\newtheorem{theorem}{Theorem}
\begin{theorem}[Floquet's theorem]
\label{theorem1}
Let $\frac{\mathrm{d}\vec{x}}{\mathrm{d}t}=A(t)\vec{x}$, be a first order linear differential system, where $A(t)$ is $T$-periodic matrix such that $A(t+T)=A(t), \forall \:t$. This system has at least one non-trivial solution $\vec{x}=\vec{\chi}(t)$ such that 
\begin{equation}
\label{theo 1}
 \vec{\chi}(t+T)=\mu \vec{\chi}(t),\quad \forall \:t 
\end{equation}

where $\mu$ is a \textbf{characteristic number} or \textbf{Floquet multiplier}.
\end{theorem}

If $\Phi(t)$ is a fundamental solution matrix of system $\vec{\dot x}=A(t) \vec{x}$, then $\Phi(t+T)$ is also a fundamental matrix, and there exists a non-singular matrix $C$ such that, 
\begin{subeqnarray}
\slabel{matrix C1}
\Phi(t+T)&=&C\Phi(t),\quad \forall \:t\\
\slabel{matrix C2}
C&=&\Phi^{-1}(t)\:\Phi(t+T).
\end{subeqnarray}

where \textbf{characteristic numbers}, $\mu'\text{s}$, are the eigenvalues of $C$. We can obtain the matrix $C=\Phi(T)$ for initial conditions at $t=0$ such that $\Phi(0)={\boldsymbol I}$, where ${\boldsymbol I}$ is the identity matrix.
\begin{theorem}
\label{theorem2}
For system $\vec{\dot x}=A(t) \vec{x}$, where $A(t)$ is $T$-periodic, with \textbf{characteristic numbers} $\mu_1,\mu_2,\dots\mu_n$, the product of the \textbf{characteristic numbers} is obtained as: 
\begin{equation}
\label{theo 2}
 \mu_1\mu_2\dots\mu_n=\mathrm{exp}\left(\int_0^T \mathrm{tr} \{A(t)\}\right).
\end{equation}
\end{theorem}

Now, we define the \textbf{characteristic exponent}, $\rho$ of the system as $e^{\rho T}=\mu$.
\begin{theorem}
\label{theorem3}
Let $C$ has $n$ distinct eigenvalues, $\mu_i$ and corresponding $\rho_i$, i=1,2,...,n. Then system $\vec{\dot x}=A(t) \vec{x}$ has $n$ linearly independent solutions of the form:
\begin{equation}
\label{theo 3}
\vec{\chi_i}(t)=e^{\rho_i t}\vec{P_i}(t)
\end{equation}

here $\vec{P_i}(t)$ are the periodic vector functions with period $T$.
\end{theorem}
Clearly, \textbf{characteristic exponents} $e^{\rho_i t}$, will determine the behaviour of the solutions (\ref{theo 3}).

Now, we apply above theorems to study the nature of solutions of mathieu equation (\ref{a0 eq3}). We can rewrite equation (\ref{a0 eq3}) as: 
\begin{equation}
     \label{mathieu 1}
     \Ddot{a}+\left( \gamma +\alpha + \beta \cos{\tau} \right)a=0,
 \end{equation}
 
 where, 
\begin{equation}
     \label{mathieu var}
     \gamma = \frac{f^2 \cos^2{(\theta)}}{\omega^2};\quad \alpha = \frac{N^2 \sin^2{(\theta)}}{\omega^2};\quad \beta = F\frac{N^2 \sin^2{(\theta)}}{\omega^2}.
\end{equation}
 
 We begin by defining $a=X$ and $\Dot{a}=Y$, so equation (\ref{mathieu 1}) can be expressed as a first-order system:
\begin{equation}
\setlength{\arraycolsep}{5pt}
\renewcommand{\arraystretch}{1.3}
\left[
\begin{array}{ccccc}
  \Dot{X}    \\
  \displaystyle
    \Dot{Y}   \\
 \end{array}  \right] = \left[
\begin{array}{ccccc}
  0 & 1   \\
  \displaystyle
    -(\gamma+\alpha+\beta\cos{\tau})  &  0   \\
 \end{array}  \right] \left[
\begin{array}{ccccc}
  X    \\
  \displaystyle
    Y   \\
 \end{array}  \right].
\label{system}
\end{equation}

In the notation of Theorem \ref{theorem1},
\begin{equation}
\setlength{\arraycolsep}{5pt}
\renewcommand{\arraystretch}{1.3}
A(\tau) = \left[
\begin{array}{ccccc}
  0 & 1   \\
  \displaystyle
    -(\gamma+\alpha+\beta\cos{\tau})  &  0   \\
 \end{array}  \right] .
\label{matrix A}
\end{equation}

Here, $A(\tau)$ is periodic with period $2\pi$ and $\mathrm{tr}\{A(\tau)\}=0$. As our system is of size $2\times2$, so we have two \textbf{characteristic numbers} $\mu_1$ and $\mu_2$. Therefore, from Theorem \ref{theorem2} 
\begin{equation}
\label{product mu}
\mu_1\mu_2=\mathrm{e}^0=1 .
\end{equation}

For the following initial condition
\begin{equation}
\setlength{\arraycolsep}{5pt}
\renewcommand{\arraystretch}{1.3}
\Phi(0) = \left[
\begin{array}{ccccc}
  1 & 0   \\
  \displaystyle
    0  &  1   \\
 \end{array}  \right] ,
\label{matrix phi}
\end{equation}

 we obtain the matrix $C=\Phi(T)$ by numerically integrating (with MATLAB$^{\circledR}$) the system (\ref{system}) from $\tau=0$ to $\tau=2\pi$. The eigen values or \textbf{Characteristic numbers} are the solutions of quadratic characteristic equation of $C$
\begin{equation}
\label{cha eq1}
\mu^2-(\text{sum of roots})\mu + \text{(product of roots)}=0,
\end{equation}

 and using the equation (\ref{product mu}), we get
\begin{equation}
\label{cha eq2}
\mu^2-\phi(\gamma,\alpha,\beta)\mu + 1=0,
\end{equation}

here, $\phi(\gamma,\alpha,\beta)$ represents the sum of roots.
The solutions $\mu_1$ and $\mu_2$ of the equation (\ref{cha eq2}) are 
\begin{equation}
\label{roots}
\mu_1,\mu_2=\frac{1}{2}\left(\phi\pm\sqrt{\left(\phi^2-4\right)}\right).
\end{equation}

Thus, different values of $\phi$ will determine the behaviour of solutions as follows:
\begin {enumerate}
\item When $\phi>2$: $\mu_1$, $\mu_2$ are both real and positive with $\mu_1\mu_2=1$ (from equation \ref{product mu}), one of them, say $\mu_1>1$ and $\mu_2<1$. The corresponding \textbf{characteristic exponents} are real and have the form $\rho_1=\sigma>0$, $\rho_2=-\sigma<0$ (because $\rho_2=\ln({\mu_2})/T<0$). So, the general solution from Theorem \ref{theorem3} can be written as: 
\begin{equation}
\label{phi gt2}
X(\tau)=c_1e^{\sigma \tau}P_1(\tau)+c_2e^{-\sigma t}P_2(\tau),
\end{equation} 

 where $\sigma>0$, thus $\mathrm{e}^{\sigma \tau}$ corresponds to the exponential growth in time and solution becomes \textbf{unstable} with \textbf{harmonic} in nature.\\
\item When $\phi<-2$: $\mu_1$, $\mu_2$ are both real and negative with $\mu_1\mu_2=1$. The form of general solution is: 
\begin{equation}
\label{phi lt2}
X(\tau)=c_1e^{\left(\sigma+\frac{1}{2}\mathrm{i}\right) \tau}P_1(\tau)+c_2e^{\left(-\sigma +\frac{1}{2}\mathrm{i}\right) \tau}P_2(\tau).
\end{equation} 

Here, $\mathrm{e}^{\sigma \tau}$ leads to the exponential growth of solution and eventually becomes \textbf{unstable}, with \textbf{sub-harmonic} response.\\
\item When $-2<\phi<+2$: $\mu_1$, $\mu_2$ are complex, we have $\rho_1=\mathrm{i}\nu$ and $\rho_2=-\mathrm{i}\nu$, thus we get the solution  
\begin{equation}
\label{phi bt 2,-2}
X(\tau)=c_1e^{\left(\mathrm{i}\nu\right) \tau}P_1(\tau)+c_2e^{\left(-\mathrm{i}\nu\right) \tau}P_2(\tau).
\end{equation} 

The solutions are bounded at all times and oscillatory but not necessarily periodic, and lies in \textbf{stable parameter region}. \\
\item When $\phi=+2$: \:$\mu_1=\mu_2=1$, one solution is periodic with period $2\pi$ and lies on the \textbf{harmonic tongue} of \textbf{stability curve} $\phi=2$ .\\
\item When $\phi=-2$: \:$\mu_1=\mu_2=-1$, one solution is periodic with period $4\pi$ and lies on the \textbf{sub-harmonic tongue} of \textbf{stability curve} $\phi=2$ .
\end {enumerate}

\section{WKB \textbf{(Wentzel–Kramers–Brillouin}) approximation of equation \ref{mphi eq} }\label{appB}
%u3 fluc12 
We consider the equation (\ref{m eq}) and assume that $N(x_3)=\sqrt{-2\mathcal{A}g_0\left(\partial \langle C \rangle_H/\partial x_3\right)}$ (see equation (\ref{strat fre})) is a slowly varying function such that its fractional change over a vertical wavelength is much less than unity, therefore $m(x_3)$ is also a slowly varying function. With such slowly varying properties of the medium (here $N$), we can approximate the WKB solution \citep[for details see][pp. 743-745]{kundu2015fluid} of the equation (\ref{mphi eq}) by assuming a solution of the form
\begin{equation}
 \label{app b1}
 \phi=A(x_3)\mathrm{e}^{i\psi(x_3)},
 \end{equation}
 
where $A$ represents the amplitude and $\psi$ the phase. We substitute this solution in the equation (\ref{mphi eq}) and equate the real and imaginary parts, and after doing some simplifications \citep[see][pp. 743-745]{kundu2015fluid}, we get 
\begin{equation}
 \label{app b2}
 A = \frac{A_0}{\sqrt{m}}\;; \quad \psi=\pm i\int_{}^{x_3} m \mathrm{d}x_3,
 \end{equation}
 
 where $A_0$ is a constant. Therefore, the WKB solution becomes
\begin{equation}
 \label{app b2}
 \phi=\frac{A_0}{\sqrt{m}}\mathrm{e}^{\pm \int_{}^{x_3} m \mathrm{d}x_3}.
 \end{equation}
 
 Substitution of this WKB solution (\ref{app b2}) in the assumed solution of $u_3$ (from equation (\ref{u3 sol1})) yields: 
 \begin{equation}
     \label{app b3}
     u_3=\frac{A_0}{\sqrt{m}}\mathrm{e}^{i \left(kx_1+lx_2 \pm \int_{}^{x_3} m \mathrm{d}x_3-\Omega t \right)}.
 \end{equation}
 
 Now, taking real parts of the above equation (\ref{app b3}), we obtain the vertical velocity 
 \begin{equation}
     \label{app b4}
     u_3=\frac{A_0} {\sqrt{m}}\cos{\left(kx_1+lx_2\pm\int_{}^{x_3} m \mathrm{d}x_3 -\Omega t\right)}.
 \end{equation}

Here, the argument of $\cos{()}$ represents the ``phase'' and $\partial(phase)/\partial{x_3}=m$, therefore $m$ is the local vertical wavenumber in vertical direction $x_3$ and we can define the dispersion relation as:
\begin{equation}
 \label{app b5}
 m^2=\frac{k^2\left(N^2-\varOmega^2\right)}{\left(\varOmega^2-f^2\right)}.
 \end{equation}
 
 \bibliographystyle{jfm}
\bibliography{jfm}

%\bibliographystyle{jfm}
%\bibliography{jfm}
%Use of the above commands will create a bibliography using the .bib file. Shown below is a bibliography built from individual items.
%\begin{thebibliography}{99}

%\expandafter\ifx\csname natexlab\endcsname\relax
%\def\natexlab#1{#1}\fi
%\expandafter\ifx\csname selectlanguage\endcsname\relax
%\def\selectlanguage#1{\relax}\fi

%\bibitem[Batchelor (1971)]{Batchelor59}
%{\sc Batchelor, G.K.} 1971 {Small-scale variation of convected quantities like temperature in turbulent fluid part1, general discussion and the case of small conductivity}, {\it J. Fluid Mech.}, {\bf 5}, pp. 3-113-133.

%\end{thebibliography}

% End of file `jfm2esam.bib'.

\end{document}